\documentclass[12pt]{article} 
\usepackage{amssymb}
\usepackage{mathrsfs}
\usepackage{amsmath,amsthm} 
\usepackage{cprotect}
\usepackage{enumitem}
\usepackage{cite}
\usepackage{youngtab}
\usepackage[colorlinks=true, linkcolor=blue,
  citecolor=blue, urlcolor=blue]{hyperref}
\usepackage[capitalize,nameinlink]{cleveref}  
\input xy
\xyoption{all}

\usepackage{color}

\newcommand{\p}{\partial}

\theoremstyle{plain}
\newtheorem{thm}{Theorem}[section]
\newtheorem{defn}{Definition}[section]
\newtheorem{conj}{Conjecture}[section]
\newtheorem{lem}[thm]{Lemma}
\newtheorem{prop}[thm]{Proposition}
\newtheorem{cor}[thm]{Corollary}

\newtheorem{remark}[thm]{Remark}

\numberwithin{equation}{section}

\numberwithin{table}{section}

\begin{document}

\begin{center}

{\large\bf Quantum $K$-theory levels in physics and math}

\vspace*{0.2in}

Irit Huq-Kuruvilla$^1$, Leonardo Mihalcea$^1$, Eric Sharpe$^2$, Hao Zhang$^2$

\begin{tabular}{cc}
{\begin{tabular}{l}
$^1$ Department of Mathematics, MC 0123\\
225 Stanger Street\\
Virginia Tech\\
Blacksburg, VA  24061
\end{tabular}}
&
{\begin{tabular}{l}
$^2$ Department of Physics MC 0435 \\
850 West Campus Drive \\
Virginia Tech \\
Blacksburg, VA 24061 
\end{tabular}}
\end{tabular}

{\tt irithk@vt.edu}, {\tt lmihalce@vt.edu}, {\tt ersharpe@vt.edu},
{\tt hzhang96@vt.edu}

\end{center}

The purpose of this paper is to describe  a 
dictionary between Chern-Simons levels in three-dimensional 
gauged linear sigma models (GLSMs) and twistings of quantum $K$-theory in mathematics,
including as special cases the (coincidentally-named) 
Ruan-Zhang levels. 
Each defines a twisting of quantum $K$-theory, and our proposed 
dictionary identifies these two twistings, in the cases of projective spaces, smooth Fano toric varieties, Grassmannians, and flag manifolds.
We verify the dictionary
by realizing the Coulomb branch equations
as symbols of certain difference operators annihilating a
twisted version of the $I$ function associated to the 
abelianized GLSM theory, and also by comparing the geometric window for Chern-Simons levels to an analogous window in mathematics.
In the process, we interpret the geometric window for the Chern-Simons levels in terms of equalities of $I$ and $J$ functions.
This provides a fuller mathematical understanding of some special cases in the physics literature.
We also make conjectures for twisted quantum $K$-theory of gerbes, following up earlier conjectures on ordinary quantum $K$-theory of gerbes.

\begin{flushleft}
September 2026
\end{flushleft}

\newpage

\tableofcontents

\newpage

\section{Introduction}

Quantum $K$-theory can be realized physically as operator product relations of
parallel Wilson lines in three-dimensional Chern-Simons matter theories,
computed as a symmetrization of Coulomb branch relations in KK reductions to two dimensions
\cite{Bullimore:2014awa,Jockers:2018sfl,Jockers:2019wjh,Jockers:2019lwe,Jockers:2021omw,Ueda:2019qhg,Koroteev:2017nab,Closset:2016arn,Closset:2017zgf,Closset:2018ghr,Closset:2019hyt,Gu:2020zpg,Gu:2022yvj,Gu:2023tcv,Gu:2023fpw,Closset:2023vos,Closset:2023bdr}.
Now, both the physical theory and the mathematical theory can be twisted:
\begin{itemize}
    \item In physics, we can twist by the Chern-Simons level. We recover the Coulomb branch relations corresponding to conventional quantum $K$-theory relations for one particular value of the Chern-Simons level, and one can choose different values of the Chern-Simons level, within a range, the `geometric window.'   Values of the level outside the geometric window describe a different geometry, one decorated by Coulomb (topological) vacua. 

    \item In mathematics, there exist twists of quantum $K$-theory (defined in \cite{cg,Tonita,wallcrossing}, see section~\ref{sss:twistings}), that correspond to considering $K$-theoretic 
    Gromov-Witten invariants with a $K$-theoretic characteristic class ${\mathfrak C}(E)$ for some element of $K$-theory $E$.  
    Certain choices of these twists correspond to the Chern-Simons levels in physics.

    In particular, Chern-Simons levels corresponding to abelian gauge factors correspond to (coincidentally named) Ruan-Zhang level structures \cite{rz,rwz}, defined by letting $\mathfrak{C}$ equal the class $\det ^{-\ell}$ for an integer $\ell$. 
\end{itemize}
See also section~\ref{sss:homog-bundles} for another statement of this conjecture.

It has previously been suggested (see e.g.~\cite{Jockers:2018sfl,Ueda:2019qhg}) 
that in some cases these two twistings are related. For Grassmannians, a 
dictionary was previously discussed in \cite[section 4.2]{Ueda:2019qhg}.
In this paper we use 
the abelian/nonabelian correspondence in quantum K theory to propose a specific dictionary
between Chern-Simons levels and (mathematical) twistings for smooth Fano GIT quotients,
which we confirm 
in the cases of projective spaces, Grassmannians, flag manifolds,
and smooth Fano toric varieties, extending the expectations above.
(We also discuss conjectures for twisted quantum K theory of gerbes.)

Let us outline this in a little more detail.
Consider a Fano GIT quotient $X = [V // G]$, where $V$ is a vector space and $G$ a reductive algebraic group.
Let $A \subset G$ be a maximal torus.  

In physics, this is described by a gauge theory, with gauge group given by the maximal compact subgroup of the reductive algebraic group $G$.  (In the rest of this paper, we will use $G$ to denote both the reductive algebraic group appearing in algebro-geometric descriptions, and also its maximal compact group, appearing as the gauge group in the physical gauge theory.)  There is a symmetric matrix of Chern-Simons levels $\kappa_{\rm eff}^{ab}$
where $a, b \in \{1,\cdots, \dim A\}$, indexing gauge group factors, and constrained by the Weyl group.
This matrix corresponds to
$G$-invariant quadratic forms on 
$\mathfrak{g}^*$, i.e., elements of $Sym^2(\mathfrak{g}^*)^G$, 
where $\mathfrak{g}^*$ is the dual of the Lie algebra of $G$. 
Furthermore, in our case $G$ is a product of $GL(k)$'s, giving standard and adjoint representations; see section \ref{sec:csmath}. 
Each of these representations determines bundles on both $V//G$ 
and $V//A$, which are used to define twisted Gromov-Witten theories on the
corresponding quasimap spaces, as explained in section \ref{sss:homog-bundles}.

Briefly, we will propose a specific dictionary identifying
each choice of Chern-Simons level, within a range of values (the `geometric window' for which the Higgs branch is $X$, without any extra Coulomb vacua), with a choice of twisting in quantum $K$-theory.

Under this correspondence, the Coulomb branch equations correspond to symbols of operators annihilating a certain abelianized $I$-function of the target GIT quotient, which is twisted by the classes corresponding to the Chern-Simons levels.

We address specific details for the case of gauge groups which split into $U(1)$ and $SU(k)$ factors. 
In this case, the resulting Chern-Simons levels come in three types: $U(1)-$levels, $U(1)\times U(1)-$levels, and $SU(k)-$levels. 
More explicitly,
for a $U(k)$ factor in $G$,
we write the Chern-Simons levels in terms of $U(1)$ and $SU(k)$ levels $\kappa_{U(1)}^{\rm eff}$ and $\kappa_{SU(k)}^{\rm eff}$ as
\begin{equation}  \label{eq:dict:su-u}
    \kappa_{\rm eff}^{ab} \: = \: \kappa_{SU(k)}^{\rm eff} \delta^{ab} \: + \:
    \frac{ \kappa_{U(1)}^{\rm eff} - \kappa_{SU(k)}^{\rm eff} }{k}.
\end{equation}
Mathematical twistings of quantum K theory are defined in definition~\ref{twistdef}.

\begin{conj}  \label{generalconj}
\begin{itemize}
    \item A $U(1)$ or $U(1)\times U(1)$ level corresponds to a (coincidentally-named) Ruan-Zhang level-structure \cite{rz,rwz} associated to the bundle coming from the largest nontrivial $U(1)-$eigenspace in the prequotient vector space, and an integer possibly depending on the other Chern-Simons levels.

 In mathematics,  we choose Ruan-Zhang level structures indexed by unordered pairs $a, b$ with $a, b \in \{1, \cdots, \dim A\}$. We specify the bundle later, but each pair $(a,b)$ is associated with an integer $\ell_{ab}$, which determines a twisting defined by taking the characteristic class ${\mathfrak C}$ defining the twisting (definition~\ref{twistdef}) to equal det$^{-\ell_{ab}}$, a power of the determinant of a certain bundle, 
 see definition~\ref{defn:level-qk}.

We conjecture that the physical Chern-Simons levels $\kappa_{\rm eff}^{ab}$ correspond to this choice of Ruan-Zhang level structures via the dictionary: 
\begin{equation}  \label{eq:level-conj}
    \kappa_{\rm eff}^{ab} \: = \: \left\{ \begin{array}{cl}
    \sum_c \ell_{ac} & a=b, 
    \\
    \ell_{ab} & a \neq b.
    \end{array} \right.
\end{equation}

    \item The $SU(k)-$level $\kappa_{SU(k)}$ is controlled by taking $\mathfrak{C}$ to be the $w$th power of the Euler class of the vector bundle associated with the sum of the roots of $SU(k)$. To be precise, there is a linear relationship between $w,\kappa_{U(1)}^{\rm eff},\kappa_{SU(k)}^{\rm eff}$, where the $U(1)$ factor comes from $GL(k)$; cf.~e.g., \S \ref{sss:homog-bundles}.
    
\end{itemize}
    
\end{conj}

In the case of toric varieties and partial flag manifolds, we 
provide mathematical statements demonstrating
this conjecture, and explicitly describe the integers appearing in the Ruan-Zhang levels. The complete mathematical proofs will be presented elsewhere.

The Grassmannians and flag manifolds may be represented in the 
form $V//G$ for $G$ of the form of a product of $U(k)$ factors. 
Then the $\kappa^{\rm eff}_{U(1)}$ Chern-Simons levels 
correspond to Ruan-Zhang levels.
(These are also sometimes known as the bare Chern-Simons level for a 
$U(1)$ theory \cite[equ'n (5.2)]{Closset:2024sle}.)
In these cases, 
\begin{equation} \label{eq:univ:grfl}   
    \left(\mbox{Ruan-Zhang level}\right) \: = \: 
    \kappa_{U(1)}^{\rm eff} \: = \:
    \left(\mbox{$U(1)$ Chern-Simons level}\right) \: + \:
    \frac{n}{2}
\end{equation}
where $n$ is the sum of the number of fundamentals and antifundamentals (charged under the given $U(1)$) defining $V$ as a representation of $G$.

We check the conjecture above in two ways:
\begin{enumerate}
    \item[(a)] Comparing Coulomb branch equations for vacua in physics to symbols of difference operators in mathematics.
    \item[(b)] Comparing `geometric windows' of allowed values of levels in both physics and mathematics, when possible.
\end{enumerate}

First, we give an overview of comparisons of Coulomb branch equations to symbols of difference operators.
The key mathematical object is the $I$ function associated to twistings
of the bundles above. This $I$-function satisfies 
difference equations $\mathcal{D}I =0$, and the symbols of 
$\mathcal{D}$ are the vacuum equations for the Coulomb 
branch of $V//G$. More precisely, the following holds. 

Let $\widehat{I}_{Ab}^\ell$ denote the twisted $I$-function 
associated to the abelian quotient $V//A$, with $\ell=(\ell_{ab})$ 
the matrix of twistings corresponding to the Chern-Simons levels. Denote by 
$\mathcal{W}=\mathcal{W}_X$ the twisted superpotential 
associated to $V//G$, coupled with Chern-Simons levels; see equation~\eqref{E:defW}.

We then establish:
\begin{enumerate}
\item[(a)] The $I$-function $\widehat{I}_{Ab}^\ell$  satisfies 
a difference equation
\begin{equation}
\mathcal{D}_a \widehat{I}_{Ab}^\ell = 0 
\end{equation}
where $\mathcal{D}_a$  is an operator which is explicitly identified.

\item[(b)]  Let $\sigma(\mathcal{D}_a)$ be the symbol of the $\tau$-difference
operator $\mathcal{D}_a$. Then the critical locus equation $\exp(\partial \mathcal{W}(X_a;q_i)/\partial \ln X_a) =1$ is equivalent to the zero locus of the symbols $\sigma(\mathcal{D}_a)= 0$ of the difference operators (after renaming the variables $P_a$ to $X_a$).

\end{enumerate}
We note that for partial flag manifolds $Fl(k_1, \ldots, k_s;n)$, the Coulomb branch equations may be thought of as a
modified version of the Bethe Ansatz equations, using the dictionary
\begin{equation}
 \ell_i=\kappa_{U(1)}^{(i)}+\frac{k_{i-1}+k_{i+1}}{2} 
\end{equation} (where for clarity we omit mixed levels).
In the untwisted case these equations arise as the integrability equations of an integrable system given by a $5$-vertex 
model \cite{GK:quantum,GKM:QKYB}. 
It would be interesting to find out such an integrable system in the twisted situation. 

The twisted $I$-functions on the spaces $V//G$ and $V//A$ are related by a quantum version of the abelian non-abelian correspondence (cf. Theorem \ref{nonab}):
\[ \phi(\widehat{I}_{V//A}^\ell) = {I}_{V//G}^\ell \/.\]

So far we have discussed one of the consistency checks, namely comparisons of Coulomb branch equations and symbols of difference operators.  Next, we turn to geometric windows.  In physics, geometric windows are defined to be ranges of Chern-Simons levels which do not introduce extra Coulomb vacua, so that the Higgs branch corresponds to the desired geometry, without any additional `dust.'  In mathematics, the geometric window is defined in terms of the relation between
the $I$-function and the $J$-function of $V//G$.
The $I$-function controls 
the quasimap Gromov-Witten theory, while the $J$-function controls 
the stable map theory. We compare the intervals for the components of $\ell$ for which $I=J$ to those giving the `geometric window' of the physical theory. We obtain the following conjecture:
\begin{conj}\label{windowconj}
\begin{enumerate}
\item[(a)] Let $X=V//A$ be a smooth, Fano, toric variety. If the effective 
    Chern-Simons levels are nonnegative, the physical geometric window corresponds to the set of levels where $I^\ell_{V//A}$ vanishes at $\tau=\infty$.
    
\item[(b)]  Let $X=V//G$ be a flag variety. Then $\ell$ is in the geometric window iff $I_X^\ell = J_X^\ell$.
\end{enumerate}
\end{conj}
Part (b) may be interpreted as the $K$-theoretic analogue of triviality of the mirror map, in the sense of Ciocan-Fontanine and Kim \cite{mirrormap}. 
Combining mirror triviality with results of Iritani, Milanov and Tonita 
\cite{IMT} suggests that symmetrizations of the Coulomb branch equations should give generators for the ideal of relations of twisted quantum K theory of flag manifolds, generalizing to the twisted case results in 
\cite{Gu:2020zpg,Gu:2022yvj,HK1,Gu:2023tcv,Gu:2023fpw,HK2}.

Evidence for all conjectures is given either using explicit calculations of the relevant $I$-functions, or formulating precise mathematical statements about the $I$ functions which are then used to compare to results from physics. Complete mathematical proofs will be provided in a follow-up paper \cite{leoirittoappear}.

Our construction is also compatible with certain mathematical and physical 
dualities. There is a biholomorphic
isomorphism of partial flag manifolds 
$Fl = Fl(k_1,\dots,k_s;n) \simeq Fl^*:=Fl(n-k_s,\dots,n-k_1;n)$. Then, by a theorem of Yan \cite{XY}, the quantum $K$-theory of $Fl$ with level structure 
$\ell$ is equivalent to that of the `dual' manifold $Fl^*$ with level structure 
$-\ell$. In physics, this corresponds to IR duality of the physical theories, 
discussed in sections \ref{sect:gr} and \ref{sect:flag}.

Finally, we also briefly discuss decomposition conjectures describing quantum $K$-theory on
gerbes \cite{Gu:2021yek,Gu:2021beo,Sharpe:2024ujm}.
We check a very special case, namely that the quantum $K$-theory of $B {\mathbb Z}_k = [{\rm point}/{\mathbb Z}_k]$
is $k^2$ copies of the quantum $K$-theory of a point, and also outline extensions of
the conjectures in \cite{Gu:2021yek,Gu:2021beo,Sharpe:2024ujm} to twisted quantum $K$-theory.

Next we give a brief outline of the organization of the paper. In section 
\ref{sec:math-prelim} we recall the mathematical definitions of 
the relevant moduli spaces, the (twisted) $I$ and $J$ functions, and
some mathematical context used throughout the paper.
In section~\ref{sect:conj}, we formulate a description of the physical theories relevant for all the mathematical twistings we discuss, for general smooth Fano GIT quotients.
In section~\ref{sect:proj} we describe both physics and mathematics 
computations
of quantum $K$-theory ring relations twisted by Chern-Simons and 
twistings, as well as allowed windows (ranges) of values 
of the levels, for the case of projective spaces, and check conjecture~(\ref{eq:level-conj}).  
In section~\ref{sect:tv} we generalize to smooth Fano toric varieties, and check conjecture~(\ref{eq:level-conj}) for chiral rings, and for geometric windows in a special case.
In section~\ref{sect:gr} we similarly check conjecture~(\ref{eq:level-conj}) for 
Grassmannians $Gr(k,n)$. In these cases, we compare the (physics) geometric window
of the CS levels to the (mathematical) geometric window
of RZ levels, obtaining the identity~(\ref{eq:univ:grfl}). In section~\ref{sect:flag} we perform similar comparisons in flag manifolds. 
In section~\ref{sect:gerbe} we briefly outline predictions for twisted quantum $K$-theory rings of gerbes, outlining a generalization of results in \cite{Gu:2021yek,Gu:2021beo,Sharpe:2024ujm}.

\section{Some mathematical preliminaries}\label{sec:math-prelim}
The goal of this section is to briefly recall some of the main mathematical objects in this paper.
We focus on the $I$ and $J$-functions associated to projective manifolds $X$, which
arise in the quasimap quantum $K$-theory \cite{kimcfm}, respectively the quantum $K$-theory with level structure (after Ruan and Zhang \cite{rz}). Our main examples of manifolds $X$ - toric varieties, Grassmannians, and flag manifolds - are all GLSMs, more precisely GIT spaces $V//G$ with $V$ a $G$-representation.   

An important feature is that we consider {\em twisted} versions of quantum K 
theories, as pioneered by Coates and Givental \cite{cg}, and by Ruan and 
Zhang \cite{rz}. The main theme of the paper is to relate 
these twisted theories to those from the quantum $K$-theory in physics. 
The twists arise from classes of homogeneous bundles on the target space $V//G$, 
and have interpretations in physics as coupling with Chern-Simons levels. We recognize that a more detailed (and rigorous) mathematical treatment will require significantly more space, and it is left for a follow-up paper. 

\subsection{Twisted $I$ and $J$ functions on GLSMs}
For $X$ a projective manifold we denote by $K(X)$ the Grothendieck 
ring of vector bundles on $X$. In our examples $X$ will also admit an action of 
a group $H$ (usually a torus); the corresponding equivariant K-ring is denoted by
$K_H(X)$, and it is an algebra over $K_H(pt) = Rep(H)$, the representation ring of $H$. 

\subsubsection{Moduli spaces}\label{sec:qmaps}
Let $d$ be an effective degree in $X$, and denote by $\overline{\mathcal{M}}_{0,n,d}(X)$ 
Kontsevich's moduli space of stable maps from curves of genus $0$ with $n$ 
marked points. It is equipped with evaluation maps 
$\mathrm{ev}_i: \overline{\mathcal{M}}_{0,n,d}(X) \to X$, for $1 \le i \le n$ 
and it carries a virtual fundamental sheaf 
$\mathcal{O}_{\overline{\mathcal{M}}_{0,n,d}(X)}^{vir}$.
These ingredients are used to define 
(stable map) K-theoretic Gromov-Witten invariants with descendants 
$\langle \alpha_1,\dots,\alpha_n\rangle_{0,n,d}$, 
where $\alpha_1, \ldots, \alpha_n\in K(X)[\tau, \tau^{-1}]$, 
and $\tau$ is an indeterminate. We will further augment our $K$-theory classes with additional variables, taken from some ground $\lambda$-algebra which we denote $\mathcal{A}$. The same definitions apply in this setting, with inputs being taken from  $K(X)[\tau, \tau^{-1}] \otimes \mathcal{A}$ with values in $\mathcal{A}$. If we work torus-equivariantly, we include the $T$-equivariant parameters into $\mathcal{A}$. We refer e.g. to the papers \cite{yp, fulton.pandharipande,kim.pandharipande} for details.

Another compactification of the space of maps $Hom_d(\mathbb{P}^1, X)$ is given by
quasimap moduli spaces. We use a version due to Ciocan-Fontanine, Kim, and 
Maulik \cite{kimcfm}. Let $V$ be a $G$ representation, where $G$ is a reductive 
group. Choosing a  character $\theta$ of $G$ determines a stability condition 
and a GIT quotient $V//G:= V//_{\theta} G$. 
We make the assumptions that the stable and semistable loci coincide, and that 
the stable locus is smooth and has no isotropy. 
For a parameter $0 < \epsilon \le \infty$ we denote by $Q^\epsilon_{0,n,d}(V//G)$
the moduli stack parametrizing genus $0$, $\epsilon$-stable quasimaps to $V//G$
of degree $d$ with $n$ marked points. Then $Q^\epsilon_{0,n,d}(V//G)$ is 
a proper Deligne-Mumford stack with a perfect obstruction theory, and it carries a virtual fundamental sheaf $\mathcal{O}^\epsilon$. The quasimap spaces have $n$ evaluation maps at the $n$ marked points, denoted by $\mathrm{ev}_i: Q^\epsilon_{0,n,d}(V//G)\to V//G$.~{The space $Q^\epsilon_{0,n,d}(V//G)$} admits a universal family 
\begin{equation}\label{E:piuniv} \pi: C^\epsilon \to Q^\epsilon_{0,n,d}(V//G)\end{equation}
and a universal {principal $G$ bundle $B \to C^\epsilon$}. Unlike the case of stable maps, 
the universal family is not constructed by adding a marked point. As for stable maps, using inputs in $K(V//G)[\tau, \tau^{-1}]$ and the evaluation maps, Tseng and You \cite{kquasimap} defined $K$-theoretic quasimap GW invariants on $Q^\epsilon_{0,n,d}(V//G)$, including gravitational descendants.

\subsubsection{Twistings and level structures} \label{sss:twistings} 

A key notion we use is that of twisted 
invariants, initially introduced for quantum cohomology by Coates and Givental in \cite{cg}. These invariants are defined for stable map quantum $K$-theory by Tonita in \cite{Tonita}. Their quasimap counterparts were defined in general by Zhang in \cite{wallcrossing}. (These are {\em twistings of type I}, in the language of Givental.) 
An insertion into a Gromov-Witten invariant can be regarded as imposing a constraint on the behavior of the stable map at the corresponding marked point. Similarly, twistings can be thought of as imposing universal constraints on the image of the entire curve. Technically, this corresponds to defining Gromov-Witten invariants for which the virtual fundamental sheaf is further twisted by classes from the target space.

Let $\mathfrak{C}$ be a $K$-theoretic characteristic class in the sense of \cite[definition 6.1]{z}, and additionally satisfying the following properties: 
\begin{itemize}
    \item Multiplicativity: 
    for a short exact sequence of vector bundles $A_1$, $A_2$, $E$ 
    \begin{equation}
        0 \: \longrightarrow \: A_1 \: \longrightarrow \: E \: \longrightarrow \: A_2 \: \longrightarrow \: 0,
    \end{equation}
    then $\mathfrak{C}(E) = \mathfrak{C}(A_1) \, \mathfrak{C}(A_2)$.
    \item Invertibility: for $E$ a vector bundle, $\mathfrak{C}(E)$ is an invertible element in $K(X)\otimes \mathcal{A}$. 
\end{itemize}

Such a class $\mathfrak{C}$ is well-defined on elements of $K(X)$ by letting $\mathfrak{C}(-A):=\frac{1}{\mathfrak{C}(A)}$ for $A$ a vector bundle, and extending by multiplicativity. 
Given an invertible multiplicative $K$-theoretic characteristic class $\mathfrak{C}$ and an element $E\in K(X)$, twisted $K$-theoretic Gromov-Witten invariants are defined by modifying the virtual structure sheaf of the appropriate moduli space with a class depending on $\mathfrak{C}$ and $E$. 

\begin{defn}\label{twistdef} The
$K$-theoretic Gromov-Witten invariants with twisting given by $(\mathfrak{C},E)$ are defined by replacing the virtual fundamental sheaf $\mathcal{O}^{vir}$ with $\mathcal{O}^{vir}\otimes T_{(\mathfrak{C},E)}$ in the definitions of the Gromov-Witten invariants. The class
$T_{(\mathfrak{C},E)}$ depends on the chosen moduli space:
\begin{itemize}
    \item (Cf.~Tonita's paper \cite{Tonita}.) For the moduli space of stable maps: 
    \begin{equation}\label{E:TCE-stable-map} T_{(\mathfrak{C},E)}:=\mathfrak{C}(ft_*\mathrm{ev}_{n+1}^*E) \/. \end{equation}
    \item (Cf.~Ruan-Zhang's paper \cite{wallcrossing}.) 
    For the quasimap moduli space the vector bundle $E$ is induced by a $G$-representation $R$:
    \begin{equation}\label{E:TCE-qmap} T_{(\mathfrak{C},E)}:=\mathfrak{C}(\pi_*(B\times^G R)) \in K(Q^\epsilon_{g,n,d}(V//G)) \/, \end{equation}
where $B$ is the universal principal $G$ bundle on the universal space $C^{\epsilon}$.
 \end{itemize}       
Multiple twistings can be taken in succession, given characteristic classes $\mathfrak{C}_i$ and $K$-theory elements $E_i$. The corresponding modification to the virtual structure sheaf is:
\begin{equation}\label{multtwistclass}
T_{((\mathfrak{C}_1,E_1),\dots,(\mathfrak{C}_k,E_k))}:= \bigotimes_i T_{(\mathfrak{C}_i,E_i)} .
\end{equation}
We refer to such a theory as being twisted by $((\mathfrak{C}_1,E_1),\dots,(\mathfrak{C}_k,E_k))$.  
\end{defn}
We consider the following main examples of characteristic classes: 
\begin{itemize}
    \item Euler classes: Choose $\mathfrak{C}$ to be the class $Eu_\lambda$, defined on a vector bundle $F$ by  
    \[Eu_\lambda(F)= 1- \lambda F^* + \lambda^2 \wedge^2 F^* - \ldots \/.\]
Here $\lambda$ will arise as an equivariant parameter.
    \item Determinants: Choose $\mathfrak{C}$ to be the $\ell$th power of the determinant $\det(F)^\ell$, where $\ell$ is any integer. 
    \item Products of the above classes.
\end{itemize}

Twistings using the determinant were introduced by Ruan-Zhang \cite{rz}, and 
the resulting invariants are referred to as quantum $K$-theoretic invariants with (Ruan-Zhang) 
level structure. We make these more explicit below.

\begin{defn}\label{defn:level-qk} Let $E_1, \ldots , E_k$ be vector bundles on $X$ and $\ell_1, \ldots, \ell_k$ be integers. The  $K$-theoretic GW invariants with level structure determined by $(E_1,\ell_1),\dots,(E_k,\ell_k)$
are the $K$-theoretic Gromov-Witten invariants with twisting given by 
$$((\det{}^{-\ell_1},E_1), \dots, (\det{}^{-\ell_k},E_k)) \/.$$ 
In this case, the virtual structure sheaves $\mathcal{O}^{vir}$ are twisted as follows, for the stable map and quasimap theory respectively:
\begin{itemize}

\item In the stable map theory, the virtual structure sheaf is tensored with
\begin{equation}\label{E:TCE-level-stable-map} \det(ft_*\mathrm{ev}_{n+1}^*E_1)^{-\ell_1} \otimes \ldots \otimes 
\det(ft_*\mathrm{ev}_{n+1}^*E_k)^{-\ell_k} \/.\end{equation}
where (recall) $ft:\overline{\mathcal{M}}_{g,n+1,d}(X) \to \overline{\mathcal{M}}_{g,n,d}(X)$
is the forgetful map.

\item In the quasimap theory, for a GIT quotient $V//G$, 
let $R_i$ be the $G$-representation giving $E_i$, and let 
$\pi_*(B \times_G R_i)$ be the corresponding (virtual) bundle on the quasimap space. 
Then $\mathcal{O}^{vir}$ is tensored with
\begin{equation}\label{E:TCE-level-qmap} \det(\pi_*(B\times_G R_1))^{-\ell_1} \otimes \ldots \otimes 
\det(\pi_*(B\times_G R_k))^{-\ell_k} \/.\end{equation}

\end{itemize}
    
\end{defn} 

\begin{remark}
    To keep notation consistent with both \cite{rz} and other literature regarding level structures, and references on general twisted invariants (e.g. \cite{Tonita}), we have two notations for level structures. A level structure given by the bundles $(E_1,\ell_1),\dots,(E_k,\ell_k)$, is, in the notation of Definition \ref{twistdef}, a twisting described by $((\det^{-\ell_1},E_1),\dots,(\det^{-\ell_k},E_k))$.
\end{remark}

If $g=0$, and in the case of the stable map invariants, it is proved in \cite{HK1} that twisted invariants define a ring structure. To define the product, one first defines the pairing:
    \[(\!(a,b)\!)^{tw}:=\sum_{d \ge 0}  \chi\Bigl(\overline{\mathcal{M}}_{0,2,d}(X); \mathrm{ev}_1^*(a) \otimes \mathrm{ev}_2^*(b) \otimes T_{(\mathfrak{C},E)}\Bigr) q^d \/. \]
Then the product $\star$ is uniquely determined by the (Frobenius) condition:
\[ (\!(a\star b, c)\!)^{tw} = \sum_{d \ge 0}  \chi\Bigl(\overline{\mathcal{M}}_{0,3,d}(X); \mathrm{ev}_1^*(a) \otimes \mathrm{ev}_2^*(b) \otimes \mathrm{ev}_3^*(c) \otimes T_{(\mathfrak{C},E)}\Bigr) q^d \/. \]

\subsubsection{(Small) $I$ and $J$ functions}\label{sec:IJ}

Given a target space $X$, and a twisting determined by a class $\mathfrak{C}$ and 
a bundle $E$ we recall next the definition of the (twisted) small $I$ and 
$J$-functions of $X$, which are certain generating functions for the quasimap 
invariants (at $\epsilon=0$), and stable map invariants, respectively. 
Let $\phi^\alpha,\phi_\alpha$ denote twisted-Poincar{\'e} dual bases for $K(X)$. We introduce the Novikov ring as the semigroup ring of the effective curve classes in $X$, generated by Novikov variables $q_i$, associated to a basis of effective curves in $H_2(X)$.
For any curve class $d=\sum_{i}d_ig_i$, we use $q^d$ as shorthand for $\prod_i q_i^{d_i}$. The notation $d>0$ means that the curve class $d$ is effective.

The small (twisted) $K$-theoretic $J$-function of $X$ is (cf.~\cite[Section 4]{Tonita}):
\begin{equation} \label{E:JXCE}
J_X^{(\mathfrak{C},E)}:=1-\tau+\sum_{d>0 }q^d\left\langle\frac{\phi_\alpha}{1-\tau L_1}\right\rangle^{(\mathfrak{C},E)}_{0,1,d} \, \phi^\alpha.
\end{equation}
Here $L_1$ is the cotangent line bundle at the marked point. The variable $\tau$ is typically denoted $q$ elsewhere; here we use $\tau$ to avoid confusion with Novikov variables. 

The quasimap counterpart of this function is the (twisted) small $I$-function, denoted $I_X^{(\mathfrak{C},E)}$ (introduced originally in \cite{kquasimap}, and with twistings in \cite[Section 2]{wallcrossing}), defined by:
\begin{equation}\label{E:IXCE}  
I_X^{(\mathfrak{C},E)}:=1-\tau+(1-\tau)(1-\tau^{-1})\sum_{d>0} \frac{q^d}{\mathfrak{C}(E)} \mathrm{ev}_{0*} \left( \frac{\mathcal{O}^{vir}_{F_{0,d}}\otimes T_{(\mathfrak{C},E)}|_{F_{0,d}}}{Eu_\tau(N_{0,d})} \right).
\end{equation} 
 Here $F_{0,d}$ is a certain subspace on the quasimap space (see section \ref{sec:qmaps}), $N_{0,d}$ is the normal bundle to $F_{0,d}$,  and 
 $\mathrm{ev}_0:F_{0,d}\to V//G$ is the evaluation map at the distinguished marked 
 point denoted $0$ earlier. 
 Note that the fixed locus $F_{0,d}$ inherits a perfect obstruction theory from the quasimap graph space, and thus has a well-defined virtual structure sheaf $\mathcal{O}^{vir}_{F_{0,d}}$. 

The definitions of both the $I$ and $J$-functions vary in the literature, with 
some differing by a factor of $(1-\tau)$. Our definition of the $J$-function agrees with \cite{Tonita}, and our definition of $I$-function differs from that of \cite{wallcrossing} by a factor of $(1-\tau)$. For the purposes of considering 
$\tau$-difference operators, we also add the following logarithmic prefactor. Let
$\{ P_i \}$ be the collection of line bundles determined by the fact that 
$-c_1(P_i)$ is dual to the Novikov variables $q_i$ for each $i$. Then we add the prefactor
$${\prod_i} P_i^{\log(q_i)/\log(\tau)} \/.$$ 
We will denote functions that have received this modification in boldface. In particular,
$$\mathbf{I}_X^{(\mathfrak{C},E)}:=\prod_i P_i^{\log(q_i)/\log(\tau)}I_X^{(\mathfrak{C},E)},$$
$$\mathbf{J}_X^{(\mathfrak{C},E)}:=\prod_iP_i^{\log(q_i)/\log(\tau)}J_X^{(\mathfrak{C},E)}\/.$$
The advantage of adding the prefactor is that it transforms polynomials $f(P_i\tau^{q_i\partial_{q_i}})$ in products of factors $P_i \tau^{q_i\partial_{q_i}}$ into polynomials $f(\tau^{q_i\partial_{q_i}})$ involving only the difference operators
$\tau^{q_i\partial_{q_i}}$, which are easier to work with:
\begin{equation} \label{E:diff-prefactor}
f\left(\tau^{q_i\partial_{q_i}}\right) \prod_i P_i^{\log(q_i)/\log(\tau)} = \prod_i P_i^{\log(q_i)/\log(\tau)} f\left(P_i \tau^{q_i\partial_{q_i}}\right).
\end{equation}

\begin{remark} The prefactor may also be interpreted as a formal Taylor expansion. More precisely,
write
\[ P_i^{\log(q_i)/\log(\tau)} = (1-(1-P_i))^{\log(q_i)/\log(\tau)} . \]
Using the formal Taylor expansion
\[ (1-x)^\alpha = \sum_{k=0}^\infty \binom{\alpha}{k}x^k \/, \quad \textrm{ where }\quad  \binom{\alpha}{k}:=\frac{\prod_{i=0}^{k-1}(\alpha-i)}{k!} \/,\] 
we substitute $\alpha=\frac{\log(q_i)}{\log(\tau)}$, and 
{$x=1-P_i$.}
Non-equivariantly, the expansion is well-defined since $P_i-1$ is nilpotent. 
When working $T$-equivariantly, the expression only becomes well-defined after 
completing the ground algebra with respect to the $T$-equivariant parameters.
\end{remark}

The primary connection between the small $J$-function and the stable-map quantum $K$-theory rings is given by the following theorem due to Iritani, Milanov, and Tonita \cite{IMT} in the untwisted case, and generalized to the twisted case in \cite{HK1}:

\begin{thm}\label{twisted-imt}
    Let $D(\tau^{q_i\partial_{q_i}},\tau)$ be a polynomial $\tau-$difference operator annihilating $\mathbf{J}_X^{(\mathfrak{C},E)}$. Then
    $$D(\hat{P}_i,1)=0\in QK^{(\mathfrak{C},E)}_{T}(X)\/,$$
    where $\hat{P}_i$ are multiplication operators quantizing the multiplication by the line bundles $P_i$, and defined as solutions to a certain Lax equation.
\end{thm}

In other words, symbols of $\tau$-difference operators annihilating $\mathbf{J}$ become (after specializing $\tau\to 1$)  relations in quantum $K$-theory.

It was observed by Anderson, Chen, Tseng and Iritani in \cite[Lemma 6+]{act} that under certain conditions, the quantizations $\hat{P}_i$ are trivial, i.e., $\hat{P}_i=P_i$. A general statement valid for twisted invariants is available in \cite{HK1}, and reads as follows:

\begin{thm}\label{unhat} 
Write $J^{(\mathfrak{C},E)}=(1-\tau)\sum_{d\geq 0} J_d$, where $J_d$ is the $q^d$-term, and fix an index $i$. If for all nonzero effective curve classes $d=(d_1,\dots,d_k)$, 
$$\lim_{\tau\to \infty} \tau^{d_i} J_d=0\/,$$
then $\hat{P}_i=P_i$.
\end{thm}

\subsection{Notational conventions}
For further use, we establish some conventions regarding twistings.
Note that, in this work, we only consider two kinds of twistings, or compositions thereof: 
\begin{enumerate}
\item Ruan-Zhang level structures. The ones appearing in this paper are determined by a set of bundles $(E_1,\dots,E_n)$ and an integer vector $\ell=(\ell_1,\dots,\ell_n)$. 

The choice of bundles will be clear from context, so we denote both the $I$ and $J$-functions, and the quantum $K$-ring of $X$ with this choice of level structure by  $I_X^\ell$, $J_X^\ell$, and $QK^\ell(X)$, respectively.

\item Twistings by the equivariant Euler class of the root bundle on the abelianization of a GIT quotient. We will denote quantum $K$-rings with this twisting by $QK^{tw}(X)$, and we denote the corresponding $I$-functions as $\widehat{I}_{Ab}$. 

\item In the presence of both twistings, we denote the corresponding rings as $QK^{tw,\ell}(X)$, and the $I$-functions as $\widehat{I}_{Ab}^\ell$.
\end{enumerate}

\subsection{Some technical results}
\label{sect:tech}

In this section we collect some technical results that will be used in the main text.
Proofs will appear in \cite{leoirittoappear}.

\subsubsection{Twisted $I$ functions}

The twisted small $I$ and $J$ functions $I^{(\mathfrak{C},E)}$, $J^{(\mathfrak{C},E)}$ were defined in equations~(\ref{E:JXCE}), (\ref{E:IXCE}).
Note that $I_X^{(\mathfrak{C},E)}$ is well defined, due to our hypothesis that $\mathfrak{C}(E)$ is invertible.

The following is the key fact needed in this paper. Variants of it are available in the literature (see, e.g., the proof of \cite[Theorem 2.6]{GY}), but it was not stated in the exact language  necessary for our purposes. 
We present only a brief outline of a proof, as a full description would take us too far afield. A full proof will appear in \cite{leoirittoappear}.

\begin{thm}\label{thm:I=J}
Let $X=V//G$ be a GIT quotient (possibly with a torus action), and recall the twisted {small} $I$ and $J$ functions {defined in
\eqref{E:IXCE} and \eqref{E:JXCE}, respectively. Write
$$I_X^{(\mathfrak{C},E)}=(1-\tau)\sum_{d\geq 0}  I_d\/.$$
Then the equality of (torus-equivariant) twisted small $I$ and $J$ functions}: 
$$I_X^{(\mathfrak{C},E)}=J_X^{(\mathfrak{C},E)}$$
 is equivalent to the following conditions:
   \begin{enumerate}
    \item $I_d$ has no poles at $\tau=0$, and,
    \item For $d=0$, $I_d=1$, otherwise $deg_\tau(I_d)<-1$. 
    \end{enumerate}
\end{thm}

The two main types of twisting we consider are Ruan-Zhang levels, and twisting by $\left((Eu_\lambda)^w, E\right)$, where $E$ is a vector bundle, $Eu_\lambda$ is the $\mathbb{C}^*$-equivariant Euler class, and $w$ is some nonnegative integer.

We state below the following result, recording how 
these twistings affect the $I$-function.
The mathematical proof of this theorem will appear elsewhere.

\begin{thm}\label{ltoric} Let $X=V//G$, and
let $I_X$ be the $I$-function of $X$, (possibly twisted by 
some previous choice of twisting). If 
$I_X=(1-\tau)\sum_d I_d$ then:

\begin{itemize}
    \item Adding a Ruan-Zhang level structure with level $\ell$ and bundle $E$ (where $E$ splits as a direct sum of line bundles $E_i$) results in the following $I$-function: 
    $$I_X^\ell:=(1-\tau)\sum_d  I_d\prod_i E_i^{-\ell \langle d,c_1(E_i)\rangle}\tau^{\ell\binom{\langle d,c_1(E_i)\rangle+1}{2}}$$

    \item Twisting with the data $\left((Eu_\lambda)^w,E\right)$ (where $E$ splits as a direct sum of line bundles $E_i$) results in the following $I$-function: 
$$
I_X^w := (1-\tau)\sum_d  I_d
\left(
\prod_i
\frac{
\prod_{m=-\infty}^{\langle c_1(E_i),d\rangle} (1 - \lambda E_i^{-1} \tau^m)
}{
\prod_{m=-\infty}^{0} (1 - \lambda E_i^{-1} \tau^m)
}
\right)^w
$$
\end{itemize}
\end{thm}

\begin{remark}
\label{levelremark} As before, assume $E= \bigoplus E_i$ splits as a sum of line bundles, with Chern roots $a_i$.
Adding the Ruan-Zhang level $\ell$ is equivalent to doing the substitution 
$$L_\ell:q^\beta \mapsto \big(\prod_i E_i^{-\ell(\beta,a_i)}\tau^{\ell \frac{((\beta,a_i)+1)(\beta,a_i)}{2}}\big)q^\beta $$
to the untwisted $I$-function. For multiple level structures, apply the substitutions sequentially.
\end{remark}

Thus, applying both twistings at the same time, choosing the bundle $E=\bigoplus_i E_i$ for the Ruan-Zhang level, and the bundle $F=\bigoplus_i F_i$ for the Eulerian twisting, yields the following modification of the $I$-function:

$$I_X^{\ell, w}:=
(1-\tau)\sum_d  I_d\left(\prod_i E_i^{-\ell \langle d,c_1(E_i)\rangle}\tau^{\ell\binom{\langle d,c_1(E_i)\rangle+1}{2}}\right)
\left(
\prod_i
\frac{
\prod_{m=-\infty}^{\langle c_1(F_i),d\rangle} (1 - \lambda F_i^{-1} \tau^m)
}{
\prod_{m=-\infty}^{0} (1 - \lambda F_i^{-1} \tau^m)
}
\right)^w
$$

\subsubsection{The Abelian/non-Abelian correspondence for $I$-functions}\label{sss:ana}

We recall the
abelian/non-abelian correspondence of Harada-Landweber \cite{har}. Let $V//G$ 
be a GIT quotient, and let $A$ be a maximal torus of $G$. Then 
there is a surjective map $\phi:K(V//A)^W\to K(V//G)$,
where $W$ denotes the Weyl group of $G$.

There are various quantum extensions of this theorem, due in the context of stable 
maps to Givental-Yan \cite{GY} (for Lagrangian cones), the first named author 
\cite{HK1} for quantum $K$-rings, and Wen \cite{wen} for quasimap $I$-functions. 
If we assume the $G$-unstable locus has codimension at least 2, the 
`abelian-nonabelian' morphism $\phi$ has a natural extension 
to quantum parameters, proved by Coates-Lutz-Shafi in \cite[Prop 2.3]{cls}. 
If one extends $\phi$ so that, in addition, $\phi(\lambda)=1$, then Theorem 1.1 of \cite{wen} states that:
\begin{equation}\label{E:phiIhat}
\phi(\hat{I}_{Ab})=I_{V//G} .
\end{equation}

\begin{remark}
The parameter $\lambda$ is necessary to ensure the class $Eu_\lambda$ is invertible. We also note that the limit $\lambda\to 1$ of $\widehat{I}^\ell_{Ab}$ may not exist. However, after applying $\phi$ to the quantum parameters, the terms with poles at $\lambda=1$ cancel out, and the resulting $I$-function is well-defined.
\end{remark}

Wen's theorem also applies to $I$-functions with additional level structure. A necessary insight is that the abelian/non-abelian correspondence also respects general twistings, expressed by the following result, to be proved in \cite{leoirittoappear}: 
\begin{thm}\label{nonab}
Let $R$ be a $G$ representation, which induces a vector bundle $E$ on $V//G$ and $\tilde{E}$ on $V//A$, then $\phi(I_{Ab}^{\mathfrak{C},\tilde{E}})=I_{V//G}^{\mathfrak{C},E}$.
\end{thm}

\subsection{The physical setup}\label{sec:physical-twists}

We now describe the specific choices of twistings that occur which describe the Coulomb branch equations: 

Let $V//G$ be a GIT quotient by a reductive group $G$. The Lie algebra of 
$G$ splits into central parts and semisimple parts. 
We initially work on the abelianization, the quotient $V//A$, where $A$ is a maximal torus in $G$. We apply the following twistings:

\begin{itemize}
\item For each $u(1)$-summand in the Lie algebra of $G$, we associate a bundle $E_i$ of $V//A$ (we describe this procedure later). For each summand, apply the level structure $(E_i,\ell_i)$.

\item For each pair of distinct $u(1)$ summands, apply the level structure $(E_i\otimes E_j,\ell_{ij})$.

\item For the $i$th semisimple summand  of $G$, we consider the `root bundle'
$\bigoplus L_{\alpha,i}$, where the summation is over all roots, and we
twist by $\left( (Eu_\lambda)^{w_i},\bigoplus L_{\alpha,i}\right)$. 
\end{itemize}
The first two will be Ruan-Zhang twistings, and the third an Eulerian twisting. For the mathematical description of the integers giving the twistings in terms of quadratic forms we refer to \ref{sss:homog-bundles}. 

\begin{defn}\label{ihat}
    Define the abelianized $I$-function $\widehat{I}_{Ab}^{\ell,w}$ to be the $I$-function of $V//A$ with the twistings described above.

\end{defn}

Setting $\ell, w$ to 0 recovers $I_{V//A}$. However, in light of the abelian/non-abelian correspondence for $I$-functions, we will consider $w$ to take a default value of $1$ on each semisimple factor, yielding the following definition:

\begin{defn}
Let $\widehat{I}_{Ab}^{\ell}$ be $\widehat{I}_{Ab}^{\ell, w}$, with $w=(w_i)$ is specialized to 1 on each semisimple factor (i.e., we take $w=(w_i)$ as a sequence of parameters with ``default value" of $1$).   
\end{defn}

The abelian/non-abelian correspondence for $I$-functions implies the following result:
\begin{thm}
Let $\phi$ be the abelian/nonabelian map introduced in section \ref{sss:ana}. 
Then \[ \phi(\widehat{I}^{\ell, w}_{Ab})=I_{V//G}^{\ell, w-1} \]
where $I_{V//G}^{\ell, w-1}$ is the $I$-function of $V//G$ twisted by the same Ruan-Zhang levels, and the same Eulerian twistings, with each $w_i$ replaced by $w_i-1$. However now the bundles are taken on $V//G$. 
\end{thm}
(A proof will appear in \cite{leoirittoappear}.)

\subsection{Mathematics of Chern-Simons levels}\label{sec:csmath}

\subsubsection{Invariant bilinear forms} 
Consider a GIT quotient $V//G$ with $G$ a connected, reductive group, and
$V$ a $G$-representation. Denote by $\mathfrak{g}$ 
the Lie algebra of $G$. 
The Chern-Simons (CS) levels are given by elements 
in $\mathrm{Sym}^2(\mathfrak{g}^*)^G$, i.e., by 
bilinear (or quadratic) forms on $\mathfrak{g}$ 
invariant under the adjoint action of $G$.

Since $G$ is reductive, one may decompose $\mathfrak{g} =\mathfrak{g}_{ss}\oplus Z(\mathfrak{g})$, where $\mathfrak{g}_{ss}$ is the semisimple part, and $Z(\mathfrak{g})$ is the center of $\mathfrak{g}$. This gives a decomposition
\begin{equation}\label{E:sym2decomp} \mathrm{Sym}^2(\mathfrak{g}^*)^G = \mathrm{Sym}^2(\mathfrak{g}_{ss}^*)^{G_{ss}} \oplus \mathrm{Sym}^2(Z(\mathfrak{g}^*)) \simeq \left(\bigoplus_{i=1}^k \mathrm{Sym}^2(\mathfrak{g}_{i}^*)^{G_i}\right) \oplus \mathrm{Sym}^2(Z(\mathfrak{g}^*)) \end{equation}
Here ${\mathfrak g}_{ss} = \oplus_{i=1}^k \mathfrak{g}_i$ is the decomposition into simple Lie algebras. By the Schur lemma, the bilinear forms are all multiples of the (canonically defined) Killing form on each simple algebra in this decomposition. There are no 
non-trivial `mixed factors' coming from 
two different simple algebras. In particular,   
\[ \mathrm{Sym}^2(\mathfrak{g}_{ss}^*)^{G_{ss}} \simeq \mathbb{C}^k \]
because for a simple Lie algebra the Killing form is the unique $G$-invariant form, up to scalars. Thus one way to classify the CS levels is by tuples corresponding to the coefficients of the semisimple and central part of the decomposition \eqref{E:sym2decomp}: 
\begin{equation}\label{E:CSseq} \kappa= (\kappa_{ss}^1, \ldots,\kappa_{ss}^k;\kappa_{cen}^1, \ldots, \kappa_{cen}^r)\end{equation}
where
\[ r = \dim \mathrm{Sym}^2(Z(\mathfrak{g}^*)) = {\dim Z(\mathfrak{g}^*) +1 \choose 2} \/.\]
We will use the bilinear forms (multiples of the Killing form) to identify ${\mathfrak g}_{ss} \simeq {\mathfrak g}_{ss}^*$.   

Fix a basis $\chi_1, \ldots, \chi_a$ of the (integral) weight lattice in $\mathfrak{a}^*$, used to identify $\mathrm{Sym}^\bullet (\mathfrak{g}^*)$ with the (Laurent) polynomial algebra $\mathbb{Z}[\chi_1^{\pm 1}, \ldots, \chi_a^{\pm 1}]$ in the given basis elements. 
The Chevalley restriction theorem (see e.g.~\cite[section 23.1, 23-appendix]{humphreys}) gives a graded isomorphism
\begin{equation}\label{E:Chev-iso}
 \mathrm{Sym}^\bullet (\mathfrak{g}^*)^G \simeq \mathrm{Sym}^\bullet (\mathfrak{a}^*)^W \end{equation}
where $W$ is the Weyl group of $(G,A)$ with $\mathfrak{a} = Lie (A)$ a Cartan subalgebra. Considering degrees $p$ on both sides gives isomorphisms
\[ \mathrm{Sym}^p (\mathfrak{g}^*)^G \simeq \mathrm{Sym}^p (\mathfrak{a}^*)^W \]
If we specialize to $p=2$, then elements in
$\mathrm{Sym}^2(\mathfrak{a}^*)$ may be represented by symmetric $rk(A) \times rk(A)$-matrices (depending on the chosen basis of integral weights of $\mathfrak{a}$). This gives a second way to represent the CS levels,
as $rk(A) \times rk(A)$ symmetric, Weyl invariant, matrices for elements in $\mathrm{Sym}^2(\mathfrak{a}^*)^W $. The two points of view will be used to relate the appearance of the CS levels in physics and mathematics. 

\subsubsection{Examples}\label{sss:examples-CS} 
We illustrate this in three cases. First take 
$G=\mathrm{GL}(k)$ for $k \ge 2$ (alternatively, one may also take its compact form $U(k)$). Its Lie algebra decomposes 
\[ \mathfrak{gl}_k = \mathfrak{sl}_k \oplus \mathbb{C} \] 
In terms of matrices, if $X \in \mathfrak{gl}_k$, then
\[ X= X_0 + \frac{tr(X)}{k} Id \]
where $X_0$ has trace $0$.
An element in $\mathrm{Sym}^2( \mathfrak{gl}_k^*)$ is of the form
\[ \kappa(X,Y) = \kappa_{SU(k)} tr(X_0 Y_0)+ \frac{\kappa_{U(1)}}{k} tr(X)tr(Y) \]
for some parameters $\kappa_{SU(k)}$ and $\kappa_{U(1)}$.
We now restrict to the case when $X,Y \in \mathfrak{t}$, 
the Cartan torus in $\mathfrak{g}$, i.e., 
$X=diag(x_1, \ldots, x_k),Y=diag(y_1,\ldots, y_k)$
are diagonal matrices. Then
\[
\kappa(X,Y) = \kappa_{SU(k)}\sum_{i=1}^k \left(x_i-\frac{tr(X)}{k}\right) \left(y_i-\frac{tr(Y)}{k}\right) + \frac{\kappa_{U(1)}}{k} \left(\sum_{i=1}^k x_i \right) \left(\sum_{i=1}^k y_i \right) 
\]

In terms of the matrix entries $\kappa_{ij}$ for $\kappa$:
\[ \kappa_{ij}= \kappa_{SU(k)} \delta_{ij} + \left(\frac{ \kappa_{U(1)}-\kappa_{SU(k)}}{k} \right) \]
In other words, we obtain a symmetric, $S_k$-invariant matrix:
\[ \begin{pmatrix} a & b & b & \cdots & b \\ b & a & b & \cdots & b \\ \vdots & \vdots & \vdots & \cdots & \vdots \\ b & b & b & \cdots & a \end{pmatrix}= \kappa_{SU(k)} Id_k + b J_k\]
where
\begin{displaymath}
    J_k \: = \: \left( \begin{array}{ccccc}
    1 & 1 & 1 & \cdots & 1 \\
    1 & 1 & 1 & \cdots & 1 \\
    \vdots & \vdots & \vdots & \cdots & \vdots \\
    1 & 1 & 1 & \cdots & 1 \end{array} \right),
\end{displaymath}
and
\begin{equation}\label{E:CSmatrix2} a=\kappa_{SU(k)}+\frac{\kappa_{U(1)}-\kappa_{SU(k)}}{k} \textrm{ and } b= \frac{ \kappa_{U(1)}-\kappa_{SU(k)}}{k} \/.\end{equation}
(Note that the row sums, and column sums, of this matrix are equal to $\kappa_{U(1)}$.)

More generally, if one considers products $G=\mathrm{GL}(k_1) \times \ldots \times \mathrm{GL}(k_s)$, then
\[ \mathrm{Sym}^2(\mathfrak{g}^*)^G = \oplus_{i=1}^s \mathrm{Sym}^2(\mathfrak{gl}_{k_i}^*)^{\mathrm{GL}(k_i)}=  \left(\oplus_{i=1}^s \mathrm{Sym}^2(\mathfrak{sl}_{k_i}^*)^{\mathrm{SL}(k_i)} \right) \bigoplus \mathrm{Sym}^2((\mathbb{C}^s)^*) \]
(Note: if some index $k_i=1$, then the corresponding factor is only counted in the central part.) As before, the decomposition $\mathfrak{gl}_{k_i}=\mathfrak{sl}_{k_i} \oplus \mathbb{C}$ may be written in terms of matrices as 
\[ X^{(i)} = X_0^{(i)}+ \frac{tr(X^{(i)})}{k_i}Id_{k_i} \/; \quad 1 \le i \le s \]
where $X_0^{(i)} \in \mathfrak{sl}_{k_i}$. Set $X= (X^{(1)}, \ldots, X^{(s)})$, and similarly for $Y, X_0, Y_0$. We obtain the decomposition of a $G$-invariant bilinear form $\kappa(X,Y)$:
\[ \kappa(X,Y)= \sum_{i=1}^s \kappa_{SU(k_i)} K(X_0^{(i)}, Y_0^{(i)}) + \sum_{1 \le i , j \le s} \kappa_{ij} tr(X^{(i)})tr(Y^{(j)}) \]
with $\kappa_{ij} = \kappa_{ji}$. Define $\kappa_{U(1)}^{(i)}$ by $\kappa_{ii}= \frac{\kappa_{U(1)}^{(i)}}{k_i}$. 
If one restricts to diagonal matrices $X,Y$, then the previous bilinear form gives a symmetric matrix 
$(\kappa^{ij})_{1\le i,j\le s}$ which is  
$S_{(k_1, \ldots, k_s)}$-invariant:
\begin{equation}\label{E:CSlevels-general}
\kappa^{ij}=
\begin{cases}
\kappa_{SU(k_i)}\,P_{k_i} + \frac{\kappa_{U(1)}^{(i)}}{k_i}\,J_{k_i} & i=j,\ k_i\ge 2\\[0.9em]
\kappa_{U(1), k_i} & i=j,\ k_i=1,\\[0.4em]
\kappa_{ij}\,J_{k_i\times k_j} & i\neq j\/,
\end{cases}
\end{equation}
where $\kappa^{ij}$ stands for a $k_i \times k_j$ block matrix, $J_{k_i \times k_j}$ is the $k_i \times k_j$-block matrix 
of $1$'s, $J_{k_i} = J_{k_i \times k_i}$, and 
$P_{k_i}= Id_{k_i} - \frac{1}{k_i} J_{k_i}$.

The larger matrix of CS levels is made up of the block $k_i \times k_j$ matrices $(\kappa^{ij})$. We will refer to entries in the CS matrix by pairs
\[ \kappa^{((i,p), (j,t))} \textrm{ where } \quad 1 \le i, \quad j \le s, \quad 1 \le p \le k_i, 
1 \le t \le k_j \/. \]
If $s=1$ we will drop the first component from the notation. We will denote by 
$(\kappa^{((i,p), (j,t))})$ the matrix of effective Chern-Simons levels; 
see equations \eqref{eq:keff:genl} or \eqref{eq:CS-levels:flag:comp2} below.

For example, consider a product $G=\mathrm{GL}(1) \times \mathrm{GL}(2)$.
Then
\[ \mathrm{Sym}^2(\mathfrak{g}^*)^G = \mathrm{Sym}^2(\mathfrak{sl}_{2}^*)^{\mathrm{SL}(2)} \bigoplus
\mathrm{Sym}^2((\mathbb{C}^2)^*) \]
Then
\[ \kappa(X,Y)= \kappa_{SU(2)} K(X_0^{(2)}, Y_0^{(2)}) + \sum_{1 \le i , j \le 2} \kappa_{ij} tr(X^{(i)})tr(Y^{(j)}) \]
and $\kappa_{12} =\kappa_{21}$, $\kappa_{U(1)} = \kappa_{11}, \kappa_{U(1), 2} = 2 \kappa_{22}$. The resulting $3 \times 3$ symmetric matrix is:
\begin{equation}
\label{E:FL3} (\kappa^{ij}) = \begin{pmatrix} \kappa_{U(1)} & \kappa_{12} & \kappa_{12} \\
\kappa_{21} & \frac{\kappa_{SU(2)} + \kappa_{U(1), 2}}{2} & \frac{\kappa_{U(1), 2} - \kappa_{SU(2)}}{2} \\ 
\kappa_{21} & \frac{\kappa_{U(1), 2} - \kappa_{SU(2)}}{2}& \frac{\kappa_{SU(2)} + \kappa_{U(1), 2}}{2} 
\end{pmatrix}
\end{equation}

To illustrate further, consider 
$G= \mathrm{GL}(3) \times \mathrm{GL}(2) \times \mathrm{GL}(1) \times \mathrm{GL}(3)$. 
The Weyl group is $S_3 \times S_2 \times S_1 \times S_3$, 
and it has order $72$. The invariant matrix is
\[
\begin{pmatrix}
\kappa_{SU(3)}\,P_3+\kappa_{U(1), 1}\,\frac{1}{3}J_3
& \kappa_{12}\,J_{3\times2}
& \kappa_{13}\,J_{3\times1}
& \kappa_{14}\,J_{3\times3}
\\[0.4em]
\kappa_{21}\,J_{2\times3}
& \kappa_{SU(2)}\,P_2+\kappa_{U(1), 2}\,\frac{1}{2}J_2
& \kappa_{23}\,J_{2\times1}
& \kappa_{24}\,J_{2\times3}
\\[0.4em]
\kappa_{31}\,J_{1\times3}
& \kappa_{32}\,J_{1\times2}
& \kappa_{U(1), 3}
& \kappa_{34}\,J_{1\times3}
\\[0.4em]
\kappa_{41}\,J_{3\times3}
& \kappa_{42}\,J_{3\times2}
& \kappa_{43}\,J_{3\times1}
& \kappa_{SU(3)}\,P_3+\kappa_{U(1), 4}\,\frac{1}{3}J_3
\end{pmatrix}.
\]
The matrix depends on $3+ 4+ {4 \choose 2}=13$ CS levels,
fitting with the general form of a  
symmetric $S_3 \times S_2 \times S_1 \times S_3$-invariant 
matrix of the form (with blocks emphasized):
\[
\begin{pmatrix}
\boxed{\begin{matrix}
a_1 & b_1 & b_1\\
b_1 & a_1 & b_1\\
b_1 & b_1 & a_1
\end{matrix}}
&
\boxed{\begin{matrix}
c_{12} & c_{12}\\
c_{12} & c_{12}\\
c_{12} & c_{12}
\end{matrix}}
&
\boxed{\begin{matrix}
c_{13}\\
c_{13}\\
c_{13}
\end{matrix}}
&
\boxed{\begin{matrix}
c_{14} & c_{14} & c_{14}\\
c_{14} & c_{14} & c_{14}\\
c_{14} & c_{14} & c_{14}
\end{matrix}}
\\[1em]

\boxed{\begin{matrix}
c_{21} & c_{21} & c_{21}\\
c_{21} & c_{21} & c_{21}
\end{matrix}}
&
\boxed{\begin{matrix}
a_2 & b_2\\
b_2 & a_2
\end{matrix}}
&
\boxed{\begin{matrix}
c_{23}\\
c_{23}
\end{matrix}}
&
\boxed{\begin{matrix}
c_{24} & c_{24} & c_{24}\\
c_{24} & c_{24} & c_{24}
\end{matrix}}
\\[1em]

\boxed{\begin{matrix}
c_{31} & c_{31} & c_{31}
\end{matrix}}
&
\boxed{\begin{matrix}
c_{32} & c_{32}
\end{matrix}}
&
\boxed{\begin{matrix}
a_3
\end{matrix}}
&
\boxed{\begin{matrix}
c_{34} & c_{34} & c_{34}
\end{matrix}}
\\[1em]

\boxed{\begin{matrix}
c_{41} & c_{41} & c_{41}\\
c_{41} & c_{41} & c_{41}\\
c_{41} & c_{41} & c_{41}
\end{matrix}}
&
\boxed{\begin{matrix}
c_{42} & c_{42}\\
c_{42} & c_{42}\\
c_{42} & c_{42}
\end{matrix}}
&
\boxed{\begin{matrix}
c_{43}\\
c_{43}\\
c_{43}
\end{matrix}}
&
\boxed{\begin{matrix}
a_4 & b_4 & b_4\\
b_4 & a_4 & b_4\\
b_4 & b_4 & a_4
\end{matrix}}
\end{pmatrix},
\]
where for example
\begin{displaymath}
    c_{14} \: = \: \kappa_{14}, \quad c_{24} \: = \: \kappa_{24}, \quad c_{34} \: = \: \kappa_{34},
\end{displaymath}
and other entries can be worked out similarly.

Finally, we consider the case when $G=\mathrm{Sp}(2n)$ is the symplectic group. This is a simple group, thus its center is trivial, and there is only $1$ CS level associated to it. The Lie algebra of the Cartan torus is $n$-dimensional, and the Weyl-group invariant matrix is a scalar multiple of the identity matrix.

\subsubsection{Homogeneous bundles on $V//G$}\label{sss:homog-bundles}
Fix a maximal torus $A \subset G$ and basis $\chi_1, \ldots, \chi_r$ 
on the character lattice of $A$. We explain next how the Chern-Simons 
levels give twistings of the Gromov-Witten theory of quasimap 
spaces. Recall from section \ref{sss:twistings} that a 
twisting is determined by a pair $(E, \ell)$, where $E$ 
is a $G$-representation, and $\ell$ an integer. We abuse 
notation and denote again by $E$ the corresponding 
bundle on $V//G$.

In this paper we only consider the cases when 
$G= GL(k_1) \times \ldots \times GL(k_s)$, where each $k_i \ge 1$. 
For such $G$ there is a natural bijection between indices 
$1 \le i \le s$ of $GL$/central factors
in $\mathfrak{g}^*$ and those of a basis in $Z(\mathfrak{g}^*)$.
Recall the block matrix $(\kappa^{(i,p),(j,t)})$
of CS levels from the previous section.
We now associate representations and integers to this data.
\begin{itemize}
\item For each $1 \le i \le s$, define $(E_i, \ell_i)$ where 
$E_i \simeq \mathbb{C}^{k_i}$ is the standard representation of $GL(k_i)$
and 
\[ \ell_i = \sum_{p=1}^{k_i} \kappa^{(i,1),(i,p)}- \sum_{j \neq i} k_j \kappa^{(i,1),(j,1)}\]
i.e., the sum of the entries in any of the rows of the 
corresponding $k_i \times k_i$ diagonal block matrix,
minus the contributions from the mixed blocks.
This will give a twist of Ruan-Zhang type.

\item For each pair $(i,j)$ with $i< j$ define 
$(E_{ij}, \ell_{ij}) = (E_i \otimes E_j, \kappa^{(i,1),(j,1)})$. 
In other words, the level $\ell_{ij}$ is (any of) the coefficients 
that arise on the $(i,j)$ off-diagonal block matrix of the CS levels. 
This will give a twist of Ruan-Zhang type.

\item For each $1 \le i \le s$ such that $k_i >1$ define $(F_i, w_i)$ where
$F_i = \bigoplus L_{\alpha,i}$ is the `root bundle,' defined by a subset of the adjoint representation, summing 
over all roots $\alpha$ for $\mathfrak{sl}_{k_i}$. The corresponding twist, denoted by $w_i$, is equal to 
\[ w_i = -\kappa^{(i,1),(i,k_i)} \]
Thus $-w_i$ is any off diagonal entry of the $i$th diagonal block of the CS levels matrix. 
The corresponding pair will give an Eulerian twist.
\end{itemize}
We stress that throughout sections~\ref{sss:examples-CS} and~\ref{sss:homog-bundles} all Chern--Simons
levels are effective Chern--Simons levels; we suppress the subscript
${\rm eff}$ in these two sections. Thus the matrix
$(\kappa^{((i,p),(j,t))})$ above corresponds to the matrix
$(\kappa_{\rm eff}^{ab})$ appearing in equation~\eqref{eq:keff:genl}.

\section{Superpotential for smooth Fano GIT quotients}   \label{sect:conj}

Consider a smooth Fano GIT quotient of the form $[ V//G ]$, for $V$ a vector space and $G$ a reductive algebraic group of rank $k= {rk}(G)$. 
(In this paper, $G$ will be a product of $GL$'s, but the conjecture extends to more general $G$.)
In this section we will describe the twisted one-loop effective superpotential.

In physics, we work with a $G$ gauge theory, where $G$ now\footnote{For notational sanity, throughout this paper and depending upon context, we use $G$ to denote both a reductive algebraic group and also its maximal compact subgroup.  The former is typically relevant for mathematics discussions; the latter, for physics discussions.} denotes the maximal compact subgroup of the reductive algebraic group above.  The quantum K theory relations are determined by an effective twisted superpotential on the Coulomb branch, which we describe below.
We assume in addition that $V$ has an action of a torus $T$ of dimension $n$, commuting with the action of (the maximal compact of) $G$. Let $A \subset G$ be a maximal torus, and 
fix a basis for $V$ consisting of eigenvectors common to both $T$ and $A$. We also fix a Borel subalgebra of the Lie algebra of $G$. With respect to this data, we can formally express the effective twisted superpotential on the Coulomb branch as
\begin{eqnarray}\label{E:defW}
    \mathcal{W} & = & \frac{1}{2} \sum_{a, b = 1}^{ \dim A} \kappa_{\rm eff}^{ab} (\ln X_a)(\ln X_b) + \sum_{a=1}^{\dim A} (\ln \phi(q_{ {\rm phys}, a}))(\ln X_a) 
    \nonumber \\
    & &
    + \sum_{i=1}^{\dim V} \mathrm{Li}_2(X^{\rho_i}/\Lambda^{\eta_i})
    + \sum_{a=1}^{\dim A} \left( i \pi \sum_{\alpha_p > 0} 
    \alpha_p^a \right) (\ln X_a)
    \label{eq:univ:superpotential}
\end{eqnarray}
where:
\begin{itemize} \item The variables $X_1, \ldots, X_{\dim A}$ correspond to a basis of the character lattice of $A$; geometrically, they give $T$-equivariant
line bundles on the abelian space $V//A$.

\item The variables $\Lambda_1, \ldots, \Lambda_{\dim T}$ correspond to a basis of the character lattice of $T$, and physically are determined by twisted masses $m_s$ as $\Lambda_s = \exp(2 \pi i m_s)$.

\item Since $V$ is both a $T$ and an $A$ module, it decomposes as a sum of one dimensional modules $V =\bigoplus_{i=1}^{\dim V} V_i$.
For each $i$, denote the $A$-weight of $V_i$ by 
$\rho_i= (\rho_i^1, \ldots, \rho_i^{\dim A})$ and the $T$-weight
by $\eta_i= (\eta_i^1, \ldots, \eta_i^{\dim T})$.
(In the literature, the weights corresponding to $A$ are referred to as {\em gauge} symmetries, and those for $T$ as {\em flavor} symmetries.) With this notation, set 
\[ X^{\rho_i} = \prod_{a=1}^{\dim A} X_a^{\rho_i^a} \/; \qquad \Lambda^{\eta_i} = \prod_{s=1}^{\dim T} \Lambda_s^{\eta_i^s}\]

\item The Novikov variables $q_{V//A}:=(q_1, \ldots, q_k)$ give a basis for $H_2(V//A)$ and $\phi$ is the abelian-nonabelian map 
\[ \phi: K_T(V//A)^W[\tau,\tau^{-1}][\![{\bf q}_{V//A}]\!] \longrightarrow K_T(V//G)[\tau,\tau^{-1}][\![{\bf q}_{V//G}]\!]\]
restricted to the Novikov variables. The map $\phi$ applied to Novikov variables is calculated as follows. Since the unstable locus is codimension 2, Coates-Lutz-Shafi in \cite[Prop.~2.3]{cls} show that abelian/non-abelian map identifies $H^2(V//G)$ with the $W$-invariants of $H^2(V//A)$. Dualizing this map gives a map from the effective curve classes on $V//A$ to the effective curve classes on $V//G$, and defines the associated map on Novikov variables. 
\item 
The Cartan decomposition of $G$ gives
\[ Lie(G) = Lie(A) \oplus \left(\bigoplus_{p \in P} \mathfrak{g}_{\alpha_p}\right) 
\]
where $\mathfrak{g}_{\alpha_p}$ are the root subspaces. We denote by $P$ the index set of roots, and by $\alpha_p^a$ a coordinate of the root $\alpha_p$, that is $\alpha_p=(\alpha_p^1, \ldots, \alpha_p^{\dim A})$.
The fixed Borel subalgebra of $Lie(G)$ determines which roots are considered positive.

\item The symmetric matrix of coefficients $\kappa^{ab}_{\rm eff} = \kappa^{ba}_{\rm eff}$ of the quadratic terms are the effective Chern-Simons levels.  (More formally, we identify the Chern-Simons levels with ${\rm Sym}^2( {\mathfrak g}^*)^G \cong {\rm Sym}^2( {\mathfrak a}^*)^W$, as discussed in section~\ref{sec:csmath}.)
The effective levels are sometimes known as the bare Chern-Simons levels, see e.g.~\cite[equ'n (5.2)]{Closset:2024sle},
and in that context are related to ordinary levels by 
\begin{equation} \label{eq:keff:genl}
\kappa_{\rm eff}^{ab} \: = \: \kappa_{\rm eff}^{ba} \: = \:
\kappa^{ab} +
\frac{1}{2}\sum_{\nu=1}^{\dim V} \rho_{\nu}^a \rho_{\nu}^b.
\end{equation}
Ordinary quantum K-theory is obtained in the special case that
\begin{equation}
    \kappa^{ab} \: = \: \kappa^{ba} \: = \: 
    - \frac{1}{2}\sum_{i=1}^{\dim V} \rho_i^a \rho_i^b
    + \frac{1}{2} \sum_{p \in P} \alpha^a_p \alpha^b_{p},
\end{equation}
implying in this special case that
\begin{equation}  \label{eq:CS:ordinary-qk}
    \kappa_{\rm eff}^{ab} \: = \: \kappa_{\rm eff}^{ba} \: = \:
    + \frac{1}{2} \sum_{p \in P} \alpha^a_p \alpha^b_{p},
\end{equation}
where the $\alpha_p^a$ are the components of the root vectors of $G$,
applying\footnote{In the notation of that reference, the weight vectors have R-charge $0$ and
the root vectors have R-charge $2$. This is the reason for the sign difference between the terms in equation~(\ref{eq:keff:genl}).}
\cite[equ'n (2.6)]{Gu:2023tcv}.

\item Finally, the sum in the last term in the superpotential is a sum over positive roots; for example, for the case $G = GL(k)$, 
it can be shown that
\begin{equation}
    \sum_{\alpha_p > 0} \alpha_p^a \: \equiv \: (k-1)  \: \mod \, 2
\end{equation}
for every $a \in \{1, \cdots k\}$.
The superpotential is independent of the choice of Borel subalgebra defining the positive roots.
\end{itemize}

The superpotential is a function of $\sigma$'s ($\sigma_a \propto \ln X_a$, as described in e.g.~\cite{Gu:2020zpg,Gu:2023tcv,Gu:2022yvj,Gu:2023fpw}) defined on (a subset of) ${\mathbb C}^{{\rm rank}\, G}/W$ where $W$ is the Weyl group of $G$, and as such, consistency with the Weyl group will force identifications amongst many of the parameters
$\kappa_{\rm eff}^{ab}$, $q_a$ in general, as discussed in section~\ref{sec:csmath}.

More generally, there will be levels for $U(1)$ and $SU(k)$ factors, and also mixed
$U(1)-U(1)$ levels for different $U(1)$ gauge factors, but no $U(1)-SU(k)$ or $SU(k_1)-SU(k_2)$ levels, as those would involve Chern-Simons terms such as $A \wedge dA$ with a different trace on each factor, and the trace of a single $SU(k)$ generator vanishes.

The Coulomb branch relations (which we conjecture imply the quantum K-theory ring relations) are derived from the critical locus of the superpotential~(\ref{eq:univ:superpotential}) as
\begin{equation}
    \exp\left( \frac{\partial {\cal W} }{\partial \ln X_a} \right) \: = \: 1 \: \: \: \mbox{ for each } a.
\end{equation}

\section{Projective space}  \label{sect:proj}

As a warm-up exercise, we will start with a discussion of the $N$-dimensional projective space, denoted by $\mathbb{P}^N$.
We summarize physics results, specifically for the
Wilson line ring relation and a constraint on Chern-Simons levels arising in order to describe ${\mathbb P}^N$ itself, without additional topological vacua.
We will then describe analogous mathematics results, and finally, compare the physics and math results
to the prediction of section~\ref{sect:conj}.

\subsection{Physics results}

\subsubsection{Chiral ring}

We describe ${\mathbb P}^N$ by
a $U(1)$ gauge theory with $N+1$ chiral fields of charge $1$. The twisted effective superpotential~(\ref{eq:univ:superpotential}) specializes to
\begin{equation}\label{eq:pN:W} 
    {\cal W} = \frac{1}{2}  \kappa_{\rm eff} (\ln X)^2 
    + (\ln q_{\rm phys}) (\ln X) + \sum_{i=0}^{N} \mathrm{Li}_2(X/\Lambda_i).
\end{equation}
where $\Lambda_i = \exp(2\pi i m_i)$, and
Li$_2$ denotes the dilogarithm function, with the property
\begin{equation}
    \frac{d}{d \ln x} {\rm Li}_2(x) \: = \: - \ln(1-x).
\end{equation}
In addition, in this case, applying~(\ref{eq:keff:genl}), we have
\begin{equation}
    \kappa_{\rm eff} \: = \: \kappa + \frac{N+1}{2},
\end{equation}
for $\kappa$ the ordinary $U(1)$ Chern-Simons level.

The Coulomb branch equation is derived from the critical locus of the superpotential~(\ref{eq:pN:W}) as
\begin{equation}
    \exp\left( \frac{\partial {\cal W} }{\partial \ln X} \right) \: = \: 1.
\end{equation}
Here,
\begin{equation}
    \exp\left( \frac{\partial {\cal W} }{\partial \ln X} \right)
    \: = \: \frac{ q_{\rm phys} X^{ \kappa + (N+1)/2} }{ \prod_i (1-X/\Lambda_i) },
\end{equation}
hence the Coulomb branch equation is
\begin{equation}   \label{eq:pN:phys:ring:m}
    X^{\kappa + (N+1)/2} q_{\rm phys} = \prod_{i=0}^N \left(1 - \frac{X}{\Lambda_i} \right).
\end{equation}

\subsubsection{Geometric window}

As noted in \cite{Intriligator:2013lca}, for Chern-Simons levels outside of a range, there can be additional (topological) vacua, which spoil the intended geometric interpretation of a GLSM.  We refer to the allowed range for which there are no such additional vacua as the `geometric window.'

A detailed computation of the allowed range of Chern-Simons levels is given in \cite[section 3.2]{Closset:2023jiq}, 
which for completeness we briefly review below.

For the projective space ${\mathbb P}^N$, defined by $N+1$ chiral multiplets of charge $+1$,
the semi-classical vacuum equations are
\begin{equation}
\begin{aligned}
    \sigma \phi_i &= 0, \quad i = 1, \dots, N+1\\
    \sum_{i=1}^{N+1} \phi_i^\dagger \phi_i &= \frac{1}{2\pi} \left(\xi + \kappa_{U(1)} \sigma + \frac{N+1}{2} |\sigma| \right).
\end{aligned}
\end{equation}
(We assume that the FI parameter $\xi > 0$.)

The ordinary semiclassical Higgs vacua arise for $\sigma = 0$, and describe ${\mathbb P}^N$. 
The additional (topological) vacua, when they exist, are defined by 
$\sigma \neq 0$.
These are nontrivial solutions to
\begin{equation}
    \xi + \left(\kappa_{U(1)} + {\rm sign}(\sigma) \frac{N+1}{2}\right) \sigma = 0.
\end{equation}
Note that when $|\kappa_{U(1)}|= (N+1) / 2$, there are no solutions for $\sigma$ when $\xi>0$.
Let us then consider the case where $|\kappa_{U(1)}| \neq (N+1) / 2$.
\begin{itemize}
    \item If $\sigma >0$, we have \begin{equation}
        \sigma = - \frac{\xi}{\kappa_{U(1)} + \frac{N+1}{2}}, \quad \kappa_{U(1)} + \frac{N+1}{2} <0.
    \end{equation}
    \item If $\sigma < 0$, we have
    \begin{equation}
        \sigma = - \frac{\xi}{\kappa_{U(1)} - \frac{N+1}{2}}, \quad \kappa_{U(1)} - \frac{N+1}{2}>0.
    \end{equation}
\end{itemize}

Thus, we see that the range of levels $\kappa_{U(1)}$ for which there are no topological vacua, the geometric window, is
\begin{equation}    \label{eq:pN:interval}
    -\frac{N+1}{2} \leq \kappa_{U(1)} \leq  + \frac{N+1}{2}.
\end{equation}

So far we have used physics methods to compute the chiral ring and geometric window.
We now proceed to explain these formulae starting from a mathematics perspective, on a twisted quantum $K$-ring of $\mathbb{P}^N$.

\subsection{Math results}

\subsubsection{Operators and Coulomb branch equations}
Let $X=\mathbb{P}^N$, and consider the twisted quantum $K$-theory.
In this case the only contribution is coming from the
one-dimensional center. Thus there is a single (central/Ruan-Zhang) level $\ell$, and one line bundle $\mathcal{S}$,
which by convention we take to be $\mathcal{S}=\mathcal{O}(-1)$; see 
Definition \ref{defn:level-qk}.

Since $\mathbb{P}^N$ is already a GIT quotient by a torus, there is no 
abelianization procedure necessary, and the abelianized function 
$\widehat{I}_{Ab}^\ell$ is equal to non-abelian one 
$I_{\mathbb{P}^N}^\ell$:
\begin{equation}
I_{\mathbb{P}^N}^\ell:=(1-\tau)\sum_{d\geq 0} q^d\frac{{\mathcal S}^{\ell d} \tau^{\ell \binom{d}{2}}}{\prod_{i=0}^N\prod_{j=1}^d(1-{\mathcal S}\tau^j/\Lambda_i)}.
\end{equation}

The corresponding function with the logarithmic prefactor added looks as follows:

\begin{equation}
\mathbf{I}_{\mathbb{P}^N}^\ell:=\mathcal{S}^{\log(q)/\log(\tau)}(1-\tau)\sum_{d\geq 0} q^d\frac{{\mathcal S}^{\ell d} \tau^{\ell \binom{d}{2}}}{\prod_{i=0}^N\prod_{j=1}^d(1-{\mathcal S}\tau^j/\Lambda_i)}.
\end{equation}

Consider the difference operator $\tau^{q\p_{q}}$ which sends $q \mapsto \tau q$.

\begin{prop}\label{prop:Pn-diff}
  $\mathbf{I}_{\mathbb{P}^N}^\ell$ satisfies the following difference equation:
  \begin{equation}\label{E:Pn-diff}\prod_{i=0}^N(1-\Lambda_i^{-1}\tau^{q\p_{q}})\mathbf{I}_{\mathbb{P}^N}^\ell 
  = 
  q  \tau^{\ell  q \p_{q}} \mathbf{I}_{\mathbb{P}^N}^\ell \/. \end{equation}

\end{prop}

\begin{proof} Using the equation \eqref{E:diff-prefactor} above, 
we observe that
$$\tau^{q\partial_q}\mathcal{S}^{\log(q)/\log(\tau)}=\mathcal{S}^{\log(q)/\log(\tau)}\mathcal{S}\tau^{q\partial_q} \/.$$

Thus the desired proposition is equivalent to the following $\tau$-difference equation for $I_{\mathbb{P}^N}^\ell$ (without the logarithmic prefactor):

  \begin{equation}\label{E:Pn-diff:nolog}\prod_{i=0}^N(1-\mathcal{S}\Lambda_i^{-1}\tau^{q\p_{q}})I_{\mathbb{P}^N}^\ell 
  = 
  q \mathcal{S}^\ell \tau^{\ell  q \p_{q}} I_{\mathbb{P}^N}^\ell \/. \end{equation}

This follows from examining the recursive relationship between the $q^d$ terms of $I_{\mathbb{P}^N}^\ell$, which we detail below:

\begin{eqnarray}
\lefteqn{
\prod_i \left(1-\mathcal{S}\Lambda_i^{-1}\tau^{q\p_{q}}\right) I_{\mathbb{P}^N}^\ell
} \nonumber 
\\
& = &
(1-\tau) \sum_{d\geq 0}\prod_i \left(1-\tau^d {\mathcal S}/\Lambda_i\right)\frac{q^d {\mathcal S}^{\ell d}\tau^{\ell\binom{d}{2}}}{\prod_i\prod_{j=1}^{d}(1-\tau^j {\mathcal S}/\Lambda_i)}
\\
& = &
(1-\tau)\sum_{d\geq 1}\frac{q^d {\mathcal S}^{\ell d}\tau^{\ell\binom{d}{2}}}{\prod_i\prod_{j=1}^{d-1}(1-\tau^j{\mathcal S}/\Lambda_i)}
\\
& = &
(1-\tau) \sum_{d\geq 1}\tau^{\ell (d-1)} {\mathcal S}^{\ell }\frac{q^d {\mathcal S}^{\ell (d-1)}\tau^{\ell\binom{d-1}{2}}}{\prod_i\prod_{j=1}^{d-1}(1-\tau^j {\mathcal S}/\Lambda_i)}
\\
& = &
(1-\tau)\mathcal{S}^{\ell} q\tau^{\ell q\p_{q}} \sum_{d\geq 1}\frac{q^{d-1} {\mathcal S}^{\ell (d-1)}\tau^{\ell\binom{d-1}{2}}}{\prod_i\prod_{j=1}^{d-1}(1-\tau^j {\mathcal S}/\Lambda_i)}
\\
& = &
q \mathcal{S}^\ell \tau^{\ell q\p_{q}}I_{\mathbb{P}^N}^\ell.
\end{eqnarray}
\end{proof}
The difference equation \eqref{E:Pn-diff} may be rewritten as $\mathfrak{D} \mathbf{I}^\ell_{\mathbb{P}^N} =0$, where 
\[ \mathfrak{D} = \prod_{i=0}^N \left(1-\frac{1}{\Lambda_i}\tau^{q \partial_q}\right) - q (\tau^{q \partial_q})^\ell \/. \]
The symbol of this $\mathfrak{D}$ is obtained by making the substitution
\[  \tau^{q \partial_q} \mapsto \mathcal{S} \/. \]
which yields
\begin{equation}  \label{pN:symbol}
    \prod_i (1-\mathcal{S}/\Lambda_i) = q \mathcal{S}^\ell \/.
    \end{equation}
If one identifies $X = \mathcal{S}$ then one obtains the Coulomb branch equations 
\eqref{eq:pN:phys:ring:m}, for some chosen value of Chern-Simons levels, determined by $\ell$.

\subsubsection{Mathematical geometric window}

In this discussion we will briefly review the mathematical geometric window for ${\mathbb P}^N$.
These are the set of levels $\ell$ such that $I^{\ell} = J^{\ell}$.

To compute that range, we use
a result proven
by Givental and Yan, see 
\cite[Cor. 7.2]{GY}. For the convenience of the reader, we
include a proof, based on Theorem \ref{thm:I=J} below.
\begin{thm}\label{pN:window} Let $J^\ell_{\mathbb{P}^N}$ be the small $J$-function associated to the quantum $K$-theory with (Ruan-Zhang) level structure $\ell$. Then 
$I^\ell_{\mathbb{P}^N}=J^\ell_{\mathbb{P}^N}$ if and only if
\begin{equation}  \label{eq:pN:math:range}
    -1<\ell\leq N+1.
\end{equation}
\end{thm}
This gives the mathematical geometric window for ${\mathbb P}^N$.

\begin{proof}
Denote the $q^d$-term in $I_{\mathbb{P}^N}^\ell / (1-\tau)$ by $I_d$. 
We need to check that $\ell$ is in the given interval 
if and only if $I_d$ satisfies the two constraints in the Theorem \ref{thm:I=J}.

First, observe that $I_0=1$. The condition that 
$I_d$ has no poles at $\tau=0$ is equivalent to $\ell \ge 0$. 
We now calculate the degree $\deg_{\tau} I_d$, for $d>0$. We have:
\begin{equation}\label{E:taudeg-I}
    - {\rm deg}_{\tau}(I_d)=(N+1)\binom{d+1}{2}-\ell \binom{d}{2}= (N+1)\binom{d+1}{2}-\ell \binom{d+1}{2}+\ell d \/.
\end{equation}
This quantity is greater than 1 exactly when $\ell \le N+1$, finishing the proof. 
\end{proof}

\subsubsection{Relations of the quantum $K$-ring with level structure}
Proposition \ref{prop:Pn-diff} and Theorem \ref{pN:window} can be used to establish the following result about $QK_T^{\ell}(\mathbb{P}^N)$, the quantum $K$-theory with level structure of
$\mathbb{P}^N$.  These relations (and a similar proof) first appeared in \cite[\S 8.2]{HK1}.

\begin{cor}\label{cor:twqk-Pn} Let $-1 < \ell \le N+1$. Then the following equality
holds in ${QK}_T^{\ell}(\mathbb{P}^N)$:
\begin{equation}  \label{eq:pN:math:ring}
    \prod_i (1-\mathcal{S}/\Lambda_i) = q \mathcal{S}^\ell \/.
    \end{equation}
Furthermore, there is an isomorphism
of $K_T(\mathrm{pt})[\![q]\!]= \mathbb{Z}[\Lambda_0^{\pm 1}, \cdots, \Lambda_N^{\pm 1}][\![q]\!]$-algebras
\[ \frac{K_T(\mathrm{pt})[X][\![q]\!]}{\left< \prod_{i=0}^{N} \left(1 - X/\Lambda_i\right)- X^{\ell} q\right>} 
\: \longrightarrow \:
QK_T^{\ell}(\mathbb{P}^N) \/; \quad X \mapsto \mathcal{S}.\]
\end{cor}

\subsection{Dictionary between physics and mathematics}  \label{sect:proj:dict}

For the projective space $\mathbb{P}^{N}$, our conjecture~(\ref{eq:level-conj}) for the dictionary between the physical Chern-Simons level $\kappa$ and the Ruan-Zhang level $\ell$ specializes to
\begin{equation}  \label{eq:pN:dictionary}
    \ell = \kappa_{\rm eff} = \kappa + \frac{N+1}{2},
\end{equation}
which is of the form stated in the introduction in~(\ref{eq:univ:grfl}).

It is straightforward to check that with this dictionary, and the identification $X = {\mathcal S} $, the physical Coulomb branch
equations~(\ref{eq:pN:phys:ring:m}) with twisted masses match the symbols 
of difference operators corresponding to twisted equivariant quantum K 
theory ring relation arising in mathematics~(\ref{eq:pN:math:ring}),
with $q_{\rm phys} = q$.
In addition, the allowed ranges of the Chern-Simons levels and 
Ruan-Zhang parameters match.  Recall that the allowed values of the
Chern-Simons levels are~(\ref{eq:pN:interval}), namely
\begin{equation}
    -\frac{N+1}{2} \le \kappa \le \frac{N+1}{2},
\end{equation}
which under the dictionary~(\ref{eq:pN:dictionary}), matches 
the allowed range of Ruan-Zhang levels $\ell$ given in equation~(\ref{eq:pN:math:range}):
\begin{equation}
    0 \le \ell \le N+1.
\end{equation}

This gives a proof of Conjecture \ref{generalconj} for projective spaces.

As another check, in equation~(\ref{eq:CS:ordinary-qk}) we predicted that ordinary quantum K theory is recovered when,
in the case of a projective space,
\begin{equation}
    \kappa_{\rm eff} \: = \: 0,
\end{equation}
or equivalently $\kappa = - (N+1)/2$.  It is straightforward to verify that in this case,
the Coulomb branch equations~(\ref{eq:pN:phys:ring:m}) reduce to those of ordinary quantum K theory,
as discussed in e.g.~\cite[section 2.2]{Gu:2020zpg}.

\section{Smooth Fano toric varieties}  \label{sect:tv}

In principle, the same Coulomb branch methods can be applied to smooth Fano toric varieties.  In this section we shall outline results,
which will enable us to build a dictionary between Chern-Simons levels $\kappa^{ab}_{\rm eff} = \kappa^{ba}_{\rm eff}$,
and twistings $\ell_{ab} = \ell_{ba}$ in mathematics.

\subsection{Physics results}

First, we give the Coulomb branch relations for a Fano toric variety.

Describe the toric variety in the form ${\mathbb C}^n //  ({\mathbb C}^{\times} )^k$, where the $a$th ${\mathbb C}^{\times}$ acts on
the $i$th factor in ${\mathbb C}^n$ with weight $\rho_i^a$.
 The effective twisted superpotential is given by
\begin{equation}
\begin{split}
\mathcal{W} = \frac{1}{2} \kappa^{ab}(\ln X_a)(\ln X_b) + \sum_a (\ln q_{ {\rm phys}, a})(\ln X_a) + \sum_{i} \mathrm{Li}_2(X^{\rho_i}) + \frac{1}{4}\sum_i(\ln X^{\rho_i})^2,
\\
= \frac{1}{2}\left(\kappa^{ab} + \frac{1}{2}\sum_{\nu} \rho_{\nu}^a \rho_{\nu}^b\right) (\ln X_a)(\ln X_b) + \sum_a (\ln q_{ {\rm phys}, a})(\ln X_a) + \sum_i \mathrm{Li}_2(X^{\rho_i})  . 
\end{split}
\end{equation}
Let 
\begin{equation}
    \kappa^{ab}_\text{eff} = \frac{1}{2} \left( \kappa^{ab} + \kappa^{ba} \right) + \frac{1}{2}\sum_{\nu} \rho_{\nu}^a \rho_{\nu}^b,
\end{equation}
then the Coulomb branch equation is given by
\begin{equation}  \label{eq:tv:bethe}
\left(\prod_{b=1}^k X_b^{\kappa_\text{eff}^{ab}}\right) q_{ {\rm phys}, a} 
\: = \:
\prod_{i}(1- X^{\rho_i})^{\rho_i^a}~, \quad a = 1, 2,\cdots, k~,
\end{equation}
where $X^{\rho_i} = \prod_a X_a^{\rho_i^a}$.

Working equivariantly, we turn on twisted mass $m_i$ for each chiral multiplet, and let $\Lambda_i = \exp(2\pi i m_i)$. 
The effective twisted superpotential is
$$
\mathcal{W} = \frac{1}{2}\left(\kappa^{ab} + \frac{1}{2}\sum_{\nu} \rho_{\nu}^a \rho_{\nu}^b\right) (\ln X_a)(\ln X_b) + \sum_a (\ln q_{ {\rm phys}, a})(\ln X_a) + \sum_i \mathrm{Li}_2(X^{\rho_i}/\Lambda_i)~.
$$
The Coulomb branch equation is
\begin{equation}  \label{eq:Fano-TV-Coulomb}
\left(\prod_{b=1}^k X_b^{\kappa_\text{eff}^{ab}}\right) q_{ {\rm phys}, a} 
\: = \:
\prod_{i}(1- X^{\rho_i}/\Lambda_i)^{\rho_i^a}~, \quad a = 1, 2,\cdots, k~.
\end{equation}

For Fano toric varieties with levels inside geometric windows, this should give the
twisted quantum K theory.

As a specific example, consider ${\mathbb P}^n$ with a single smooth blowup.  This is a Fano toric variety, and can be described explicitly
as ${\mathbb C}^{n+2}// ( {\mathbb C}^{\times})^2$, with the weights given below:
\begin{center}
    \begin{tabular}{ccccc}
    $x_0$ & $x_1$ & $\cdots$ & $x_n$ & $y$ \\ \hline
    $1$ & $1$ & $\cdots$ & $1$ & $0$ \\
    $1$ & $0$ & $\cdots$ & $0$ & $1$
    \end{tabular}
\end{center}

From the table above, the  Coulomb branch equations~(\ref{eq:tv:bethe}) reduce to
\begin{eqnarray}
    \left(\prod_{b=1}^2 X_b^{\kappa_\text{eff}^{1 b}}\right) q_{ {\rm phys}, 1} 
    & = &
    (1 - X_1 X_2) (1- X_1)^n,
    \\
     \left(\prod_{b=1}^2 X_b^{\kappa_\text{eff}^{2 b}}\right) q_{ {\rm phys}, 2} 
    & = &
    (1 - X_1 X_2) (1-X_2).
\end{eqnarray}

Let $\ell_1 = \kappa^{11}_\text{eff}$, $\ell_2 = \kappa^{22}_\text{eff}$, and set $\kappa^{12}_\text{eff} = 0 = \kappa^{21}_\text{eff}$. The Coulomb branch relations are given by
\begin{eqnarray}
    X_1^{\ell_1} q_{ {\rm phys}, 1} 
    & = &
    (1 - X_1 X_2) (1- X_1)^n,    \label{eq:blpn:1}
    \\
     X_2^{\ell_2} q_{ {\rm phys}, 2} 
    & = &
    (1 - X_1 X_2) (1-X_2).    \label{eq:blpn:2}
\end{eqnarray}
Rather than try to give a thorough (and lengthy) derivation of the geometric windows,
we will estimate it using the following trick.  Note that for $0 \leq \ell_1 \leq n+1$, equation~(\ref{eq:blpn:1}) is a polynomial of degree $n+1$ in $X_1$, but for $\ell_1$ outside that range, the number of zeroes changes.  In general terms, we expect those roots to have a geometric interpretation, so if the number of roots changes, then, we are likely no longer describing the same geometry.  For this reason, we estimate that $0 \leq \ell_1 \leq n+1$.
Applying a similar crude reasoning to equation~(\ref{eq:blpn:2}) and $\ell_2$, we are led to estimate that the geometric window for this example should be given by levels in the range
\begin{equation}
    0\leq \ell_1 \leq n+1, \quad 0 \leq \ell_2 \leq 2~.
\end{equation}
A thorough analysis would take into account the fact that equations~(\ref{eq:blpn:1}) and (\ref{eq:blpn:2}) are linked, which we have not done in the estimate above.
We will see later nearly the same range arising from a thorough mathematical analysis.

\subsection{Math results}

\subsubsection{$K$-theory of toric varieties}

Let $X$ be a  smooth, projective, Fano toric variety of dimension $n$. It is expressible as $\mathbb{C}^N//A$ where $A$ is a torus of rank $k$. The variety $X$ is acted on by the torus $T:=(\mathbb{C}^*)^N$ via the standard action on $\mathbb{C}^N$, with equivariant parameters denoted $\Lambda_i$ in $K$-theory, and $\lambda_i$ in cohomology.

Choose an integer basis $p_a$ of torus characters identifying 
the Lie algebra of $A$ with $\mathbb{R}^k$. 
Define the line bundles $P_a$ to be the bundles on $X$ 
associated to the \emph{duals} of the bundles associated to 
these characters. (So $-c_1(P_a)=p_a$, where $p_a$ is regarded 
as a cohomology class in $H^2(X)$).

We define Novikov variables $q_a\in q^{H_2(X)}$ to be the Novikov variables associated to classes dual to $p_a$. (Note that these do not necessarily correspond to effective curve classes.) The equivariant $K$-theory of $X$ is generated by $U_i$, the line bundles associated to rays of the toric fan. Explicitly, one may write
$$U_i(P)=\prod_a P_a^{U_i^a} \/$$
for appropriate integers $U_i^a$. This generalizes the non-equivariant cohomology formula, where $[D_i]=\sum_{a} U_i^a p_a$.

\begin{remark}
    Two different $U_i$'s can have the same class in $K$-theory, as in projective space, where each fixed locus is a hyperplane. However, the $T$-action on each bundle is different, which will appear in their $T$-equivariant Euler classes.
\end{remark}

The equivariant $K$-theory ring $K_T(X)$ has the following description using the $K$-theoretic Stanley-Reisner/Kirwan relations (cf. e.g., \cite{VV:higherK,anderson.payne:operationalK}): 
$$\mathbb{C}[\Lambda^{\pm 1}][P^{\pm 1}]/\Big< \prod_{i\in I} (1-\Lambda_i^{-1} U_i(P))=0\text{ whenever } \cap_{i\in I} D_i=\emptyset \Big>.$$

\subsubsection{$I$-functions for general toric varieties}

By a theorem of Givental \cite[page 6, ``Linear sigma-models'' section]{GivV}, 
\begin{equation}
\label{itoric}
I_{\mathbb{C}^N//A}=(1-\tau)\sum_{d_a\in \mathbb{Z}} \prod_a q_a^{d_a}\prod_{i}\frac{\prod_{m=-\infty}^0(1-\Lambda_i^{-1} U_i(P)\tau^m)}{\prod_{m=-\infty}^{(d,D_i)}(1-\Lambda_i^{-1}U_i(P)\tau^m)}.
\end{equation}

\begin{remark}
This sum here is taken over all curve classes, not just the effective ones, however for $d$ representing a non-effective curve class, the corresponding term in the $I$-function will vanish via the Kirwan relations. 
\end{remark}

Recall that $\mathbf{I}_{\mathbb{C}^N//A}$ denotes the $I$-function with the added logarithmic prefactor.  For each $1 \le a \le rk~ H_2(X)$, this satisfies the equations:
\begin{equation}
\label{difftoric}
q_a\mathbf{I}_{\mathbb{C}^N//A}=\prod_{j}\frac{\prod_{m=-\infty}^{(D_j,1_a)-1}(1-\tau^{-m}\Lambda_j^{-1}U_j(\tau^{q\partial_q}))}{\prod_{m=-\infty}^{-1}(1-\tau^{-m}\Lambda_j^{-1}U_j(\tau^{q\partial_q}))}\mathbf{I}_{\mathbb{C}^N//A}.
\end{equation}

For each (unordered) pair $(a,b)$, with $1 \le a,b \le rk~H_2(X)$ choose an integer $\ell_{ab}$. (Because the pairs are unordered,
note that $\ell_{ab} = \ell_{ba}$.) 
Let us now introduce a Ruan-Zhang level structure given by the bundles $(P_a(P_b^{1-\delta_{ab}}),\ell_{ab})$. (In other words, when $a\neq b$, we use the bundle $P_aP_b$ with level $\ell_{ab}$, if $b=a$, we use the bundle $P_a$ with level $\ell_{aa}$). 
The resulting $I$-function is (using remark~\ref{levelremark}):
\begin{eqnarray}
\label{full-level-itoric}
I^\ell_{\mathbb{C}^N//A} & = & 
(1-\tau)\sum_{d_a\in \mathbb{Z}} \left(\prod_a P_a^{\ell_{aa}d_a}\tau^{\ell_{aa}\binom{d_a}{2}}\prod_{a<b}(P_aP_b)^{\ell_{ab}(d_a+d_b)}\tau^{\ell_{ab}\binom{d_a+d_b}{2}}\right)
\\
& & \hspace*{0.5in}  \cdot
\prod_{i}\frac{\prod_{m=-\infty}^0(1-\Lambda_i^{-1}U_i(P)\tau^m)}{\prod_{m=-\infty}^{(d,D_i)}(1-\Lambda_i^{-1}U_i(P)\tau^m)}.
\nonumber
\end{eqnarray}

After adding logarithmic prefactors, we have
\begin{thm}
\begin{equation}
\label{full-level-difftoric}
q_a\prod_a\tau^{\ell_{aa}q_a\partial_{q_a}}\prod_{b\neq a} \tau^{\ell_{ab}q_a\partial_{q_a}q_b\partial_{q_b}}\mathbf{I}_{\mathbb{C}^N//A}^\ell=\prod_{j}\frac{\prod_{m=-\infty}^{(D_j,1_a)-1}(1-\tau^{-m}\Lambda_j^{-1}U_j(\tau^{q\partial_q}))}{\prod_{m=-\infty}^{-1}(1-\tau^{-m}\Lambda_j^{-1}U_j(\tau^{q\partial_q}))}\mathbf{I}^\ell_{\mathbb{C}^N//A}.
\end{equation}
\end{thm}

\begin{proof}

Let the $q^d$-term of the (untwisted) $I$-function (without prefactor), be denoted $I_d$. Let the corresponding term in the $I$-function with level structure be denoted $I_d^\ell$. We have the recursive relation
$$\frac{I_{d}}{I_{d-1_a}}=q_a \prod_i\frac{\prod_{m=-\infty}^{(D_i,d-\mathbf{1}_a)}(1-\tau^m \Lambda_i^{-1}U_i)}{\prod_{m=-\infty}^{(D_i,d)}(1-\tau^m\Lambda_i^{-1}U_i)}.$$

Applying the twisting corresponding to the level structure introduces a multiplicative factor to each $q^d$, so this relation is multiplied by the ratio of these factors, which is
$$P_a^{\ell_{aa}}\tau^{\ell_{aa}(d_a-1)}\prod_{b\ne a}(P_aP_b)^{\ell_{ab}}\tau^{\ell_{ab}(d_a+d_b-1)}$$

So the final relation is
$$\frac{I^\ell_{d}}{I^\ell_{d-1_a}}=q_a\left(P_a^{\ell_{aa}}\tau^{\ell_{aa}(d_a-1)}\prod_{b\ne a}(P_aP_b)^{\ell_{ab}}\tau^{\ell_{ab}(d_a+d_b-1)}\right)\prod_i\frac{\prod_{m=-\infty}^{(D_i,d-\mathbf{1}_a)}(1-\tau^m\Lambda_i^{-1}U_i)}{\prod_{m=-\infty}^{(D_i,d)}(1-\tau^m\Lambda_i^{-1}U_i)}.$$

This relation corresponds to the desired $\tau-$difference equation. 
\end{proof}

These equations give symbols
\begin{equation}   
\label{full-level-symbtoric}
P_a^{\ell_{aa}}\prod_{b\neq a} (P_bP_a)^{\ell_{ab}}q_a=\prod_i (1-\Lambda_i^{-1}U_i(P))^{U_i^a}.
\end{equation}

We also consider the special case $\ell_{ab}=0$ for $a\neq b$, as this will be useful for providing evidence for our conjecture on geometric windows. In this case, if we denote $\ell_{aa}$ by $\ell_a$, we obtain
\begin{equation}
\label{special-level-symbtoric}
P_a^{\ell_{a}}q_a=\prod_i (1-\Lambda_i^{-1}U_i(P))^{U_i^a}.
\end{equation}

\subsubsection{Blowup of $\mathbb{P}^{n}$ at one point}

Now let $X$ be the blowup of $\mathbb{P}^{n}$ at one point. 
We can describe this as the GIT quotient ${\mathbb C}^{n+1}_{\phi_0, \cdots, \phi_{n}} \times {\mathbb C}_z // ({\mathbb C}^{\times})^2$, with $(\mathbb{C}^\times)^2$ weights written as column vectors as in the following table:
\begin{center}
    \begin{tabular}{ccc}
    $\phi_0$ & $\phi_1, \cdots, \phi_{n}$ & $z$ \\ \hline
    $1$ & $1$ & $0$ \\
    $1$ & $0$ & $1$
    \end{tabular}
\end{center}
From the GIT quotient description 
there are $n+2$ torus parameters 
$\Lambda_i$, $2$ character 
line bundles $P_i$, and $n+2$ toric divisors. We have $n$ toric 
divisors $D_1,\dots, D_{n}$, where $D_i$ 
has conormal bundle $U_i=P_1$ (with weight associated to the torus parameter $\Lambda_i$) 
The remaining divisors $D_0,D_{n+1}$ have conormal 
bundles $U_0=P_1P_2$ (weight $\Lambda_0$) and $U_{n+1}=P_2$ (weight $\Lambda_{n+1}$.) 

As such, the $I$-function is
\begin{eqnarray*}
I_X  &= &
(1 - \tau) \sum_{d_1, d_2 \in \mathbb{Z}} 
q_1^{d_1} q_2^{d_2} 
\left(\prod_{i=1}^{n}
\frac{
  \prod_{m=-\infty}^0 (1 - P_1 \tau^m/\Lambda_i)
}{
  \prod_{m=-\infty}^{d_1} (1 - P_1 \tau^m / \Lambda_i)
}\right)
\\
& & 
\cdot
\frac{
  \prod_{m=-\infty}^0 (1 - P_2 \tau^m/\Lambda_{n+1})
}{
  \prod_{m=-\infty}^{d_2} (1 - P_2 \tau^m / \Lambda_{n+1})
}
\cdot
\frac{
  \prod_{m=-\infty}^0 (1 - P_1 P_2 \tau^m / \Lambda_{0})
}{
  \prod_{m=-\infty}^{d_1 + d_2} (1 - P_1 P_2 \tau^m / \Lambda_{0})
}.
\end{eqnarray*}

\begin{remark}
One can check via the Stanley-Reisner relations that the terms vanish outside the locus where $d_1\geq 0$ and $d_1+d_2\geq 0$. This locus corresponds to the Mori cone $M(X)$ of effective curve classes on $X$. Note that $d_2$ may not necessarily be positive. 
\end{remark}

Using the above remark, we can simplify the description of this function to be
\begin{eqnarray*}
I_X  
&= &
(1 - \tau) \sum_{(d_1, d_2) \in M(X)} 
q_1^{d_1} q_2^{d_2} 
\frac{1}
{\prod_{i=1}^n\prod_{m=1}^{d_1} (1 - P_1 \tau^m / \Lambda_i) }
\\
& & \hspace*{0.75in}
\cdot
\frac{1}
{\prod_{m=1}^{d_1 + d_2} (1 - P_1 P_2 \tau^m / \Lambda_{0})}
\frac{
\prod_{m=-\infty}^0 (1 - P_2 \tau^m/\Lambda_{n+1})
}{
    \prod_{m=-\infty}^{d_2} (1 - P_2 \tau^m / \Lambda_{n+1})
}.
\end{eqnarray*}

 Now, we use this example to develop evidence for Conjecture \ref{windowconj}. We ignore $U(1)\times U(1)$ levels and restrict ourselves to the level structures given by $(P_1,\ell_1),(P_2,\ell_2)$. We will test our conjecture only for these levels, as this is where computation is most practical. The corresponding $I$-function is
\begin{eqnarray*}
I_X^\ell & = &
(1 - \tau) \sum_{(d_1, d_2) \in M(X)} 
P_1^{\ell_1 d_1} P_2^{\ell_2 d_2} 
\tau^{\ell_1 \binom{d_1}{2} + \ell_2 \frac{(1 - d_2)(-d_2)}{2}} 
q_1^{d_1} q_2^{d_2}
\frac{
1
}{
\prod_{i=1}^n \prod_{m=1}^{d_1} \left(1 - \frac{P_1 \tau^m}{\Lambda_i}\right)
}
\\
& & \hspace*{1.5in}
\cdot
\frac{
1
}{
\prod_{m=1}^{d_1 + d_2} \left(1 - \frac{P_1 P_2 \tau^m}{\Lambda_{0}}\right)
}
\cdot
\frac{
\prod_{m=-\infty}^0 (1 - \frac{P_2 \tau^m}{\Lambda_{n+1}})
}{
\prod_{m=-\infty}^{d_2} \left(1 - \frac{P_2 \tau^m}{\Lambda_{n+1}}\right)
}.
\end{eqnarray*}
\begin{remark}
The different exponents in $\tau$ account for the fact that $d_1 \geq 0$, 
but the same is not necessarily true for $d_2$, and we have
\[
\frac{(1 - d_2)(-d_2)}{2}
=
\begin{cases}
\binom{d_2}{2}, & d_2 \geq 0,\\[6pt]
\binom{-d_2 + 1}{2}, & d_2 \leq 0.
\end{cases}
\]
\end{remark}

Restricting to nonnegative values of the levels $\ell_a$, conjecture \ref{windowconj} regarding the geometric window is equivalent to the statement that $I^{\ell}(\tau)-(1-\tau)$ vanishes at $\tau=\infty$.  Let $I^\ell_X=(1-\tau)\sum_{d}I_d$, then the conjecture is equivalent to (for nonzero $d$ in the Mori cone)
\begin{displaymath}
{\rm deg}_\tau(I_d)< -1.
\end{displaymath}

We will now attempt to understand the asymptotic data in order to determine when this conjecture holds. 

Let $\sigma(a)=\left( {\rm sgn}(a)^2-{\rm sgn}(a) \right) / 2$, i.e.~it is 1 if $a<0$ and $0$ otherwise. 

\begin{lem}
For $d$ in the Mori cone, we have:
    $${\rm deg}_\tau(I_d)= \ell_1\binom{d_1}{2}+\ell_2\binom{|d_2|+\sigma(d_2)}{2}-n\binom{d_1+1}{2}-\binom{d_1+d_2+1}{2}-\binom{d_2+1}{2}.$$
\end{lem}
\begin{proof}
Note that the coefficients of $I_d$ lie in a nontrivial ring, and thus could cause cancellations. Computing the degree ``naively" is justified as the highest $\tau-$degree terms in the numerator and denominator are products of invertible classes and thus nonzero. 
\end{proof}

\begin{lem}
For $\ell_2=0$, the allowed range of $\ell_1$ is
\begin{displaymath}
0\leq \ell_1\leq n.
\end{displaymath}
\end{lem}
\begin{proof}
For $d\neq 0$ in the Mori cone, we have
 $${\rm deg}_\tau(I_d)\leq  n\binom{d_1}{2}-n\binom{d_1+1}{2}-\binom{d_1+d_2+1}{2}-\binom{d_2+1}{2}.$$
Now,
 $n\binom{d_1}{2}-n\binom{d_1+1}{2}=-nd_1\leq 0$, and is equal to 0 only when $d_1=0$. In that case, $d_2>0$, so the remaining terms sum to at most $-2$. Hence $(1-\tau)I_d$ vanishes at $\tau=\infty$. 

\end{proof}

\begin{remark}
For the case $\ell_1=n+1$, we have:
\begin{displaymath}
 {\rm deg}_\tau(I_d)=  \binom{d_1}{2}+n\binom{d_1}{2}-n\binom{d_1+1}{2}-\binom{d_1+d_2+1}{2}-\binom{d_2+1}{2}.
 \end{displaymath}

This function will eventually be positive.
 \end{remark}

\begin{lem}
For $\ell_1=0$, the allowed range of $\ell_2$ is
$$0\leq \ell_2\leq \operatorname{min}(n-1,2).$$
\end{lem}
\begin{proof}
For $d\neq 0$ in the Mori cone, we have
 $${\rm deg}_\tau(I_d)= \ell_2\binom{|d_2|+\sigma(d_2)}{2}-n\binom{d_1+1}{2}-\binom{d_1+d_2+1}{2}-\binom{d_2+1}{2}.$$

 If $d_2\geq 0$, the degree is negative when $\ell_2\leq 2$, since the last two terms are each at least $\binom{d_2}{2}$ in magnitude. Otherwise, the first term dominates for sufficiently large $d_2$. 

 If $d_2<0$, the Mori cone condition forces $d_1\geq -d_2$, thus the degree is bounded above by:

 $$\ell_2\binom{d_1+1}{2}-n\binom{d_1+1}{2}$$

 Since we require the degree to be strictly less than $-1$, we must have $\ell_2<n-1$.

\end{proof}

Hence we have
\begin{thm}
    The individual geometric windows for the levels $\ell_1,\ell_2$ are
    $$0\leq \ell_1\leq n, \quad 0\leq \ell_2\leq \operatorname{min}(n-2,2).$$
\end{thm}

\subsection{Dictionary between physics and mathematics}

For smooth Fano toric varieties, we apply the conjectural equation~(\ref{eq:level-conj}) to predict the following dictionary between the physical Chern-Simons levels $\kappa^{ab}_{\rm eff}$ and
the twistings $\ell_{ab}$:
\begin{equation}  \label{eq:level-reln}
    \kappa^{ab}_{\rm eff} \: = \: \left\{ \begin{array}{cl}
    \sum_c \ell_{ac} & a = b, \\
    \ell_{ab} & a \neq b.
    \end{array} \right.
\end{equation}
Comparing the Coulomb branch equations~(\ref{eq:Fano-TV-Coulomb}) to the symbols of $\tau$-difference operators~(\ref{full-level-symbtoric}), 
we see immediately that with the dictionary above, the ring relations 
match, as expected, with $q_{ {\rm phys}, a} = q_a$.
We also checked geometric windows in the special case of a blowup of ${\mathbb P}^{n}$ at a point, for which the estimate of the geometric window in physics
is nearly the same as that computed mathematically under the dictionary above, providing further evidence for the dictionary~(\ref{eq:level-conj}).

\section{Grassmannians}  \label{sect:gr}

In this section, we will repeat the same analysis for the more general case of Grassmannians.  We will first outline physics results for quantum $K$-theory rings and allowed ranges of Chern-Simons levels for which there are no extraneous topological vacua.  Then, we will give mathematical results for the Coulomb branch equations,
and allowed ranges of the Ruan-Zhang parameters.
Finally, we will note that a simple relationship preserves both the Coulomb branch structure and the allowed range of levels in the geometric windows.

\subsection{Physics results}

\subsubsection{Chiral ring}

Consider a Grassmannian $Gr(k,n)$, described by a $U(k)$ gauge theory with $n$ chiral fields in the fundamental representation. On the Coulomb branch, the effective twisted superpotential~(\ref{eq:univ:superpotential}) specializes to
\begin{equation}
\begin{aligned}
    {\cal W} ={}& \frac{1}{2} \sum_{a,b=1}^k \kappa_{\rm eff}^{ab} \, (\ln X_a) (\ln X_b) + (\ln(-1)^{k-1} q_{\rm phys}) \sum_{a=1}^k (\ln X_a)\\
    &+ \sum_{a=1}^k \sum_{i=1}^n \mathrm{Li}_2(X_a/\Lambda_i).
\end{aligned}
\end{equation}
where, applying~(\ref{eq:keff:genl}) and using the fact that for each of the $n$ fundamentals,
$\rho_{r, a}^b = \delta^a_b$, ($1 \leq r \leq n$,) we have
\begin{equation}
    \kappa_{\rm eff}^{ab} \: = \: \kappa^{ab} + \frac{n}{2}\delta_{ab}.
\end{equation}
Due to the Weyl symmetry, we have two independent Chern-Simons levels, $\kappa_{SU(k)}$ and $\kappa_{U(1)}$, with
\begin{equation}  \label{eq:grass:kappa-defns}
    \kappa^{ab} = \kappa_{SU(k)} \delta_{ab} + \frac{\kappa_{U(1)} - \kappa_{SU(k)}}{k}.
\end{equation}
The Coulomb branch equations are given by
\begin{equation}  \label{eq:grass:phys:qk}
    (-1)^{k-1} q_{\rm phys} X_a^{\kappa_{U(1)} + \frac{n}{2} } \left(\prod_{b=1}^k \frac{X_b}{X_a}\right)^{\frac{\kappa_{U(1)} - \kappa_{SU(k)} }{k}} = \prod_{i=1}^n \left(1 - \frac{X_a}{\Lambda_i}\right), \quad a = 1, 2, \dots, k.
\end{equation}

\subsubsection{Geometric window}

Just as for projective spaces, the possible Chern-Simons levels that are consistent with the geometry being a Grassmannian -- that give theories with no additional topological vacua -- are constrained.
In this section we will apply the results of\footnote{
It may be helpful to note that our $\kappa_{SU(k)}$ is denoted $k$ in
\cite{Closset:2023jiq}, and similarly their $\ell$ is our $(\kappa_{U(1)} - \kappa_{SU(k)})/k$.  Also, in applying their computations, we are restricting to the special case that their $\ell = -1$.
} \cite[section 4]{Closset:2023jiq} to sketch the derivation of the geometric window for Grassmannians, in the special case that
\begin{equation}  \label{eq:Gr:window:constr}
    \frac{\kappa_{U(1)} - \kappa_{SU(k)}}{k} \: = \: -1,
\end{equation}
or equivalently
\begin{equation}  \label{eq:gr:window:constr}
    \kappa_{U(1)} = \kappa_{SU(k)} - k.
\end{equation}
(We restrict to this range for simplicity; a similar restriction also arises in the corresponding
mathematics computation.  We leave more general cases for future work.)
We do not rederive the vacuum structure or Witten index computations, referring instead to the detailed analysis in~\cite[section 4]{Closset:2023jiq}.

Consider a $U(k)$ gauge theory with $n$ fundamental chiral multiplets, with Chern-Simons levels $(\kappa_{SU(k)}, \kappa_{U(1)})$.  

As shown in \cite[section 4]{Closset:2023jiq}, supersymmetric vacua fall into four classes: Higgs, topological, Higgs-topological, strongly-coupled. Only the Higgs vacua correspond to the geometric Grassmannian phase, where the FI parameter is positive. The other vacua must be absent in order for the GLSM to admit a purely geometric interpretation.

Topological vacua arise when all vector multiplet scalars acquire nonzero vevs. Their contribution to the Witten index is encoded in equations (4.34) and (4.40) in \cite{Closset:2023jiq}. It can be easily derived that they are absent in the range
\begin{equation}
-k-\frac{n}{2} < \kappa_{SU(k)} \leq k + \frac{n}{2}~.
\end{equation}

Hybrid Higgs-topological vacua arise when part of the vector multiplet scalars acquire nonzero vevs. Their contribution to the Witten index is given by equation (4.29) in \cite{Closset:2023jiq}. They are absent when
\begin{equation}
    - \frac{n}{2} \leq \kappa_{SU(k)} \leq \frac{n}{2} + 1~.
\end{equation}

The contribution of the strongly-coupled vacua to the Witten index is given by equation (4.41) in \cite{Closset:2023jiq}. They are absent when
\begin{equation}
    \kappa_{SU(k)} \neq - \frac{n}{2}~.
\end{equation}

Combining the above constraints, we conclude that all non-geometric vacua are absent precisely when
\begin{equation}   \label{eq:gr:gw}
    - \frac{n}{2} < \kappa_{SU(k)} \leq \frac{n}{2}+1~.
\end{equation}
This interval therefore defines the geometric window for $Gr(k,n)$.

From the constraint~(\ref{eq:gr:window:constr}), the Chern-Simons level $\kappa_{U(1)}$ is also constrained.
Using the constraint, we see that the bounds on $\kappa_{SU(k)}$ also require
\begin{equation}
    - \left( k + \frac{n}{2} \right) < \kappa_{U(1)} \leq -k + \frac{n}{2} + 1.
\end{equation}
In the case $k=1$, this reproduces the geometric window for a projective space given in~(\ref{eq:pN:interval}),
using the quantization of Chern-Simons levels (so that the strict lower inequality moves to the next allowed value).

\subsubsection{Dualities}

In passing, let us discuss the role of IR dualities in this context.
From \cite[equ'n (5.51)]{Closset:2023vos},
for $|\kappa_{SU(k)}| < n/2$, there is a duality between the following three-dimensional $\mathcal{N}=2$ supersymmetric theories:
\begin{itemize}
    \item a $U(k)_{\kappa_{SU(k)}, \kappa_{SU(k)} - k}$ Chern-Simons theory with $n$ matter fields in the fundamental,
    \item a $U(n-k)_{-\kappa_{SU(k)}, - \kappa_{SU(k)} - (n-k)}$ Chern-Simons theory with $n$ antifundamentals.
\end{itemize}
This provides a physical relationship between the three-dimensional theories for the Grassmannians $Gr(k,{\mathbb C}^n) \cong Gr(n-k,({\mathbb C}^n)^*)$. For $|\kappa_{SU(k)}| < n/2$,
this is consistent with the geometric window~(\ref{eq:gr:gw}).
We shall see an analogous duality arising in the mathematics of twisted quantum $K$-theory.

Now let us consider how equivariant parameters $\Lambda_i$ are related by the duality.
If a theory has a flavor symmetry group $G_F$, then we can add background vector multiplets, whose scalar components have a VEV denoted $m_F$, and which are valued in the Cartan subalgebra of $G_F$. Then a chiral multiplet $\Phi_i$ in some representation of $G_F$ has twisted mass  $m_i = \omega_i(m_F)$, where $\omega_i$ is the weight of the flavor representation. We define  $\Lambda_i = \exp(2\pi i m_i)$.

By the above argument, for a $U(n)$ flavor symmetry, we turn on fugacities $m_F = (m_1, m_2, \cdots, m_n)$, breaking $U(n)$ to its Cartan $U(1)^n$. The twisted mass for a chiral $\Phi^i$ in its fundamental representation is $m_i$ and the twisted mass for a chiral $\tilde{\Phi}_i$ in its antifundamental representation is $-m_i$. 
As a result, in passing to the dual theory, $\Lambda$ is replaced by $\Lambda^{-1}$.
Furthermore, we can utilize the residual Weyl symmetry $S_n$ to permute the equivariant parameters.
To match mathematics conventions, we will take
$$
\Lambda_i \mapsto \Lambda_{n+1-i}^{-1}.
$$

In passing, the Coulomb branch equation for the dual theory is 
\begin{equation}  \label{eq:gr:dualcoulomb}
    (-1)^{n-k-1}\breve{q}_{\rm phys} \breve{X}_a^{-\kappa_{SU(k)} - (n-k) + \frac{n}{2}} \left(\prod_{b=1}^{n-k} \frac{\breve{X}_b}{\breve{X}_a}\right)^{-1} = \prod_{i=1}^n \left(1 - \frac{{\Lambda}_i}{\breve{X}_a}\right)^{-1}, \quad a = 1,2, \cdots, n-k .
\end{equation}

\subsection{Math results}
\subsubsection{The Grassmannian and its Abelianization}

The Grassmannian $Gr(k,n)$ is given by the GIT quotient ${\rm Hom}(\mathbb{C}^k,\mathbb{C}^n)//\mathrm{GL}(k)$. Its abelianization is the GIT quotient ${\rm Hom}(\mathbb{C}^k,\mathbb{C}^n)//(\mathbb{C}^*)^k$, with each factor of $\mathbb{C}^*$  scaling the columns of the $n\times k$ matrix. The abelianization is isomorphic to $(\mathbb{P}^{n-1})^k$.

The fundamental representation $\mathbb{C}^k$ of $\mathrm{GL}(k)$ induces a homogeneous bundle on $Gr(k,n)$, namely the tautological subbundle $\mathcal{S}$. The same representation, regarded as a $T$-representation, induces the bundle $\widetilde{\mathcal{S}}$ on $(\mathbb{P}^{n-1})^k$, which splits as a direct sum of the tautological line bundles $\mathcal{O}(-1)$ pulled back from each factor $\mathbb{P}^{n-1}$. We denote these bundles by $P_a$, for $1 \le a \le k$.

The line bundles $P_a$ play an important role in the
abelian/non-abelian correspondence of Harada-Landweber \cite{har}.
There is a surjective map $\phi:K((\mathbb{P}^{n-1})^k)^W\to K(Gr(k,n))$,
where $W \simeq S_k$ denotes the Weyl group of $\mathrm{GL}(k)$, acting 
by permutations on the factors.
Then $\phi$ preserves symmetric functions in the classes of $P_a$, and maps
the elementary symmetric functions $e_i(P_1, \ldots, P_k)$ to exterior powers 
$\wedge^i \mathcal{S}$ of $\mathcal{S}$.

\subsubsection{The two Chern-Simons parameters}

The setup that recovers the Coulomb branch equations is actually more general than what is discussed above. It actually involves two parameter-dependent twistings, depending on parameters $\ell$ and $w$. 

We consider $\widehat{I}_{Ab}^{\ell, w}$, the $I$-function of the abelianization of the Grassmannian $Gr(k,n)$ after applying the following series of twistings.
\begin{enumerate}
    \item Let $E:=\bigoplus_a P_a$. The first twisting is a Ruan-Zhang level structure given by 
    \[ (E, \ell) \]
    \item An Eulerian twisting given by $( (Eu_\lambda)^{w},F)$, where $F$ is defined to be 
    $$\bigoplus_{a\neq b} \, \frac{P_a}{P_b}$$. 
\end{enumerate}

In the notation of definition \ref{twistdef}, this twisting is given by 
$$\left( \left( {\rm det}^{-\ell} ,E \right),\left( (Eu_\lambda)^w,F \right) \right)\/.$$

Using the twisting formulas given in \Cref{ltoric}, we obtain the following formula for $\widehat{I}_{Ab}^{\ell, w}$:
$$\widehat{I}_{Ab}^{\ell, w}
    :=
    (1-\tau) \sum_{d_a}\frac{\prod_{a=1}^k q_a^{d_a} P_a^{\ell d_a}\tau^{\ell\binom{d_a}{2}}}{\prod_{j=1}^n\prod_{a=1}^k\prod_{m=1}^{d_a} (1-\tau^m P_a/\Lambda_j)}
    \prod_{a\neq b} \left(\frac{ \prod_{m=-\infty}^{d_a-d_b} \left(1-\tau^m\lambda\frac{P_a}{P_b}\right) }{ \prod_{m=-\infty}^{0}\left(1-\tau^m\lambda\frac{P_a}{P_b}\right) }\right)^w.$$

Applying $\phi$ yields the following.
    \begin{equation}
    I_{Gr}^{\ell,w}
    =
    (1-\tau) \sum_{d\geq 0}\sum_{\sum_ad_a=d}\frac{\prod_{a=1}^k q^{d_a} P_a^{\ell d_a}\tau^{\ell\binom{d_a}{2}}}{\prod_{j=1}^n\prod_{a=1}^k\prod_{m=1}^{d_a} (1-\tau^m P_a/\Lambda_j)}
    \left(\prod_{a\neq b}\frac{\prod_{m=-\infty}^{d_a-d_b} (1-\tau^m\frac{P_a}{P_b})}{\prod_{m=-\infty}^{0}(1-\tau^m\frac{P_a}{P_b})}\right)^w.
    \end{equation}

We will henceforth shorten the notation for this function to $I^{\ell, w}$.

\begin{remark} The Givental and Lee's quantum K ring is obtained by setting $w=1$ and $\ell=0$.
\end{remark}
To analyze the geometric window, we will set $w=1$ and let the level $\ell$ vary. To this aim, denote by
\[ \widehat{I}_{Ab}^{\ell}:= \widehat{I}_{Ab}^{\ell,1}\/; \quad I_{Gr}^{\ell}= I_{Gr}^{\ell,1} \/.\]
Then $I_{Gr}^{\ell}=\phi(\widehat{I}_{Ab}^{\ell})$ is the $I$-function of the Grassmannian with Ruan-Zhang level structure given by the tautological bundle 
$\mathcal{S}$ and the integer $\ell$.
In this context, $P_a$'s (abusively) represent the $K$-theoretic Chern roots of the tautological bundle $\mathcal{S}$. Note that the abelianized $I$ functions are symmetric in $P_a$'s, 
thus they determine well-defined rational functions of $\tau$ with coefficients in $K_T(\mathrm{Gr}(k,n))[\![q]\!]$. 
We also note that the $I$-function was calculated elsewhere by Dong-Wen in \cite[equation (44)]{DW}, and by Givental-Yan in \cite[Cor 7.2]{GY} (in this case, in the context of stable map theory).

As in the case of projective space, $I^\ell$ will be used to get 
information about the geometric window of the levels. However, unlike 
for projective spaces, it does not appear to satisfy any 
interesting $\tau-$difference equations. Instead, the $\tau$-difference 
equations whose symbols match the Coulomb branch equations come from 
the abelianized function $\widehat{I}_{Ab}^\ell$.

As in the case of projective space, we first modify the function to include a logarithmic prefactor:
$$\widehat{\mathbf{I}}_{Ab}^{\ell, w}:=\prod_i P_i^{\log(q_i)/\log(\tau)}\widehat{I}_{Ab}^{\ell, w} .$$
\begin{thm}
The abelianized $I$-function $\widehat{\mathbf{I}}_{Ab}^{\ell, w}$ satisfies the following $\tau$-difference equations:
\begin{eqnarray}\label{E:diffeqGr}
\lefteqn{
\prod_{b\neq a} (1-\tau\lambda \tau^{q_b\partial_{q_b}-q_a\partial_{q_a}})^w \prod_i (1-\tau^{q_a\partial_{q_a}}/\Lambda_i)\widehat{\mathbf{I}}_{Ab}^{\ell, w}} 
\nonumber \\
& = &
q_a(\tau^{q_a\partial_{q_a}})^\ell
\prod_{b\neq a} (1-\tau\lambda \tau^{q_a\partial_{q_a}-q_b\partial_{q_b}})^w\widehat{\mathbf{I}}_{Ab}^{\ell,w} \/,
\end{eqnarray}
for $1 \le a \le k$. 
\end{thm}
\begin{proof}

This result translates to the following equations without logarithmic prefactors:
\begin{eqnarray}
\lefteqn{
\prod_{b\neq a} (1-\tau\lambda P_b P_a^{-1}\tau^{q_b\partial_{q_b}-q_a\partial_{q_a}})^w \prod_i (1-P_a\tau^{q_a\partial_{q_a}}/\Lambda_i)\widehat{I}_{Ab}^{\ell, w}} 
\nonumber \\
& = &
q_a(P_a \tau^{q_a\partial_{q_a}})^\ell
\prod_{b\neq a} (1-\tau\lambda P_aP_b^{-1}\tau^{q_a\partial_{q_a}-q_b\partial_{q_b}})^w\widehat{I}_{Ab}^{\ell, w}.
\end{eqnarray}
We will prove the above version, to make notation easier to read.
Let $\widehat{I}_d$ be the $q^d$-term of 
$\frac{\widehat{I}_{Ab}^{\ell,w}}{1-\tau}$. Here $d$ is a vector with 
$k$ entries labelled $d_a$. Let $1_a$ denote the $a$th standard basis vector in 
$\mathbb{Z}^k$.
These terms satisfy the following recurrence relations. For $d_a>0$, we have
\begin{equation}
    \widehat{I}_d=q_aP_a^\ell \tau^{\ell(d_a - 1)}\frac{\prod_{b\neq a} \left(1-\tau^{d_a-d_b}\lambda\frac{P_a}{P_b}\right)^w }{ \prod_{j=1}^n(1-\tau^{d_a}P_a/\Lambda_j)\prod_{b \neq a} \left(1-\tau^{d_b-(d_a-1)}\lambda\frac{P_b}{P_a} \right)^w }
    \widehat{I}_{d-1_a}.
\end{equation}

Equivalently,
\begin{eqnarray}
\lefteqn{
    \prod_{j=1}^n \left(1-\tau^{d_a}P_a/\Lambda_j \right) \prod_{b \neq a} \left(1-\tau\lambda\tau^{d_b-d_a}\frac{P_b}{P_a} \right)^w
    \widehat{I}_d
    } \nonumber
    \\
    & = &
    q_aP_a^\ell\tau^{\ell (d_a-1)} \prod_{b \neq a} \left(1-\tau\lambda\frac{P_a}{P_b}\tau^{(d_a-1)-d_b} \right)^w 
    \widehat{I}_{d-1_a}.
\end{eqnarray}

This recurrence relation is equivalent to 
\begin{eqnarray}
\lefteqn{
\prod_{b\neq a} (1-\tau\lambda P_b P_a^{-1}\tau^{q_b\partial_{q_b}-q_a\partial_{q_a}})^w \prod_i (1-P_a\tau^{q_a \partial_{q_a}} / \Lambda_i) \widehat{I}_d}
\nonumber \\
& = &
q_a(P_a \tau^{q_a\partial_{q_a}})^\ell
\prod_{b\neq a} (1-\tau\lambda P_aP_b^{-1}\tau^{q_a\partial_{q_a}-q_b\partial_{q_b}})^w
\widehat{I}_{d-1_a}.
\end{eqnarray}

Summing over these equations for all $d$ yields the desired result.
\end{proof}

Taking symbols on each side of the equation \eqref{E:diffeqGr} yields 
\begin{equation} \label{gr:symbol}
\prod_{b\neq a}(1-\lambda P_b P_a^{-1})^w \prod_i (1-P_a/\Lambda_i) = q_a P_a^\ell \prod_{b\neq a} (1-\lambda P_a P_b^{-1})^w.
\end{equation}

Specializing $\lambda\to 1$ and $q_a\to q$, as prescribed by 
the quantum version of the abelian-nonabelian correspondence map $\phi$, 
and rearranging, yields:
\begin{equation}\label{eq:grass:math:bethe}
(-1)^{w(k-1)} (P_a)^\ell q\prod_b \left( \frac{P_a}{P_b} \right)^w
=
\prod_i (1-P_a/\Lambda_i).
\end{equation}

Thus we have established the following theorem:
\begin{thm}\label{grass:correspondence}
After changing $P_i$ to $X_i$, the symbol of the difference 
equations annihilating $\widehat{I}_{Ab}^{\ell, w}$ coincides with the Coulomb 
branch equations \eqref{eq:grass:phys:qk}, for a specific choice of Chern-Simons terms.
\end{thm}
In analogy to the study of the ordinary (untwisted) quantum $K$-theory, we make the following conjecture:
\begin{conj} Interpret the variables $X_a$ as the exponentials of the Chern roots of the tautological bundle $\mathcal{S}$. Let 
\[    -k < \ell \le n-k +1.\]
Then, for $w=1$, symmetric combinations of the equations 
\eqref{eq:grass:math:bethe} generate the ideal of relations in 
$QK^{\ell}_T(\mathrm{Gr}(k,n))$, the twisted quantum $K$-theory ring of $Gr(k,n)$, with twisting determined by $det^{-\ell}(\mathcal{S})$.
\end{conj}

In passing, we note that Sinha-Zhang \cite[Thm.~4.1]{sz} use a very slightly different
geometric window, given by
\begin{equation}
    -k \leq \ell \leq n-k.
\end{equation}
We leave the reconciliation for future work.

We note that the range of the level $\ell$ will appear below in relation to the physical 
geometric window; see equation \eqref{E:geom-window-Gr}.
For $\ell=0$, this conjecture was proved 
in \cite{Gu:2022yvj}.

\subsubsection{Mathematical geometric window}

In this section we specialize to the case that $w=1$, i.e., we only use the Eulerian twisting that appears in the abelian/nonabelian correspondence. 
We show below that, as in the projective space before, $I^\ell$ is equal to
the small $J$-function. 
The physical geometric window is realized mathematically via this condition, and the 
actual range is described by the following theorem of Givental-Yan (cf. \cite[Corollary 7.2]{GY}):

\begin{thm}[Givental-Yan \cite{GY}] Let $\ell $ be in the range 
\begin{equation}
\label{range}
    -k< \ell\leq n-k+1.
\end{equation}
Then  $I^\ell$ is equal to the small $J$-function of the quantum $K$-theory of $Gr(k,n)$ with level structure given by $\mathcal{S}$ and $\ell$.
\end{thm}
For projective spaces, this bound is sharp.

\begin{remark}
    The window cannot be directly read from $\widehat{I}_{Ab}^\ell$, since this function contains poles at $\tau=0$ for negative levels. 
\end{remark}

\subsubsection{Grassmann duality}
For this section, we also restrict to the case $w=1$ for simplicity. 
The Grassmannian $Gr(k,\mathbb{C}^n)$ is canonically isomorphic to the dual Grassmannian $Gr(n-k,(\mathbb{C}^n)^*)$, the Grassmannian of $n-k$-dimensional subspaces of $(\mathbb{C}^n)^*$, via the map sending $V\subset \mathbb{C}^n$ to $(\mathbb{C}^n/V)^*$. However, to describe how Ruan-Zhang levels behave under replacing $k$ with $n-k$, we instead use the (non-canonical) \emph{level-rank duality}, which we described as in e.g., \cite[Section 3.3]{GKM:QKYB}. (The use of the word `level' here is unrelated to the Ruan-Zhang level.)

Our choice of torus-equivariant parameters $\Lambda_i$ is equivalent to choosing an ordered basis for $\mathbb{C}^n$. Identifying $\mathbb{C}^n$ with $(\mathbb{C}^n)^*$
by sending the standard basis $\{e_i \}$ to the dual basis in 
opposite order $\{ e_{n+1-i}^* \}$,
one identifies $Gr(n-k,(\mathbb{C}^n)^*)$ with $Gr(n-k,\mathbb{C}^n)$; 
see, e.g., \cite[Section 3.3]{GKM:QKYB}. We denote by $\mu$ the resulting isomorphism.
This is $T$-equivariant, and it sends the equivariant parameter
$\Lambda_i$ to $\Lambda_{n+1-i}^{-1}$.
Via pullback, $\mu$ induces an isomorphism on $T$-equivariant $K$-theory, and of effective curve classes. However, the Grassmannian $Gr(n-k,\mathbb{C}^n)$ has a different GIT quotient interpretation (and thus a different abelianization), and a natural question 
is how the $I$-functions are related under this duality. 

Henceforth, we refer to $Gr(k,\mathbb{C}^n)$ and $Gr(n-k,\mathbb{C}^n)$ as 
$Gr$ and $Gr^*$ respectively. To fix notation, we denote the tautological 
bundle on $Gr^*$ by $\breve{\mathcal{S}}$ with $K$-theoretic Chern roots $\breve{P}_a$. 
The level-rank duality map $\mu$ satisfies: 
$$\mu^*(\breve{\mathcal{S}})=(\mathbb{C}^n/\mathcal{S})^*\in K_T(Gr) \/.$$
The morphism $\mu$ also induces an isomorphism between the 
moduli spaces of stable maps with targets $Gr$ and $Gr^*$. By a theorem of Yan \cite[Thm.~8]{XY} in the stable map theory, this isomorphism identifies twisted virtual structure sheaf corresponding to the level structure given by $\mathcal{S},\ell$ with a scaling of the sheaf corresponding to $\breve{\mathcal{S}}^*,-\ell$. Furthermore, this 
identification respects the small $J$-functions. 
Yan's observation yields the following corollary, which was proven directly by Dong and Wen in \cite[Thm.~1.1]{DW}:
\begin{cor}\label{level_corr}
    For $\ell$ in the range $-k < \ell <n-k$:
    $$I^\ell_{Gr}=\mu^*I_{Gr^*}^{-\ell} .$$
Here the levels are taken with respect to the bundles $\mathcal{S}$ and $\breve{\mathcal{S}}^*$, respectively. 
\end{cor}

\begin{remark}
    We note that the duality only applies to a subset of the range of the geometric window,
    $-k < \ell \leq n-k+1$, by equation~(\ref{range}).
   This subset is described by the condition that $\ell$ must lie in the geometric window for $Gr$, and $-\ell$ must lie in the geometric window for $Gr^*$.
\end{remark}

\begin{thm}
Corollary \ref{level_corr}, referred to in \cite{XY} and \cite{DW} as \emph{level correspondence} reflects the physical IR duality, in other words, starting from the (abelianized) $I$-function of $Gr^*$ with level structure given by $(\breve{S}^*,-\ell)$ and considering symbols of difference operators annihilating it, yields the Coulomb branch equations that are IR dual from the ones obtained from $Gr$, with the level structure $\mathcal{S}^{\ell}$. 
\end{thm}

\begin{proof}
We use the dual Grassmannian to give a mathematical derivation of 
the IR-dual Coulomb branch equations.

We do this by performing the exact same procedure as before, but using the dual Grassmannian with the corresponding level structure. The corresponding abelianized $I$-function is denoted $\widehat{I}_{Ab^*}^{-\ell}$. (Here $\ell$ is still the level on the original Grassmannian.) It is defined by
\begin{equation}
    \widehat{I}_{Ab^*}^{-\ell}
    :=
    (1-\tau) \sum_{d_a}\frac{\prod_{a=1}^{n-k} \breve{q}_a^{d_a} \breve{P}_a^{-\ell d_a}\tau^{-\ell\binom{d_a+1}{2}}}{\prod_{j=1}^n\prod_{a=1}^{n-k}\prod_{m=1}^{d_a} (1-\tau^m \breve{P}_a/\Lambda_j^{-1})}
    \prod_{a\neq b}\frac{\prod_{m=-\infty}^{d_a-d_b} \left(1-\tau^m\lambda\frac{\breve{P}_a}{\breve{P}_b}\right)}{\prod_{m=-\infty}^{0}\left(1-\tau^m\lambda\frac{\breve{P}_a}{\breve{P}_b}\right)}.
\end{equation}

The differences are as follows: Since we are taking the dual, the torus parameters are replaced with their inverses, and the exponent of $\tau$ is $-\ell\binom{d_a+1}{2}$ instead of $-\ell \binom{d_a}{2}$. The corresponding finite-difference equations are
\begin{eqnarray}
\lefteqn{
\prod_{b\neq a} \left(1-\tau\lambda \breve{P}_b \breve{P}_a^{-1}\tau^{\breve{q}_b\partial_{\breve{q}_b}-\breve{q}_a\partial_{\breve{q}_a}}\right)
\prod_i \left(1-\breve{P}_a\tau^{\breve{q}_a\partial_{\breve{q}_a}}/\Lambda_i^{-1}\right)
\widehat{I}_{Ab^*}^{-\ell}}
\nonumber \\
& \hspace*{1.5in} = &
(\breve{P}_a \tau^{\breve{q}_a\partial_{\breve{q}_a}})^{-\ell}\breve{q}_a
\prod_{b\neq a} \left(1-\tau\lambda \breve{P}_a\breve{P}_b^{-1}\tau^{\breve{q}_a\partial_{\breve{q}_a}-\breve{q}_b\partial_{\breve{q}_b}}\right)\widehat{I}_{Ab^*}^{-\ell}.
\end{eqnarray}

Taking symbols yields
\begin{equation}
\prod_{b\neq a}\left(1-\lambda \breve{P}_b \breve{P}_a^{-1}\right)
\left(\prod_i (1-\breve{P}_a\Lambda_i)\right) 
= \breve{P}_a^{-\ell}\breve{q}_a  
\prod_{b\neq a} \left(1-\lambda  \breve{P}_a \breve{P}_b^{-1}\right).
\end{equation}

Applying the map $\phi$ corresponds to the specialization $\lambda\mapsto 1$, $\breve{q}_a\mapsto \breve{q}$, yielding:
\begin{equation}
\label{neglevelbethe}
    (-1)^{n-k-1} \breve{P}_a^{-\ell} \breve{q} \left(\prod_{b=1}^{n-k} \frac{\breve{P}_b}{\breve{P}_a}\right)^{-1} = \prod_{i=1}^n \left(1 - \breve{P}_a\Lambda_i\right), \quad a = 1, 2, \dots, n-k.
\end{equation}
After taking a reciprocal of the equation, this is equivalent to the physics relation \eqref{eq:gr:dualcoulomb} for the IR dual theory
provided that we modify the 
expressions according to the map
\begin{equation} \label{eq:dual:dictionary}
    \breve{P}_a \mapsto \breve{X}_a^{-1}, \quad \breve{q} \mapsto \breve{q}^{-1}.
\end{equation}
\end{proof}
The reason for the exponent on $\breve{X}$ in Equation~(\ref{eq:dual:dictionary}) is that physics utilizes a GIT quotient presentation of
$Gr(n-k,({\mathbb C}^n)^*)$, whereas math is utilizing $Gr(n-k,{\mathbb C}^n)$, in order to match existing theorems in the literature.  The reason for the exponent on $\breve{q}$ is the same, as in physics the IR duality maps fundamentals to antifundamentals, so that the sign of the FI parameter flips.

\subsection{Dictionary between physics and mathematics}

\subsubsection{Prediction}

Next, we extract the prediction of conjecture~(\ref{eq:level-conj}) for this case.

To do so, we need to relate the twistings $\ell$, $w$ of this section to $\ell_{ab}$ used in~(\ref{eq:level-conj}). Corresponding to $\ell$, $w$, here are two Weyl-invariant elements of $\ell_{ab}$, one for $a = b$ (independent of the choice of $a$) and the other for $a \neq b$ (again otherwise independent of indices).  We can relate them by comparing the expression~(\ref{eq:grass:math:bethe}) for the ring in terms of $\ell$, $w$, to that for the toric case~(\ref{full-level-symbtoric}).  Those rings will have the same form if we identify
\begin{eqnarray}
    -w & = & \ell_{ab} \: \: \: \mbox{ for } a \neq b,
    \\
    \ell + (k-1) w & = & \ell_{aa} + \sum_{b \neq a} \ell_{ab} \: = \: \ell_{aa} - (k-1) w.
\end{eqnarray}
After a little algebra, 
we identify
\begin{equation}
    \ell_{ab} \: = \: \left\{ \begin{array}{cl}
    -w & a \neq b,
    \\
    \ell +2(k-1) w & a = b.
    \end{array} \right.
\end{equation}

With that identification, it is straightforward to solve~(\ref{eq:level-conj}), namely,
\begin{equation} 
    \kappa_{\rm eff}^{ab} \: = \: \left\{ \begin{array}{cl}
    \ell_{ab} & a \neq b,
    \\
    \sum_c \ell_{ac} & a=b, 
    \end{array} \right.
\end{equation}
for $\kappa_{\rm eff}^{ab}$ in terms of $\ell$, $w$.  One finds
\begin{equation}  \label{eq:gr:pred:int1}
    \kappa_{\rm eff}^{ab} \: = \: \left\{ \begin{array}{cl}
    -w & a \neq b, 
    \\
    \ell + (k-1) w & a = b,
    \end{array} \right.
\end{equation}

Now, from the definition~(\ref{eq:keff:genl}) and the expansion~(\ref{eq:grass:kappa-defns}),
\begin{eqnarray}
\kappa_{\rm eff}^{ab} & = & \kappa_{\rm eff}^{ba} \: = \:
\kappa^{ab} +
\frac{1}{2}\sum_{i=1}^{\dim V} \rho_i^a \rho_i^b \: = \: \kappa^{ab} + \frac{n}{2} \delta^{ab},
\\
& = &
\left( \kappa_{SU(k)} + \frac{n}{2} \right) \delta_{ab} + \frac{\kappa_{U(1)} - \kappa_{SU(k)}}{k} .
\label{eq:Gr:kappas2}
\end{eqnarray}
from which~(\ref{eq:gr:pred:int1}) implies
\begin{equation}
   \frac{\kappa_{U(1)} - \kappa_{SU(k)}}{k} \: = \: -w, \: \: \:
   \kappa_{SU(k)} + \frac{n}{2} +  \frac{\kappa_{U(1)} - \kappa_{SU(k)}}{k}
   \: = \: \ell + (k-1) w.
\end{equation}

After some algebra, we have
\begin{equation}  \label{eq:grass:dictionary}
    \ell = \kappa_{U(1)} + \frac{n}{2}, \quad \frac{\kappa_{U(1)} - \kappa_{SU(k)}}{k} = - w.
\end{equation}

Equation~(\ref{eq:grass:dictionary}) is our prediction for the dictionary between the Chern-Simons levels in physics and the mathematical twistings of the quantum K-theory ring of the Grassmannian $Gr(k,n)$, generalizing results for projective spaces.

From~(\ref{eq:CS:ordinary-qk}), we also predict that one recovers ordinary quantum K theory in the special case that
\begin{equation}
    \kappa_{\rm eff}^{ab} \: = \: \kappa_{\rm eff}^{ba} \: = \:
    + \frac{1}{2} \sum_{p \in P} \alpha^a_p \alpha^b_{p} \: = \: + \frac{1}{2} (2 k \delta^{ab} - 2 )
    \: = \: k \delta^{ab} - 1,
\end{equation}
where we have used
\begin{equation}
    \alpha^a_p \: = \: \alpha^a_{\mu \nu} \: = \: - \delta^a_{\mu} + \delta^a_{\nu}
\end{equation}
for $\mu, \nu \in \{1, \cdots k\}$ (see e.g.~\cite[section 4.1]{Gu:2018fpm}).
From~(\ref{eq:dict:su-u}), this means ordinary quantum K theory arises for
\begin{equation}
    \kappa_{U(1)} \: = \: - n/2, \: \: \: \kappa_{SU(k)} \: = \: k - n/2,
\end{equation}
or equivalently
\begin{equation}
    \ell \: = \: 0, \: \: \: w \: = \: +1.
\end{equation}

\subsubsection{Comparison of chiral rings, geometric windows}

Comparing the Coulomb branch equations~(\ref{eq:grass:phys:qk}) to the symbols of $\tau$-difference operators~(\ref{gr:symbol}),
we see immediately that with the dictionary~(\ref{eq:grass:dictionary}) above, the equivariant ring relations also match, so long as we also make
the minor identification
\begin{equation}
    q_{\rm phys} \: = \: (-1)^{(w-1)(k-1)} q .
\end{equation}

In addition, the allowed ranges of the Chern-Simons levels and Ruan-Zhang parameters also match for the case $w=1$.
Recall that the allowed values of the Chern-Simons levels are
\begin{equation}
    -\frac{n}{2} < \kappa_{SU(k)} \le \frac{n}{2} + 1.
\end{equation}
Under the dictionary~(\ref{eq:grass:dictionary}), this matches the allowed range of the Ruan-Zhang
parameter $\ell$, which is
\begin{equation}\label{E:geom-window-Gr}
    -k < \ell \le n-k +1.
\end{equation}
(Computations of geometric windows for $w \neq 1$ are considerably more complicated,
so as we are only performing consistency checks, we restrict to this case.)

This demonstrates Conjecture \ref{generalconj} for Grassmannians.
Furthermore, as has already been noted, this is compatible with duality.

\section{Flag manifolds}   \label{sect:flag}

In this section we will repeat the same analysis for the even more general case of flag manifolds.
As before, we will first outline the physics results for Coulomb branch relations and allowed ranges of Chern-Simons levels for which there are no extraneous topological vacua.  Then, we will give mathematical results for the Coulomb branch equations and allowed ranges of the Ruan-Zhang parameters.  Finally, we will note that a simple relationship preserves both the Coulomb branch structure and the allowed range of levels in the geometric windows.

\subsection{Physics results}

Consider the gauged linear sigma model for a flag manifold $Fl(k_1, \dots, k_s; n)$. The gauge group of the theory is $U(k_1) \times U(k_2) \times \cdots \times U(k_s)$. We have chiral superfields $\phi_{(i)}^{~(i+1)}$ that are in the representation $(\mathbf{k}_i, \overline{\mathbf{k}_{i+1}})$ of $U(k_i) \times U(k_{i+1})$, for $i = 1, 2, \cdots, s-1$,
where ${\bf k}$ denotes the fundamental representation of $U(k)$, and $\overline{\bf k}$ the antifundamental.  
We also have $n$ chiral superfields, that are in the fundamental representation of $U(k_s)$. 

\subsubsection{Chiral ring}

First, let us consider the physical description of untwisted (ordinary) quantum $K$-theory.
For that case, we fix the Chern-Simons levels such that,
for the $i$th step, the relevant twisted superpotential is
\begin{equation}
\begin{aligned}
    {\cal W}_i ={}& \frac{k_i}{2}  \sum_{a_i = 1}^{k_i} \left(\ln X^{(i)}_{a_i}\right)^2 - \frac{1}{2} \left(\sum_{a_i = 1}^{k_i}\ln X^{(i)}_{a_i}\right)^2 + \ln \left((-1)^{k_i - 1} q_{ {\rm phys}, i}\right) \sum_{a_i = 1}^{k_i} \ln X^{(i)}_{a_i}\\
    & + \sum_{a_i = 1}^{k_i} \sum_{a_{i-1} = 1}^{k_{i-1}} \mathrm{Li}_2 \left(\frac{X^{(i-1)}_{a_{i-1}} }{X^{(i)}_{a_i}}\right) + \sum_{a_i = 1}^{k_i} \sum_{a_{i+1} = 1}^{k_{i+1}}\mathrm{Li}_2\left(\frac{X^{(i)}_{a_i}}{X^{(i+1)}_{a_{i+1}}}\right).
\end{aligned}
\end{equation}
This holds for $i = 1, 2, \cdots, s$, if we identify $X^{(s+1)}$ with the equivariant parameters
$\Lambda$,  and $k_{s+1} = n$, $k_0 = 0$. The Coulomb branch equation then is given by
\begin{equation}
    (-1)^{k_i - 1} q_{ {\rm phys}, i} \left(\prod_{b_i = 1}^{k_i} \frac{X^{(i)}_{b_i}}{X^{(i)}_{a_i}}\right)^{-1} = \frac{\prod\limits_{a_{i+1} =1}^{k_{i+1}} \left(1 - X^{(i)}_{a_i}/X^{(i+1)}_{a_{i+1}} \right) }{\prod\limits_{a_{i-1}=1}^{k_{i-1}} \left(1 - X^{(i-1)}_{a_{i-1}}/ X^{(i)}_{a_i}\right) }, \quad a_i = 1, 2, \cdots, k_i.
\end{equation}

Next we restore more general Chern-Simons levels.  Briefly, 
\begin{itemize}
    \item Each $U(k_i)$ gauge factor gives rise to both $U(1)$ and $SU(k_i)$ levels.  These are simultaneously encoded as Weyl-invariant elements of a matrix, as
    \begin{equation}\label{eqn:CS-levels}
    \begin{aligned}
    \kappa^{(i)\, \text{eff}}_{a_i, b_i} &= \kappa^{(i)}_{a_i, b_i} + \frac{1}{2}(k_{i-1}+k_{i+1}) \delta_{a_i, b_i}~,
    \\
     \kappa^{(i)}_{a_i,b_i} & = \kappa^{(i)}_{SU(k_i)} \delta_{a_i, b_i} + \frac{\kappa^{(i)}_{U(1)} - \kappa^{(i)}_{SU(k_i)}}{k_i}, \quad a_i, b_i = 1, 2, \dots, k_i.
     \end{aligned}
    \end{equation}
    \item In addition, there can also be mixed $U(1)-U(1)$ levels between the determinants of different $U(k_i)$ gauge factors.  We will denote these levels, between the $U(1)$ factors of $U(k_i)$, $U(k_j)$, $i \neq j$ by
    $\kappa_{\rm eff}^{(i,j)}$.
    
    Note that most of $\kappa^{(i,j)}_{\rm eff}$ do not receive one-loop corrections because quantum corrections arise due to matter fields charged under both $U(k_i)$ and $U(k_j)$. In our case, this only happens when $j=i+1$ or $j=i-1$. The explicit form of the correction can be derived, but as it will not be pertinent to our discussion here, we will just use $\kappa_{\rm eff}^{(i,j)}$ directly. 
\end{itemize}

These can be encoded in a single matrix of Chern-Simons levels $\kappa_{\rm eff}^{(i,a_i), (j,b_j)}$,
with $i, j \in \{1, \cdots, s\}$, $a_i \in \{1, \cdots, k_i\}$,
$b_j \in \{1, \cdots, k_j\}$, as
\begin{eqnarray}  \label{eq:CS-levels:flag:comp1}
    \kappa_{\rm eff}^{(i,a_i), (j,b_j)} 
    & = &
    \delta^{ij} \kappa^{(i) {\rm eff}}_{a_i, b_i} \: + \: \kappa_{\rm eff}^{(i,j)},
    \\
    & = & \delta^{ij} \left( \kappa^{(i)}_{a_i, b_i} + \frac{1}{2} \left( k_{i-1} + k_{i+1} \right) \delta_{a_i, b_i} \right) \: + \: \kappa_{\rm eff}^{(i,j)},
    \label{eq:CS-levels:flag:comp2}
\end{eqnarray}
in the convention that $\kappa_{\rm eff}^{(i,j)} = 0$ if $i=j$.

Then, the part of the effective twisted superpotential relevant to the $U(k_i)$ factor is given by\footnote{This is the part of the superpotential relevant to differentiating at the $i$th gauge factor, hence, the mixed term is not to be summed over $i$ so as to avoid double-counting quadratic terms.}
\begin{equation}
\begin{aligned}
    {\cal W}_i 
    = {}& \frac{1}{2} \sum_{a_i, b_i = 1}^{k_i}\kappa^{(i) \,\text{eff}}_{a_i, b_i} \left(\ln X^{(i)}_{a_i}\right) \left(\ln X^{(i)}_{b_i}\right)
    + \sum_{j \neq i}\kappa^{(i,j)}_{\rm eff} \sum_{a_i = 1}^{k_i} \sum_{b_j = 1}^{k_j}(\ln X^{(i)}_{a_i}) (\ln X^{(j)}_{b_j})
    \\
    &+ \ln \left((-1)^{k_i - 1} q_{ {\rm phys}, i}\right) \sum_{a_i = 1}^{k_i} \ln X^{(i)}_{a_i}
    + \sum_{a_i = 1}^{k_i} \sum_{a_{i-1} = 1}^{k_{i-1}} \mathrm{Li}_2 \left(\frac{X^{(i-1)}_{a_{i-1}} }{X^{(i)}_{a_i}}\right) + \sum_{a_i = 1}^{k_i} \sum_{a_{i+1} = 1}^{k_{i+1}}\mathrm{Li}_2\left(\frac{X^{(i)}_{a_i}}{X^{(i+1)}_{a_{i+1}}}\right)~.
\end{aligned}
\end{equation}

The Chern-Simons levels that give ordinary quantum K-theory are
$\kappa_{\rm eff}^{(i,j)} = 0$ for $i \neq j$, and
\begin{equation}
\begin{aligned}
    \kappa^{(i)}_{SU(k_i)} = k_i - \frac{k_{i-1}}{2} - \frac{k_{i+1}}{2},\quad
    \frac{\kappa^{(i)}_{U(1)} - \kappa^{(i)}_{SU(k_i)}}{k_i} = -1.
\end{aligned}
\end{equation}
For general Chern-Simons levels, the Coulomb branch equation is given by
\begin{eqnarray} 
\lefteqn{
    (-1)^{k_i - 1} q_{ {\rm phys}, i} \left(X^{(i)}_{a_i}\right)^{\kappa^{(i)}_{U(1)} + \frac{k_{i-1}}{2} + \frac{k_{i+1}}{2} } \left(\prod_{j \neq i} \prod_{b_j = 1}^{k_j} (X^{(j)}_{b_j})^{\kappa_{\rm eff}^{(i,j)}}\right)\left(\prod_{b_i = 1}^{k_i} \frac{X^{(i)}_{b_i}}{X^{(i)}_{a_i}}\right)^{\frac{\kappa^{(i)}_{U(1)} - \kappa^{(i)}_{SU(k_i)}}{k_i}} 
    } \nonumber
    \\
   & \hspace*{2.5in} = &
    \frac{\prod\limits_{a_{i+1} =1}^{k_{i+1}} \left(1 - X^{(i)}_{a_i}/X^{(i+1)}_{a_{i+1}} \right) }{\prod\limits_{a_{i-1}=1}^{k_{i-1}} \left(1 - X^{(i-1)}_{a_{i-1}}/ X^{(i)}_{a_i}\right) }~.
    \label{eq:flag:phys:bethe}
\end{eqnarray}
for $a_i = 1, 2, \cdots, k_i$.

\subsubsection{IR dualities and windows}

We can also consider IR dualities in this case. (We will assume that there are no mixed $U(1)-U(1)$ levels, for simplicity.
We will also assume that the levels lie in a certain range, namely the geometric window, and we will come back to this later.) There is a duality between the following 3d $\mathcal{N}=2$ theories
\begin{itemize}
    \item a $U(k_1)_{\kappa_1, \kappa_1 - k_1} \times U(k_2)_{\kappa_2, \kappa_2 - k_2} \times \cdots \times U(k_s)_{\kappa_s, \kappa_s - k_s}$ gauge theory
    with bifundamentals transforming in $(\mathbf{k}_i, \overline{\mathbf{k}_{i+1}})$ of $U(k_i) \times U(k_{i+1})$ for $i = 1, 2, \cdots, s-1$, and $n$ chiral superfields in the fundamental representation of $U(k_s)$. (Here, $\kappa_i$ is understood as the $SU(k_i)$ level of the gauge group factor $U(k_i)$ for $i = 1, 2, \cdots, s$.)
    \item a $U(n-k_s)_{-\kappa_s, -\kappa_s - (n-k_s)} \times U(n- k_{s-1})_{-\kappa_{s-1}, - \kappa_{s-1} - (n-k_{s-1})} \times \cdots \times U(n-k_1)_{-\kappa_1, -\kappa_1 - (n-k_1)}$ gauge theory with bifundamentals transforming in $(\overline{\mathbf{n-k_i}}, \mathbf{n-k_{i-1}} )$  in $U(n-k_i) \times U(n-k_{i-1})$ for $i = s, s-1, \cdots, 2$, and $n$ chiral superfields in the antifundamental representation of $U(n-k_1)$.
\end{itemize}

This gives the physics description of the following duality in math
\begin{equation} \label{eq:flag-duality}
    Fl(k_1, k_2, \cdots, k_s; \mathbb{C}^n) \cong Fl(n-k_s, n-k_{s-1}, \cdots, n-k_1; (\mathbb{C}^n)^*). 
\end{equation}
Notice that this is more complicated than the case of Grassmannian because if we look at the middle step, a $U(k_i)$ gauge factor, there are not only fields in fundamental representation, but also fields in antifundamental representation involved. We haven't really covered the dual theory of a $U(k)$ gauge theory with both fundamental fields and antifundamental fields involved, though this is covered in \cite{Closset:2023vos}. The dual theory is more complicated, and involves a gauge singlet meson, and a superpotential. The strategy we adopt here is that we dualize the unitary group factors in a certain order such that during each step, there are only fundamentals of the unitary group factor we try to dualize, and we can simply use the duality we discussed in the Grassmannian case.

Let us illustrate this using the example of a two-step flag $Fl(k_1, k_2; n)$. The original theory is a $U(k_1)_{\kappa_1, \kappa_1 - k_1} \times U(k_2)_{\kappa_2, \kappa_2 - k_2}$ gauge theory with one chiral superfield transforming in the bifundamental representation $(\mathbf{k}_1, \overline{\mathbf{k}_2})$ of $U(k_1) \times U(k_2)$, and $n$ chiral superfields in the fundamental representation of $U(k_2)$. The dualizing steps are as follows
\begin{itemize}
    \item We first dualize $U(k_1)_{\kappa_1, \kappa_1 - k_1}$ with $k_2$ fundamentals, where we view $U(k_2)$ as a flavor symmetry group, and $\kappa_2$ a flavor Chern-Simons level. The dual to this factor is a $U(k_2 - k_1)_{-\kappa_1, -\kappa_1 - (k_2 - k_1)}$ gauge theory with $k_2$ chiral superfields transforming in the antifundamental representation of $U(k_2 - k_1)$. These fields should also transform in the fundamental representation of $U(k_2)$, and the flavor level should be modified as $\kappa_2 \to \kappa_1 + \kappa_2$
    \cite{cyrilpriv}.
    Altogether\footnote{
    In passing, we note a curious mathematical relationship.  In addition to the biholomorphic relation~(\ref{eq:flag-duality}), there also exist other flag manifolds related by diffeomorphisms which are not necessarily biholomorphic, obtained by dualizing single factors, as discussed in \cite[section 2.4]{Donagi:2007hi}.  Curiously, the intermediate theories obtained in the series of dualities above have the same form as the diffeomorphic (but not biholomorphic) flag manifolds discussed in \cite[section 2.4]{Donagi:2007hi}, except for the fact that some matter representations are dualized. To be clear, because the matter representations do not coincide, we are not claiming an IR duality relating merely diffeomorphic flag manifolds; rather, we are merely noting that those diffeomorphic flag manifolds almost appear.  
    } we have a $U(k_2 - k_1)_{-\kappa_1, -\kappa_1 - (k_2- k_1)} \times U(k_2)_{\kappa_1+\kappa_2 , \kappa_1 + \kappa_2 - k_2}$ gauge theory with one chiral superfield transforming in the bifundamental representation $(\overline{\mathbf{k_2 - k_1}}, \mathbf{k_2} )$ of $U(k_2- k_1) \times U(k_2)$, and $n$ chiral superfields in the fundamental representation of $U(k_2)$.
    \item Now if we look at $U(k_2)$, there are now only fields in the fundamental representation of $U(k_2)$, and there are $(k_2-k_1) + n$ of them. Dualizing this $U(k_2)_{\kappa_1 + \kappa_2, \kappa_1 + \kappa_2 - k_2}$ gauge theory gives a $U(n-k_1)_{-(\kappa_1 + \kappa_2), - (\kappa_1 + \kappa_2) - (n - k_1)}$ gauge theory with $(k_2 - k_1) + n$ antifundamentals, among which $(k_2 - k_1)$ of them transform in the fundamental representation of $U(k_2 - k_1)$.
    The flavor level of $U(k_2- k_1)$ also transforms as $-\kappa_1 \to -\kappa_1 + (\kappa_1 + \kappa_2) = \kappa_2$. Now we have a $U(k_2 - k_1)_{\kappa_2, \kappa_2 - (k_2-k_1)} \times U(n-k_1)_{-(\kappa_1 + \kappa_2), -(\kappa_1 + \kappa_2) - (n-k_1)}$ gauge theory with one chiral superfield transforming in the bifundamental representation $(\mathbf{k_2 - k_1}, \overline{\mathbf{n-k_1}})$ of $U(k_2 - k_1) \times U(n-k_1)$, and $n$ chiral superfields in the antifundamental representation of $U(n-k_1)$.
    \item Finally we dualize the first gauge factor $U(k_2 - k_1)_{\kappa_2, \kappa_2 - (k_2-k_1)}$. There are $n-k_1$ fields in its fundamental representation. The dual theory is a $U(n-k_2)_{-\kappa_2, -\kappa_2 - (n-k_2)}$ gauge theory with $n-k_1$ antifundamentals which also transform in the fundamental representation of $U(n-k_1)$. The flavor level associated with $U(n-k_1)$ transforms as $-(\kappa_1 + \kappa_2) \to -(\kappa_1 + \kappa_2) + \kappa_2 = - \kappa_1$. Now we have a $U(n-k_2)_{-\kappa_2, -\kappa_2 - (n-k_2)} \times U(n-k_1)_{-\kappa_1, -\kappa_1 - (n-k_1)}$ gauge theory with one chiral superfield transforming in the bifundamental representation $(\overline{\mathbf{n-k_2}}, \mathbf{n-k_1})$ of $U(n-k_2) \times U(n-k_1)$, and $n$ chiral superfields in the antifundamental representation of $U(n-k_1)$.
\end{itemize}

The reader should note that in each step above, the duality transformation was only valid for Chern-Simons levels in a particular range (which for ordinary Grassmannians nearly coincided
with the geometric window).  In the example above, the first duality was defined for
\begin{equation}  \label{eq:flag:r1}
    | \kappa_1 | < k_2/2,
\end{equation}
the second step was defined for
\begin{equation}   \label{eq:flag:r2}
    | \kappa_1 + \kappa_2 | < \frac{1}{2} \left(k_2-k_1 + n \right),
\end{equation}
and the third step was defined for
\begin{equation}  \label{eq:flag:r3}
    | \kappa_2 | < \frac{1}{2} \left( n-k_1 \right).
\end{equation}
The inequality~(\ref{eq:flag:r2}) is easily checked to be a consequence of the other two,
hence we require
\begin{equation}
    | \kappa_1 | < \frac{ k_2}{ 2 },  \: \: \: 
    | \kappa_2 | < \frac{1}{2} \left( n-k_1 \right).
\end{equation}
We do not have an explicit expression for the geometric window for a flag manifold,
but it is natural to conjecture that the geometric window is given by a partial closure of the region defined by the inequalities above, and their generalizations to other flag manifolds, 
just as we saw for ordinary
Grassmannians.

For a more general flag manifold $Fl(k_1, k_2, \cdots, k_s; {\mathbb C}^n)$,
\begin{itemize}
    \item We dualize at successive nodes starting with $U(k_1)$, $U(k_2)$, and progressing to $U(k_s)$.  At each step, the node at which one dualizes has only fundamentals.
    \item Next, we start at the node that was originally $U(k_1)$, now $U(k_2-k_1)$, and dualize up to the next-to-rightmost node -- ending at what used to be the $U(k_{s-1})$ node.
    \item We then recursively repeat, each step ending one node before the last of the previous cycle.
\end{itemize}
At each stage, the number of nodes to dualize shrinks by one.
Dualizations end when the shrinking number of nodes reaches zero.
For example, in the two-step flag manifold $Fl(k_1, k_2; {\mathbb C}^n)$,
we first dualized successively at the $U(k_1)$ and $U(k_2)$ nodes,
then, after that, we dualized what used to be the $U(k_1)$ node, but not what used to be the
$U(k_2)$ node. Based on computations in examples, this seems to require that the Chern-Simons levels obey
\begin{equation}  \label{eq:ineq-main}
    | \kappa_i | < \frac{1}{2} \left( k_{i+1} - k_{i-1} \right)
\end{equation}
in general.  (These inequalities emerge when dualizing at the first node, and inequalities arising when dualizing at other nodes seem to be weaker statements, implied by~(\ref{eq:ineq-main}).) As above, we conjecture that a partial closure of the region defined by these inequalities gives the geometric window.

In passing, the Coulomb branch equation for the dual theory is
\begin{equation}
\label{eq:phys-flag-dual}
    (-1)^{n-k_i -1} \breve{q}_{ {\rm phys}, i} \left(\breve{X}^{(i)}_{a_i}\right)^{-\kappa_i +k_i -\frac{1}{2}k_{i-1} - \frac{1}{2}k_{i+1}} \left(\prod_{b_i = 1}^{n-k_i} \frac{\breve{X}^{(i)}_{b_i} }{\breve{X}^{(i)}_{a_i}}\right)^{-1} = \frac{\prod\limits_{a_{i+1} = 1}^{n-k_{i+1}}\left(1 - \breve{X}^{(i)}_{a_i}/\breve{X}^{(i+1)}_{a_{i+1}}\right) }{\prod\limits_{a_{i-1} =1 }^{n-k_{i-1}} \left(1 - \breve{X}^{(i-1)}_{a_{i-1}}/\breve{X}^{(i)}_{a_i}\right) }~.
\end{equation}

\subsection{Math results}
\subsubsection{The flag and its Abelianization}

For the flag manifold $Fl:= Fl(k_1, \dots, k_s; n)$, the key bundles we consider are the tautological subbundles $\mathcal{S}_i$, corresponding to the fundamental representations of $GL(k_i)$ in the GIT quotient description.

The abelianization of the flag is a toric variety, given by a 
tower of $(\mathbb{P}^{k_i-1})^{k_{i-1}}$-bundles, with base $(\mathbb{P}^{n-1})^{k_s}$. 
The relationship between the $K$-theory of the abelianization and the $K$-theory of the flag itself is similar to the Grassmannian case. As in the Grassmannian case, if we denote the $a$th copy of $\mathcal{O}(-1)$ 
on $(\mathbb{P}^{k_{i+1}-1})^{k_{i}-1}$ as $P_a^i$, the abelian/nonabelian correspondence for classical $K$-theory sends $e_r(P_1^i, \cdots, P^i_{k_i})$ to $\wedge^r\mathcal{S}_i$. 
 The Weyl group is $\prod_i S_{k_i}$ and the $i$th factor permutes the bundles $P_a^i$.

\subsubsection{Operators and Coulomb branch equations}

From the GIT quotient description, we have $s$ $U(1)$ levels, $s$ nonabelian levels (one $SU(k_i)$ for each $i$), and $(s^2-s)/2$ mixed levels (for each pair of different $U(1)$-levels). 

These correspond to the following twistings of the quantum $K$-theory of the abelianization:

\begin{itemize}
    \item $U(1)$-levels: Ruan-Zhang level structures given by the bundles $\bigoplus_a P^i_a$ and integers $\ell_i$. On the flag itself, the corresponding bundles are $\mathcal{S}_i$.
    \item $SU$-levels: Twistings given by $((Eu_\lambda)^{w_i},\bigoplus_{a\neq b} (P^i_a / P^i_b ) )$. On the flag, this corresponds to the twisting $( (Eu_\lambda)^{w_i-1},\mathcal{A}_i )$, where $\mathcal{A}_i$ is the bundle induced by the adjoint representation on $GL(k_i)$. As discussed, the reason for subtracting 1 from $w_i$ is that one part of the twisting is used in the abelian/non-abelian correspondence itself. 

    \item Mixed $U(1)\times U(1)$-levels: Ruan-Zhang level structures given by the bundles $(\bigoplus_a P^i_a)\otimes (\bigoplus_b P^j_b)$ and integers $\ell_{i,j}$. On the flag itself, the corresponding bundles are $\mathcal{S}_i\otimes \mathcal{S}_j$. 
\end{itemize}
We will compute the abelianized $I$-function $\widehat{I}_{Ab}^{\ell,w}$. To preserve readability, we set all $\ell_{i,j}=0$, and deal with mixed levels in a later section.

Thus in this case $\widehat{I}_{Ab}^{\ell,w}$ is the $I$-function of the abelianization of the flag after applying the following series of twistings:
\begin{enumerate}
    \item Let $E_i:=\bigoplus_a P_a^i$. The first twisting is a Ruan-Zhang level structure given by 
    $$\left( \left(  E_1,\ell_1 \right), \left( E_2, \ell_2 \right), \dots \right).$$ 
    \item A sequence of Eulerian twistings given by $((Eu_\lambda)^{w_i},F_i)$, where $F_i$ is defined to be 
    $$\bigoplus_{a\neq b} \, \frac{P_a^i}{P_b^i}.$$    
\end{enumerate}

In the notation of definition \ref{twistdef}, this twisting is given by 
$$\left( \left( {\rm det}^{-\ell_1} ,E_1 \right), \dots, \left( {\rm det}^{-\ell_s},E_s \right),\left( (Eu_\lambda)^{w_1},F_1 \right),\dots \right).$$
Using the twisting theorems described previously, we have:

\begin{eqnarray}
\widehat{I}_{Ab}^{\ell,w}
& = & (1-\tau) \sum_{d\geq 0} \prod_{i,a} (q^i_{a})^{d^i_{a}}(P^i_a)^{\ell_i d^i_a}\tau^{\ell_i\binom{d^i_a}{2}}
 \\
& & \times
\frac{
\prod_{i=1}^s\prod_{a\neq b}^{1\leq a,b\leq k_i}\widehat{\prod}_{l=1}^{d^i_{a}-d^i_{b}}
\left(1-\lambda\frac{P^i_{a}}{P^i_{b}}\tau^l\right)^{w_i}
}{
\prod_{i=1}^{s-1}\prod_{1\leq b\leq k_{i+1}}^{1\leq a\leq k_i}\widehat{\prod}_{l=1}^{d^i_{a}-d^{i+1}_b}(1-\frac{P^i_{a}}{P^{i+1}_b}\tau^l)\cdot\prod_{1\leq b\leq n}^{1\leq a\leq k_s}\widehat{\prod}_{l=1}^{d^s_{a}}
\left(1-\frac{P^s_{a}}{\Lambda_{b}}\tau^l\right)
}.
\nonumber
\end{eqnarray}
We use the notation  $\widehat{\prod}_{i=1}^k f_i$ to mean 
\begin{equation} \frac{\prod_{i=-\infty}^k f_i}{\prod_{j=-\infty}^0 f_j}, \nonumber \end{equation} 
for the purpose of keeping the equation legible.

After applying the logarithmic prefactor, we obtain
\begin{eqnarray}
\widehat{\mathbf{I}}_{Ab}^{\ell, w}
& := & \prod_{i,a}(P^i_a)^{\log(q^i_a)/\log(\tau)}(1-\tau) \sum_{d\geq 0} \prod_{i,a} (q^i_{a})^{d^i_{a}}(P^i_a)^{\ell_i d^i_a}\tau^{\ell_i\binom{d^i_a}{2}}
 \\
& & \times
\frac{
\prod_{i=1}^s\prod_{a\neq b}^{1\leq a,b\leq k_i}\widehat{\prod}_{l=1}^{d^i_{a}-d^i_{b}}(1-\lambda\frac{P^i_{a}}{P^i_{b}}\tau^l)^{w_i}
}{
\prod_{i=1}^{s-1}\prod_{1\leq b\leq k_{i+1}}^{1\leq a\leq k_i}\widehat{\prod}_{l=1}^{d^i_{a}-d^{i+1}_b}(1-\frac{P^i_{a}}{P^{i+1}_b}\tau^l)\cdot\prod_{1\leq b\leq n}^{1\leq a\leq k_s}\widehat{\prod}_{l=1}^{d^s_{a}}
\left(1-\frac{P^s_{a}}{\Lambda_{b}}\tau^l\right)
}.
\nonumber
\end{eqnarray}

We can find the $\tau$-difference equations solved by this function in essentially the same manner as for the Grassmannian, yielding the following:
\begin{thm} For $1\leq i\leq s,1\leq a\leq k_i$, we have:
\begin{equation}
\frac{q^i_a\left( \tau^{q^i_a\partial_{q^i_a}} \right)^{\ell_i}
\prod_b\left(1-\tau^{q^{i-1}_b\partial_{q^{i-1}_b}-q^i_a\partial_{q^i_a}} \right)
\prod_{b\neq a} \left(1-\tau\lambda \tau^{q^i_a\partial_{q^i_a}-q^i_b\partial_{q^i_b}}\right)^{w_i}}{\prod_{b\neq a} \left(1-\tau\lambda \tau^{q^i_b\partial_{q^i_b}-q^i_a\partial_{q^i_a}}\right)^{w_i}
\prod_b \left(1-\tau^{q^i_a\partial_{q^i_a}-q^{i+1}_b\partial_{q^{i+1}_b}}\right)}\widehat{\mathbf{I}}_{Ab}^{\ell,w}= \widehat{\mathbf{I}}_{Ab}^{\ell,w}.
\nonumber
\end{equation}
\end{thm}  In the above, we have used the convention that $k_0=0,q^{s+1}_a=0$, and $P^{s+1}_b = \Lambda_b$
\begin{proof}

The equation is equivalent to the following equation for the $I$-function without modification
\begin{equation}
\frac{q^i_a\left(P^i_a \tau^{q^i_a\partial_{q^i_a}} \right)^{\ell_i}
\prod_b\left(1-\frac{P^{i-1}_b}{P^{i}_a}\tau^{q^{i-1}_b\partial_{q^{i-1}_b}-q^i_a\partial_{q^i_a}} \right)
\prod_{b\neq a} \left(1-\lambda \tau\frac{P_a^i}{P_b^{i}}\tau^{q^i_a\partial_{q^i_a}-q^i_b\partial_{q^i_b}}\right)^{w_i}}{\prod_{b\neq a} \left(1-\lambda \tau\frac{P^i_b}{P^i_a}\tau^{q^i_b\partial_{q^i_b}-q^i_a\partial_{q^i_a}}\right)^{w_i}
\prod_b \left(1-\frac{P^i_a}{P^{i+1}_b}\tau^{q^i_a\partial_{q^i_a}-q^{i+1}_b\partial_{q^{i+1}_b}}\right)}\widehat{I}_{Ab}^{\ell, w}= \widehat{I}_{Ab}^{\ell, w}.
\nonumber
\end{equation}

This equation is equivalent to the following relationship between the terms of the $I$-function
\begin{eqnarray}
\label{flageq}
\lefteqn{
\prod_{b\neq a} \left(1-\lambda \tau\frac{P^i_b}{P^i_a}\tau^{q^i_b\partial_{q^i_b}-q^i_a\partial_{q^i_a}}\right)^{w_i}
\prod_b \left(1-\frac{P^i_a}{P^{i+1}_b}\tau^{q^i_a\partial_{q^i_a}-q^{i+1}_b\partial_{q^{i+1}_b}}\right)\widehat{I}_{Ab}^{\ell,w}
}
\\
& = &
q^i_a\left(P^i_a \tau^{q^i_a\partial_{q^i_a}} \right)^{\ell_i}
\prod_b\left(1-\frac{P^{i-1}_b}{P^{i}_a}\tau^{q^{i-1}_b\partial_{q^{i-1}_b}-q^i_a\partial_{q^i_a}} \right)
\prod_{b\neq a} \left(1-\lambda \tau\frac{P_a^i}{P_b^{i}}\tau^{q^i_a\partial_{q^i_a}-q^i_b\partial_{q^i_b}}\right)^{w_i}\widehat{I}_{Ab}^{\ell,w}.
\nonumber
\end{eqnarray}

Writing $\widehat{I}_{Ab}^{\ell,w}=(1-\tau)\sum_{d\geq 0}\widehat{I}_d$, where $d$ is a vector of entries labelled $d^i_a$, $i\leq s$, $a\leq k_i$, we can observe that
\begin{equation}
\frac{\widehat{I}_{d}}{\widehat{I}_{d-1^i_a}} =q^i_a (P^i_a)^{\ell_i}\tau^{\ell_i (d^i_a-1)}\frac{\prod_{b\neq a} (1-\lambda P^i_a/P^i_b\tau^{d^i_a-d^i_b})^{w_i}}{\prod_{b\neq a} (1-\lambda P^i_b/P^i_a\tau^{d^i_b-(d^i_a-1)})^{w_i}}\frac{\prod_b (1-P^{i-1}_b/P^i_a \tau^{d^{i-1}_b-(d^i_a-1)})}{\prod_b (1-P^{i}_a/P^{i+1}_b \tau^{d^{i}_a-d^{i+1}_b})}.
\end{equation}

Equivalently,
\begin{eqnarray}
\lefteqn{\prod_{b\neq a} (1-\tau\lambda P^i_b/P^i_a\tau^{d^i_b-d^i_a})^{w_i}\prod_b (1-P^{i}_a/P^{i+1}_b \tau^{d^{i}_a-d^{i+1}_b})\widehat{I}_d} \\&=&q^i_a(P^i_a)^{\ell_i}\tau^{\ell_i(d^i_a-1)}\prod_{b\neq a}(1-\tau\lambda P^i_a/P^i_b\tau^{d^i_a-1-d^i_b})^{w_i}\prod_b (1-P^{i-1}_b/P^i_a \tau^{d^{i-1}_b-(d^i_a-1)})\widehat{I}_{d-1^i_a}.
\nonumber
\end{eqnarray}

However the left and right hand sides of this equation are just the operators from \eqref{flageq} applied to $\widehat{I}_d$ and $\widehat{I}_{d-1^i_a}$. Thus we conclude that
\begin{eqnarray}
\lefteqn{
\prod_{b\neq a} \left(1-\tau\lambda\frac{P^i_b}{P^i_a}\tau^{q^i_b\partial_{q^i_b}-q^i_a\partial_{q^i_a}}\right)^{w_i}
\prod_b \left(1-\frac{P^i_a}{P^{i+1}_b}\tau^{q^i_a\partial_{q^i_a}-q^{i+1}_b\partial_{q^{i+1}_b}}\right)\widehat{I}_d
}
\\
& = &
q^i_a\left(P^i_a \tau^{q^i_a\partial_{q^i_a}} \right)^{\ell_i}\prod_b\left(1-\frac{P^{i-1}_b}{P^{i}_a}\tau^{q^{i-1}_b\partial_{q^{i-1}_b}-q^i_a\partial_{q^i_a}} \right)
\prod_{b\neq a} \left(1-\lambda \tau\frac{P_a^i}{P_b^{i}}\tau^{q^i_a\partial_{q^i_a}-q^i_b\partial_{q^i_b}}\right)^{w_i}\widehat{I}_{d-1^i_a}.
\nonumber
\end{eqnarray}

Collecting terms over all $d$ concludes the proof.
\end{proof}
Taking the symbol of this equation and applying $\phi$ yields:
\begin{equation}  \label{eq:flag:math:bethe}
(-1)^{w_i(k_i-1)}q_i(P^i_a)^{\ell_i} \left(\prod_{b}\frac{P^i_a}{P^i_b} \right)^{w_i}=
\frac{\prod_{b=1}^{k_{i+1}} 1-\frac{P^i_a}{P^{i+1}_b}}{\prod_{b=1}^{k_{i-1}} 1-\frac{P^{i-1}_b}{P^i_a}}.
\end{equation}

\begin{cor}\label{flag:correspondence}
The symbols of the difference operators annihilating $\widehat{I}_{Ab}^{\ell,w}$, after replacing the $P^i_a$ with the corresponding variables $X^{(i)}_a$, are identical to the Coulomb branch equations \eqref{eq:flag:phys:bethe}, for specific choices of Chern-Simons terms. 
\end{cor}
As in the case of the Grassmannian, we expect that these equations correspond to quantum $K$-theory ring relations in the following sense:

\begin{conj}
Let the symmetric group $S_{k_i}$ act on the variables $X^{(i)}_a$ by permuting them. 
Then, for $\ell$ lying in the geometric window, after interpreting the variables $X^{(i)}_a$ as exponentials of Chern roots of $\mathcal{S}_i$,  $\prod_i S_{k_i}$-invariant combinations of equations \eqref{eq:flag:phys:bethe} generate the ideal of relations in $QK^{\ell,w}_T(Fl)$, the quantum $K$-ring of the flag manifold, twisted by the classes $(\det^{-\ell_i},\mathcal{S}_i)$, and $\left((Eu_\lambda)^{w_i-1},\mathcal{A}_i\right)$, for each $i$. 

\end{conj}

\begin{remark}
    For $\ell=(0,\dots,0)$, $w=(1,\dots,1)$ this conjecture was made by Gu-Mihalcea-Sharpe-Xu-Zhang-Zou in \cite{Gu:2023tcv} and proved by the first author in \cite{HK2}. 
    For other presentations with generators and relations of the equivariant quantum $K$-theory of partial flag manifolds, see for example \cite{MNS:QKpres,AHMOX}.
\end{remark}

\subsubsection{Mathematical geometric window}
As with Grassmannians, when discussing geometric windows, we restrict each $w_i$ to its default value of 1. 

The range of the mathematical geometric window for a general choice of level structure remains unknown. However in the case where all levels are 0 except for one level $\ell_i$, a portion of the range was calculated by Yan in \cite[Remark 16]{XY}. 

\begin{thm}[Yan]
Choose a level $\ell_i$.
If all other $\ell_j=0$, then the $I$-function for the flag with level structure is equal to the small $J$-function for $\ell$ in the range:
$$0\leq \ell_i\leq \frac{\prod_{a=i}^s (k_{a+1}-k_a)}{(n-i+1)}.$$
\end{thm}
This bound in general is not sharp, as can be seen by specializing to the case of the Grassmannian.

\subsubsection{Flag duality}

As with Grassmannians, when discussing geometric windows, we restrict each $w_i$ to its default value of 1 for simplicity. Our results can be extended to the case of different values as well. As with the case of Grassmannians, we can apply level-rank duality to give an isomorphism $\mu$ between $Fl(k_1,\dots,k_s;\mathbb{C}^n)$ and $Fl^*:=Fl(\breve{k}_1,\breve{k}_2,\dots,\breve{k}_s;\mathbb{C}^n)$,  $\breve{k}_i := n - k_{s+1-i}$, obtained by using the explicit isomorphism $\mathbb{C}^n\to (\mathbb{C}^{n})^*$ defined by sending the standard basis into the dual basis, in opposite order.

As with Grassmannians, we denote the tautological bundles on $Fl^*$ by $\breve{\mathcal{S}}_i$, their Chern roots by $\breve{P}_j^i$. Let the quantum parameter associated to $\breve{\mathcal{S}}_i$ be denoted $\breve{q}_i$.

Then, as with Grassmannians,
$\mu^*\breve{\mathcal{S}}_i=(\mathbb{C}^n/\mathcal S_{s+1-i})^*$. After imposing $\mu^*\Lambda_{n+1-r}^{-1}=\Lambda_r$, $\mu$ defines an isomorphism between $K_T(Fl)$ and $K_T(Fl^*)$.

As in the case of Grassmannians, choosing levels $\ell=(\ell_1,\dots,\ell_s)$ for the bundles $\mathcal{S}_1,\dots,\mathcal{S}_s$ is related to choosing the levels $-\ell := (-\ell_s,\dots, -\ell_1)$ for the duals of the corresponding  
tautological bundles $\breve{\mathcal S}_{i}^*$ on $Fl^*$. 
Yan's duality theorem \cite[Thm. 8]{XY} has the following consequence in this case: 

\begin{cor}\label{flag:level_corr}
For $\ell$ such that $I_{Fl}^\ell$ and $I_{Fl^*}^{-\ell}$ are both the small $J$-functions of their respective theories (which is testable by Theorem \ref{thm:I=J}), we have
$$I_{Fl}^\ell=\mu^*I_{Fl^*}^{-\ell} .$$

\end{cor}
\begin{proof}
    This is a direct consequence of a theorem of Yan \cite[Thm.~8]{XY}, noting that what Yan refers to as ``explicit descriptions,'' are in fact equal to the $I$-functions of the respective theories. 
\end{proof}

\begin{thm}
As in the case of Grassmannians, the equality of $I$-functions in Corollary \ref{flag:level_corr} corresponds to the physical IR duality with Coulomb branch equations given by Equation \eqref{eq:phys-flag-dual}.
\end{thm}

\begin{proof}
The proof is essentially the same as in the case of the Grassmannian, so we will be brief.
The corresponding abelianized $I$-function of $Fl^*$ is denoted $\widehat{I}_{Ab^*}^{-\ell}$, and has the following form:
\begin{eqnarray}
\widehat{I}_{Ab^*}^{-\ell}
& := & (1-\tau)\sum_{d\geq 0} \prod_{i,a} (\breve{q}^i_{a})^{d^i_{a}}(\breve{P}^i_a)^{-\ell_{s+1-i} d^i_a}\tau^{-\ell_{s+1-i}\binom{d^i_a+1}{2}}
\\
& & \times
\frac{\prod_{i=1}^s\prod_{a\neq b}^{1\leq a,b\leq \breve{k}_i}\widehat{\prod}_{l=1}^{d^i_{a}-d^i_{b}}(1-\lambda\frac{\breve{P}^i_{a}}{\breve{P}^i_{b}}\tau^l)}{\prod_{i=1}^{s-1}\prod_{1\leq b\leq \breve{k}_{i+1}}^{1\leq a\leq \breve{k}_i}\widehat{\prod}_{l=1}^{d^i_{a}-d^{i+1}_b}(1-\frac{\breve{P}^i_{a}}{\breve{P}^{i+1}_b}\tau^l)\cdot\prod_{1\leq b\leq n}^{1\leq a\leq \breve{k}_s}\widehat{\prod}_{l=1}^{d^s_{a}}(1-\frac{\breve{P}^s_{a}}{\Lambda^{-1}_{b}}\tau^l)}.
\nonumber 
\end{eqnarray}

By essentially the same arguments as the non-dualized flag, it satisfies the equations:
\begin{eqnarray}
\lefteqn{
\prod_{b\neq a} \left(1-\tau\lambda\frac{\breve{P}^i_b}{\breve{P}^i_a}\tau^{\breve{q}^i_b\partial_{\breve{q}^i_b}-\breve{q}^i_a\partial_{\breve{q}^i_a}}\right)
\prod_b \left(1-\frac{\breve{P}^i_a}{\breve{P}^{i+1}_b}\tau^{\breve{q}^i_a\partial_{\breve{q}^i_a}-\breve{q}^{i+1}_b\partial_{\breve{q}^{i+1}_b}}\right)\widehat{I}_{Ab^*}^{-\ell}
}
\\
& = &
\left(\breve{P}^i_a \tau^{\breve{q}^i_a\partial_{\breve{q}^i_a}} \right)^{-\ell_{s+1-i}}\breve{q}^i_a
\prod_b\left(1-\frac{\breve{P}^{i-1}_b}{\breve{P}^{i}_a}\tau^{\breve{q}^{i-1}_b\partial_{\breve{q}^{i-1}_b}-\breve{q}^i_a\partial_{\breve{q}^i_a}} \right)
\prod_{b\neq a} \left(1-\tau\lambda\frac{\breve{P}_a^i}{\breve{P}_b^{i}} \tau^{\breve{q}^i_a \partial_{\breve{q}^i_a}-\breve{q}^i_b \partial_{\breve{q}^i_b}} \right)\widehat{I}_{Ab^*}^{-\ell}.
\nonumber
\end{eqnarray}

Specializing parameters and taking symbols yield the equations (for each $i,a)$:
\begin{equation}
\prod_b \left(1-\frac{\breve{P}^i_a}{\breve{P}^{i+1}_b}\right)= (-1)^{\breve{k}_i-1}
\left(\breve{P}^i_a \right)^{-\ell_{s+1-i}}\breve{q}_i
\prod_b\left(1-\frac{\breve{P}^{i-1}_b}{\breve{P}^{i}_a} \right)
\prod_{b} \frac{\breve{P}^{i}_a}{\breve{P}^i_b}.
\end{equation}

This equation is equivalent to \eqref{eq:phys-flag-dual} after taking reciprocals and applying the following substitutions:
$$\breve{P}^i_a\mapsto (\breve{X}^i_a)^{-1}, \quad \breve{q}_i\mapsto \breve{q}_{ {\rm phys}, i}^{-1} .$$
\end{proof}

\subsubsection{Mixed levels}

As with the examples of toric varieties, the Chern-Simons levels do not just correspond to the $U(1)$ factors appearing in the GIT description, but to pairs of such factors.

We claim that turning on a general level corresponds to choosing a Ruan-Zhang level structure associated to the bundle $\mathcal{S}_i\otimes  \mathcal{S}_j$. Since the Ruan-Zhang levels are taken on the abelianization, the lift of $\mathcal{S}_i$ is $\bigoplus_a P^{i}_a$. 

We describe a choice of Ruan-Zhang in this way. Choose 
integers $\ell_{ij}$ indexed by unordered pairs of integers from $1$ to $s$. 
Each $\ell_{ij}$, for $i\neq j$, is the Ruan-Zhang level for 
the bundle $\mathcal{S}_i\otimes  \mathcal{S}_j$; for $i=j$, we use the bundle $\mathcal{S}_i$. 

The resulting abelianized $I$-function is
\begin{align}
\widehat I^{\ell,w}_{Ab}
={}&
(1-\tau)
\sum_{d\geq 0}
\prod_{i=1}^{s}\prod_{a=1}^{k_i}
(q_a^i)^{d_a^i}
(P_a^i)^{\ell_i d_a^i}
\tau^{\ell_i\binom{d_a^i}{2}}
\nonumber\\
&\times
\prod_{1\leq i<j\leq s}
\prod_{a=1}^{k_i}
\prod_{b=1}^{k_j}
(P_a^iP_b^j)^{\ell_{ij}(d_a^i+d_b^j)}
\tau^{\ell_{ij}\binom{d_a^i+d_b^j}{2}}
\nonumber\\
&\times
\frac{
\displaystyle
\prod_{i=1}^{s}
\prod_{\substack{1\leq a,b\leq k_i\\a\neq b}}
\widehat{\prod}_{l=1}^{d_a^i-d_b^i}
\left(
1-\lambda\frac{P_a^i}{P_b^i}\tau^l
\right)^{w_i}
}{
\displaystyle
\prod_{i=1}^{s-1}
\prod_{\substack{1\leq a\leq k_i\\1\leq b\leq k_{i+1}}}
\widehat{\prod}_{l=1}^{d_a^i-d_b^{i+1}}
\left(
1-\frac{P_a^i}{P_b^{i+1}}\tau^l
\right)
\prod_{\substack{1\leq a\leq k_s\\1\leq b\leq n}}
\widehat{\prod}_{l=1}^{d_a^s}
\left(
1-\frac{P_a^s}{\Lambda_b}\tau^l
\right)
}.
\end{align}

After adding the logarithmic prefactors we have the following system of $\tau$-difference equations, obtained by essentially the same methods as the toric variety and the Grassmannian case:
\begin{thm}
For $1\leq i\leq s$ and $1\leq a\leq k_i$, the abelianized
$I$-function satisfies
\begin{align}
&q_a^i
\left(
\tau^{q_a^i\partial_{q_a^i}}
\right)^{\ell_i}
\prod_{j\neq i}
\prod_{b=1}^{k_j}
\left(
\tau^{
q_a^i\partial_{q_a^i}
+
q_b^j\partial_{q_b^j}
}
\right)^{\ell_{ij}}
\nonumber\\
&\qquad\times
\frac{
\displaystyle
\prod_{b=1}^{k_{i-1}}
\left(
1-
\tau^{
q_b^{i-1}\partial_{q_b^{i-1}}
-
q_a^i\partial_{q_a^i}
}
\right)
\prod_{\substack{1\leq b\leq k_i\\ b\neq a}}
\left(
1-\tau\lambda\,
\tau^{
q_a^i\partial_{q_a^i}
-
q_b^i\partial_{q_b^i}
}
\right)^{w_i}
}{
\displaystyle
\prod_{\substack{1\leq b\leq k_i\\ b\neq a}}
\left(
1-\tau\lambda\,
\tau^{
q_b^i\partial_{q_b^i}
-
q_a^i\partial_{q_a^i}
}
\right)^{w_i}
\prod_{b=1}^{k_{i+1}}
\left(
1-
\tau^{
q_a^i\partial_{q_a^i}
-
q_b^{i+1}\partial_{q_b^{i+1}}
}
\right)
}
\widehat{\mathbf{I}}^{\ell,w}_{Ab}
=
\widehat{\mathbf{I}}^{\ell,w}_{Ab}.
\end{align}
\end{thm}
After specializing $q_a^i\to q_i, \lambda\to 1$, the symbols of these equations become:
\begin{equation} \label{eq:su-flag:math:bethe-mix}
(-1)^{w_i(k_i-1)}
q_i
(P_a^i)^{\ell_i+\sum_{j\neq i}k_j\ell_{ij}}
\prod_{j\neq i}\prod_{b=1}^{k_j}
(P_b^j)^{\ell_{ij}}
\prod_{\substack{1\leq b\leq k_i\\ b\neq a}}
\left(
\frac{P_a^i}{P_b^i}
\right)^{w_i}
=
\frac{
\displaystyle
\prod_{b=1}^{k_{i+1}}
\left(
1-\frac{P_a^i}{P_b^{i+1}}
\right)
}{
\displaystyle
\prod_{b=1}^{k_{i-1}}
\left(
1-\frac{P_b^{i-1}}{P_a^i}
\right)
}.
\end{equation}

\subsection{Dictionary between physics and mathematics}

Next, we extract and check the prediction of conjecture~(\ref{eq:level-conj}) for this case.
The conjecture relates physical Chern-Simons levels 
to an $N \times N$ matrix of twistings,
for
\begin{equation}
    N \: = \: \sum_{i=1}^s k_i,
\end{equation}
modulo the Weyl group of $U(k_1) \times \cdots \times U(k_s)$.

We denote elements of the $N \times N$ matrix of twistings by $\ell_{(i,a_i), (j,b_j)}$,
where $i, j \in \{1, \cdots, s\}$, $a_i \in \{1, \cdots, k_i\}$, and
$b_j \in \{1, \cdots, k_j\}$.
Comparing expression~(\ref{eq:su-flag:math:bethe-mix}) for the chiral ring of the flag manifold
to that for the toric case~(\ref{full-level-symbtoric}), we identify
\begin{equation}
    \ell_{(i,a_i), (j,b_j)} \: = \: \left\{
    \begin{array}{cl}
    \ell_{ij} & i \neq j,
    \\
     - w_i & i=j, \: a_i \neq b_i,
    \end{array}
    \right.
\end{equation}
and for the last case, we find an equation
\begin{equation}
    \ell_{(i,a_i), (i,a_i)} \: + \:
    \sum_{j\neq i} \sum_{b_j = 1}^{k_j} \ell_{(i,a_i), (j,b_j)} 
    \: + \:
    \sum_{b_i \neq a_i} \ell_{(i,a_i), (i,b_i)}
    \: = \: \ell_{ii} + \sum_{j\neq i} k_j\ell_{ij} + (k_i-1) w_i,
\end{equation}
which implies
\begin{equation}
     \ell_{(i,a_i), (i,a_i)} \: = \:
     \ell_{ii}  \: + \: 2 (k_i-1) w_i,
\end{equation}
where $\ell_{ii}$ is also denoted $\ell_i$.

Now, let us turn to the prediction~(\ref{eq:level-conj}).  In terms of an $N \times N$ matrix of Chern-Simons levels $\kappa_{\rm eff}^{(i,a_i), (j,b_j)}$,
the prediction is that
\begin{equation}
    \kappa_{\rm eff}^{(i,a_i),(j,b_j)} \: = \:
    \left\{ \begin{array}{cl}
    \sum_{m=1}^s \sum_{b_m=1}^{k_m} \ell_{(i,a_i),(m,b_m)} & i=j, \: \: \: a_i = b_i,
    \\
    \ell_{(i,a_i),(j,b_j)} & i \neq j \mbox{ or } a_i \neq b_j,
    \end{array} \right.
\end{equation}
which simplifies to
\begin{equation} \label{eq:flag:predict}
     \kappa_{\rm eff}^{(i,a_i),(j,b_j)} \: = \:
    \left\{ \begin{array}{cl}
    \ell_{ij} & i \neq j,
    \\
    - w_i & i=j, \: \: \: a_i \neq b_j,
    \\
    \ell_i + \sum_{j \neq i} k_j \ell_{ij} + (k_i-1) w_i & i = j \mbox{ and }a_i = b_j.
    \end{array} \right.
\end{equation}

Now, from the definition~(\ref{eq:keff:genl}) and the expansions~(\ref{eqn:CS-levels}), (\ref{eq:CS-levels:flag:comp1}),
(\ref{eq:CS-levels:flag:comp2}), 
\begin{eqnarray}
\kappa_{\rm eff}^{(i,a_i), (j,b_j)} & = & \kappa_{\rm eff}^{(j,b_j),(i,a_i)} \: = \:
\delta^{ij} \left( \kappa^{(i)}_{a_i b_j} +
\frac{1}{2}\sum_{\nu=1}^{\dim V} \rho_{\nu}^{(i,a_i)} \rho_{\nu}^{(j,b_j)} \right) + \kappa^{(i,j)}_{\rm eff} 
\\
& = & 
\delta^{ij} \left( \kappa^{(i)}_{a_i, b_j} + \frac{1}{2} \left( k_{i-1} + k_{i+1} \right) \delta_{a_i, b_i} \right) \: + \: \kappa_{\rm eff}^{(i,j)},
\\
& = &
\delta^{ij} \left(
\left( \kappa^{(i)}_{SU(k_i)} +  \frac{1}{2} \left( k_{i-1} + k_{i+1} \right) \right) \delta_{a_i,b_i} + \frac{\kappa^{(i)}_{U(1)} - \kappa^{(i)}_{SU(k_i)}}{k_i}
\right)  \: + \: \kappa_{\rm eff}^{(i,j)}. 
\nonumber
\end{eqnarray}

Comparing to the prediction~(\ref{eq:flag:predict}), we see
\begin{equation}  \label{eq:flags:dict1}
     \kappa_{\rm eff}^{(i,j)} \: = \: \ell_{ij},
     \: \: \:
      \frac{\kappa^{(i)}_{U(1)} - \kappa^{(i)}_{SU(k_i)}}{k_i} \: = \:  -w_i,
\end{equation}
and
\begin{equation}
    \kappa^{(i)}_{SU(k_i)} +  \frac{1}{2} \left( k_{i-1} + k_{i+1} \right) + 
     \frac{\kappa^{(i)}_{U(1)} - \kappa^{(i)}_{SU(k_i)}}{k_i}
     \: = \:
      \ell_i + \sum_{j \neq i} k_j \ell_{ij} + (k_i-1) w_i.
\end{equation}
After some algebra, we find the equation above simplifies to
\begin{equation} \label{eq:flags:dict2}
    \ell_i \: = \: \kappa_{U(1)}^{(i)} \: - \: \sum_{j \neq i} k_j \kappa_{\rm eff}^{(i,j)} \: + \:
     \frac{1}{2} \left( k_{i-1} + k_{i+1} \right),
\end{equation}
which together with~(\ref{eq:flags:dict1}) completes the prediction for the dictionary between physical Chern-Simons levels
$\kappa_{U(1)}^{(i)}$, $\kappa_{SU(k_i)}^{(i)}$, $\kappa_{\rm eff}^{(i,j)}$
and mathematical twistings $\ell_i$, $\ell_{ij}$, $w_i$.

This generalizes the Grassmannian dictionary~(\ref{eq:grass:dictionary}), and when mixed levels vanish, is of the form described
in the introduction in~(\ref{eq:univ:grfl}).

Comparing the Coulomb branch equations~(\ref{eq:flag:phys:bethe}) (for which a choice of gauge-flavor Chern-Simons levels was made) to the symbols of $\tau$-difference operators~(\ref{eq:su-flag:math:bethe-mix}), we see immediately that with the dictionary~(\ref{eq:flags:dict1}), (\ref{eq:flags:dict2}) above, the ring relations match, as expected, (with $q_{{\rm phys}, i} = (-1)^{(w_i-1)(k_i-1)}q_i$ as for Grassmannians,) confirming conjecture~(\ref{eq:level-conj}) in this case.  Ordinary quantum K theory is recovered in the special case that $\ell_i =0$, $\ell_{ij} = 0$, and $w_i = +1$.

\section{Predictions for gerbes}   \label{sect:gerbe}

\subsection{Conjecture}

In principle, as outlined in the introduction, similar ideas to those we have discussed here should apply to any Fano GIT quotient, including quotients realizing Deligne-Mumford stacks.  One set of examples we briefly outline here are Deligne-Mumford gerbes.  These are, in essence, $BG$ bundles over a space for finite $G$.  

Specifically, we consider ${\mathbb Z}_k$ gerbes over a space $X$.  These can be described as quotients $[P/{\mathbb C}^{\times}]$, where $P$ is a principal ${\mathbb C}^{\times}$ bundle on $X$ and the ${\mathbb C}^{\times}$ acts on the fibers of $P$ by $k$ times rotations,
leaving a trivially-acting ${\mathbb Z}_k \subset {\mathbb C}^{\times}$.
These can be described as  $B {\mathbb Z}_k$ bundles over $X$, as the classifying stack $B{\mathbb Z}_k = [{\rm pt}/{\mathbb Z}_k]$.  Physical realizations of such Deligne-Mumford stacks, their subtleties, and realizations in GLSMs 
were discussed in \cite{Pantev:2005rh,Pantev:2005zs,Pantev:2005wj}.

In \cite[section 4]{Gu:2021yek}, \cite[section 3]{Gu:2021beo}, \cite{Sharpe:2024ujm}, it was argued that
the quantum $K$-theory of a banded ${\mathbb Z}_k$ gerbe on a space $X$ should be $k^2$ copies of the quantum $K$-theory of $X$.  As noted there, this was because 
\begin{itemize}
    \item the three-dimensional theory has a (one-form) $B {\mathbb Z}_k$ symmetry, which reduces to a $B{\mathbb Z}_k \times {\mathbb Z}_k$ symmetry in two dimensions, and
    \item for the choice of Chern-Simons levels that matches ordinary quantum $K$-theory, there is no self-'t Hooft anomaly in the three-dimensional $B{\mathbb Z}_k$ symmetry.
\end{itemize}
The $B {\mathbb Z}_k$ factor gives rise to a decomposition of the effective two-dimensional theory of the Wilson line motions, the ${\mathbb Z}_k$ factor gives superselection sectors which reduce to decomposition in the low-energy limit in which quantum $K$-theory is computed, and the lack of a 't Hooft anomaly ensures that these two mechanisms operate independently.

However, a key step above \cite{Sharpe:2024ujm} was the point that there is no self-'t Hooft anomaly.  This is specific to the Chern-Simons levels arising for ordinary quantum $K$-theory.  For more general Chern-Simons levels, there will be such an anomaly.

For more general Chern-Simons levels, in general there will be a 't Hooft anomaly, so for general levels we only expect one level of decomposition, arising from the two-dimensional $B {\mathbb Z}_k$ symmetry.  In other words, we predict that the twisted quantum $K$-theory of a banded gerbe will, in general\footnote{In special cases, of course, there can be coincidences in which a subgroup may be anomaly-free, and so a larger multiplicity may arise.}, only give $k$ copies of the twisted quantum $K$-theory of the underlying space, rather than $k^2$ copies.

We summarize this analysis in the following
\begin{conj} \label{conj-gerbes}
\begin{itemize}
Let ${\mathfrak X}$ denote a ${\mathbb Z}_k$ gerbe on the Fano GIT quotient $X$.
    \item The Chern-Simons level corresponds to a Ruan-Zhang level $(E,\ell)$ on ${\mathfrak X}$ for some bundle $E \rightarrow \mathfrak{X}$. 
    \item For the choice of Ruan-Zhang / Chern-Simons level that corresponds to ordinary quantum $K$-theory, 
    \begin{equation}
        QK\left( {\mathfrak X} \right) \: = \: QK\left( \coprod_{k^2} X \right), 
    \end{equation}
    \item For general choices of level,
    \begin{equation}
        QK^\ell\left( {\mathfrak X} \right) \: = \: QK^\ell\left( \coprod_{k} X \right).
    \end{equation}
\end{itemize}
\end{conj}

To give some further insight into this conjecture, next we examine it in two examples.
First, we compute the ordinary quantum $K$-theory of a classifying stack in mathematics, and prove that the quantum $K$-theory ring has the form described in the conjecture above.
Second, we will use physical Coulomb branch methods to study the ordinary and twisted quantum $K$-theory rings of gerbes on ${\mathbb P}^{n-1}$, following the methods of 
\cite{Gu:2021yek,Gu:2021beo,Sharpe:2024ujm} which studied ordinary quantum $K$-theory on gerbes.

\subsection{Example: $B {\mathbb Z}_k$}

Next, we prove mathematically that the ordinary quantum $K$-theory of a ${\mathbb Z}_k$ gerbe is $k^2$ copies of that of the underlying space for the special case $B\mathbb{Z}_k$. We first begin by describing the ordinary $K$-theory of $B\mathbb{Z}_k$.

Recall that $K(B\mathbb{Z}_k)\cong {\rm Rep}(\mathbb{Z}_k)$, essentially tautologically, since a vector bundle over $B\mathbb{Z}_k$ is exactly a $\mathbb{Z}_k$-representation. Since $\mathbb{Z}_k$ is a finite abelian group, ${\rm Rep}(\mathbb{Z}_k)$ is isomorphic to its character ring, via the map sending a 1-dimensional representation to its character (which is just the representation itself). 

However, the $K$-theoretic Poincar\'e pairing, defined on representations as $(a,b)_K:=\chi(B\mathbb{Z}_k; a\otimes b)$, is not compatible with the standard pairing on characters. We denote this pairing by $(\cdot ,\cdot)_\chi$, defined by making the irreducible characters (which we denote $\chi_0,\dots\chi_{k-1}$, where $\chi_i$ is the character of the 1-dimensional representation with weight $i$) into an orthonormal basis. 
To see this, note that $K$-theoretic pushforward to a point corresponds to taking the dimension of the space of $\mathbb{Z}_k$ invariants, so for a nontrivial irreducible character $\chi_j$ such that $2j \not\equiv 0\mod k$, we have $(\chi_j,\chi_j)_K=0$, whereas $(\chi_j,\chi_j)_\chi=1$.

Since both pairings appear in our analysis, we will let $K(B\mathbb{Z}_k)$ denote the ring with the $K$-theory pairing, and $Rep(\mathbb{Z}_k)$ denote the ring with the character pairing. To verify part of the conjecture for $B\mathbb{Z}_k$, we need the following observation about  $K(B\mathbb{Z}_k)$:

\begin{lem}
Over $\mathbb{C}$, $(K(B\mathbb{Z}_k),(,)_K)\cong \bigoplus_{i=0}^{k-1} (K({\rm pt}),\frac{1}{k}(,)_K)$, as algebras with pairing.   The notation $\frac{1}{k}(,)_K$ means the $K$-theoretic pairing scaled by $\frac{1}{k}$.

\end{lem}
\begin{proof}
     Choose a generator $g$ of $\mathbb{Z}_k$ and label the $i$th component by $g^i$. The isomorphism is given by $V\mapsto (tr_{g^i}(V))_{i=0}^{k-1}$. 
\end{proof}

Recall that for an orbifold $\mathcal{X}$, the quantum cohomology ring is not a deformation of $H(\mathcal{X})$, but it rather deforms the Chen-Ruan cohomology of $\mathcal{X}$, which as a module is $H(I\mathcal{X})$, where $I\mathcal{X}$ is the inertia stack of $\mathcal{X}$ (the product, grading, and pairing, come from counts of degree $0$-orbicurves inside $X$, and are not standard ones in $H(I\mathcal{X})$).
Similarly, Zhang in \cite[Def.~2.13]{orb} introduced the orbifold quantum $K$-ring (whose definition we will not recall here), which does not deform $K(\mathcal{X})$, but rather is a deformation of the  \emph{full orbifold $K$-ring} introduced by Jarvis et al in \cite[Def.~9.3]{jarvis}, denoted $\mathsf{K}_{\mathrm{orb}}(\mathcal{X})$. 

To define the product on the ring $\mathsf{K}_{\mathrm{orb}}(\mathcal{X})$, we introduce the following notation. Let $II\mathcal{X}$ denote the double inertia stack of $\mathcal{X}$, its points are given by pairs $(x,(g_1,g_2))$, where $x$ is a point of $\mathcal{X}$, and $g_1,g_2$ are commuting isotropy elements. The automorphisms of such a pair are automorphisms of $x$ that commute with both group elements. There are maps $\pi_{i,j}: II\mathcal{X}\to I\mathcal{X}$ sending $(x,(g_1,g_2))\to (x,g_1^ig_2^j)$. The full orbifold $K$-ring $\mathsf{K}_{\mathrm{orb}}(\mathcal{X})$ is defined as follows:
\begin{itemize}
\item The underlying module is $\mathsf{K}_{\mathrm{orb}}(\mathcal{X})\cong K(I\mathcal{X})$.
 
\item The Poincar\'e pairing is the Mukai pairing, defined by $(a,b)_M=\chi(I\mathcal{X}; a\otimes i^*(b))$, where $i$ is the inversion map on $I\mathcal{X}$, sending $(x,g)$ to $(x,g^{-1})$. 

\item The orbifold product is given by $\mathcal{F}*\mathcal{G}:=\pi_{1,1*}(\pi_{1,0}^*\mathcal{F}\pi_{0,1}^*\mathcal{G}\otimes \Lambda_{-1}(\tilde{\mathscr{R}}))$, where $\tilde{\mathscr{R}}$ is the orbifold obstruction bundle defined by Jarvis-Kaufman-Kimura in \cite[Def.~9.2]{jarvis}. 
\end{itemize}

The proposition below confirms a special case of the Conjecture~\ref{conj-gerbes}.
\begin{prop}
For $\mathcal{X}=B\mathbb{Z}_k= [{\rm pt}/\mathbb{Z}_k]$, we have
$$(QK(\mathcal{X}),(,)_M) \cong (\bigoplus_{k^2} QK({\rm pt}), \frac{1}{k^2}(,)_{K}) \/,$$
as $\mathbb{C}$-algebras with pairing. Here $(,)_M$ is the Mukai pairing. 

\end{prop}

\begin{proof}
  
Since there are no quantum parameters, $QK(B\mathbb{Z}_k)=\mathsf{K}_{\mathrm{orb}}(B\mathbb{Z}_k)$. We will describe this ring explicitly below.

The inertia stack $IB\mathbb{Z}_k$ is $k$ copies of $B\mathbb{Z}_k$, each labeled by an element $g \in \mathbb{Z}_k$. The double inertia stack is $(k^2)$ copies of $B\mathbb{Z}_k$, labelled by pairs of elements $(g_1,g_2)$. Each component has structure sheaf denoted $\mathcal{O}_{g_1,g_2}$.
In this case, the bundle $\tilde{\mathscr{R}}=0$, so it does not appear in the formula for multiplication. For $V\in {\rm Rep}(\mathbb{Z}_k)$, $\pi_{i,j*} (V \otimes \mathcal{O}_{g_1,g_2}) = V \otimes \mathcal{O}_{g_1^ig_2^j}$.

We define $\mathbb{C}_{\mathbb{Z}_k}$ to be the vector space of class functions of $\mathbb{Z}_k$, where $1_g$ denotes the characteristic function of the element $g$.  The above discussion means that  $\mathsf{K}_{\mathrm{orb}}(\mathcal{X})$ has the following description:

\begin{itemize}
\item The underlying vector space is $K(B\mathbb{Z}_k)\otimes \mathbb{C}_{\mathbb{Z}_k}$, where $a\otimes 1_g$ is identified with the class $a \otimes \mathcal{O}_g$.

\item  The Mukai pairing is determined on simple tensors by 
$$(a\otimes f,b\otimes g)_M=(a,b)_K(f,g)_{conv} \quad \/,$$ 
where $(,)_K$ is the usual Poincar\'e pairing on $K(B\mathbb{Z}_k)$, and $(,)_{conv}$ is the convolution pairing on ${\mathbb C}_{\mathbb{Z}_k}$ given by $(a,b)_{conv}:=\sum_{g} a_gb_{g^{-1}}$.

\item Multiplication is determined by:
\begin{eqnarray}
( V_1 \otimes \mathcal{O}_{g_1}) * ( V_2 \otimes \mathcal{O}_{g_2})
& = & \pi_{1,1*}\left( \left( \sum_g ( V_1 \otimes \mathcal{O}_{g_1,g}) \right) \cdot \left( \sum_g ( V_2 \otimes \mathcal{O}_{g,g_2}) \right) \right),
\nonumber \\
& = & \pi_{1,1*} ( (V_1 \cdot V_2) \otimes \mathcal{O}_{g_1,g_2} )
= (V_1 \cdot V_2) \otimes \mathcal{O}_{g_1g_2}.
\end{eqnarray}

\end{itemize}

The identification of this space with $k^2$ copies of $K({\rm pt})$ works via the discrete Fourier transform, which assigns to a class function $a$ a representation $F(a)$, which is defined by
$$F(a)=\sum_{i} \sum_g a_g\chi_{i}(g^{-1}) \chi_i.$$
 This has the following properties:
\begin{itemize}
    \item The map $a\mapsto F(a)$ is an isomorphism with inverse defined in the following way. Let $V=\sum_i V_i \chi_i$, then $F^{-1}(V) = (1/k) \sum_{g} \sum_i V_i \chi_i(g) 1_g$.
    \item There is an equality of pairings 
    \[ (a,b)_{conv} = (1/k)(\hat{a},\hat{b})_\chi \/, \] 
    where the first pairing is the convolution pairing on 
    $\mathbb{C}_{\mathbb{Z}_k}$, and the second is the character pairing on $Rep(\mathbb{Z}_k)$. 
\end{itemize}

This result follows from general properties of the discrete Fourier transform, or can be seen explicitly in this context in \cite[p. 19]{GivIX} (the normalization used is slightly different from the one above).  
This means we can identify the module underlying full orbifold $K$-ring with ${\rm Rep}(\mathbb{Z}_k)\otimes {\rm Rep}(\mathbb{Z}_k)$, where the pairing on the second factor is scaled by $\frac{1}{k}$. The subspace generated by $\chi_i\otimes \chi_j$ is isomorphic (including at the level of the Poincar\'e pairing) to $K({\rm pt})$. It remains to check that this decomposition respects multiplication.

Choose $a,b\in K(B\mathbb{Z}_k)$ and characters $\chi_{i},\chi_{j}$, the definition of the multiplication in the full orbifold $K$-ring implies
$$(a\otimes \chi_{i})*(b\otimes \chi_{j})=\sum_g \sum_{g_1g_2=g} \frac{1}{k^2}\chi_i(g_1)\chi_j(g_2)(a*b)\mathcal{O}_g .$$
We observe that
$$\sum_{g_1g_2=g} \frac{1}{k^2}\chi_i(g_1)\chi_j(g_2)=\frac{1}{k}\chi_i(g)\delta_{i,j},$$ 
so the product is
$$(a\otimes \chi_{i})*(b\otimes \chi_{j})=(ab\otimes \chi_i \delta_{i,j}).$$
This means that the full orbifold $K$-ring decomposes (as a ring) as a direct sum 
$$\bigoplus_{i=0}^{k-1} K(B\mathbb{Z}_k)\otimes {\rm span}(\chi_i)\cong \bigoplus_{i,j=0}^{k-1} K({\rm pt}) .$$

The factor $\frac{1}{k^2}$ in front of the pairing comes from the two $\frac{1}{k}$-factors, one from the convolution pairing, the other from the decomposition of $K(B\mathbb{Z}_k)$. 
\end{proof}

\subsection{Example: Gerbes on projective spaces}

One can also see the conjecture explicitly in Coulomb branch equations in physics, as has been discussed in the case of
ordinary quantum $K$-theory in \cite{Gu:2021yek,Gu:2021beo,Sharpe:2024ujm}.  We review here a simple example, a ${\mathbb Z}_k$ gerbe over the projective space ${\mathbb P}^{n-1}$.
In the special case $n=1$, this will reduce to the result of the previous subsection.

First, we describe the gerbe over the projective space ${\mathbb P}^{n-1}$ as a GIT quotient of the vector space ${\mathbb C}^{n+1} = {\rm Spec}\, {\mathbb C}[x_1, \cdots, x_n, y]$ by $({\mathbb C}^{\times})^2$, with weights as listed in the table below:
\begin{center}
\begin{tabular}{cc}
$x_1,\cdots, x_n$ & $y$ \\ \hline
$1$ & $-m$ \\
$0$ & $k$
\end{tabular}
\end{center}
This is equivalent to a ${\mathbb C}^{\times}$ quotient of the total space of a ${\mathbb C}^{\times}$ bundle $L^{\times} \rightarrow {\mathbb P}^{n-1}$, where $L^{\times}$ is the line bundle $L = {\cal O}(-m)$ minus its zero section.

The effective twisted superpotential is given by
\begin{eqnarray}
    {\cal W} & = & \sum_{a, b = 1}^2 \frac{\kappa^{ab}}{2} (\ln X_a ) (\ln X_b) \: + \:
    \sum_{a=1}^2 (\ln q_a) (\ln X_a)
    \\
    & & \: + \: (n)\left[ {\rm Li}_2(X_1) + \frac{1}{4} (\ln X_1)^2
    \right] \: + \:
    \left[ {\rm Li}_2( X_1^{-m} X_2^k ) + \frac{1}{4} \left( \ln X_1^{-m} X_2^k \right)^2 \right],
    \nonumber
\end{eqnarray}
where
\begin{eqnarray}
    \kappa^{11} & = & -\frac{1}{2} (n + m^2) \: + \: \Delta \kappa^{11},
    \\
    \kappa^{12} & = & + \frac{m k}{2} \: + \: \Delta \kappa^{12} \:=\:  \kappa^{21}, 
    \\
    \kappa^{22} & = & - \frac{k^2}{2} \: + \: \Delta \kappa^{22}.
\end{eqnarray}
In the expressions for the Chern-Simons levels $\kappa^{ab}$ above, the first term gives the level that would match ordinary quantum $K$-theory, and the $\Delta \kappa^{ab}$ reflect differences that could arise if one considers more general Chern-Simons levels.

By computing the critical loci of the superpotential above, one finds the Coulomb branch equations 
\begin{eqnarray}
    \left(1 - X_1 \right)^n \, \left( 1 - X_1^{-m} X_2^k \right)^{-m} & = & q_1 X_1^{\Delta \kappa^{11}} X_2^{\Delta \kappa^{12}},
    \\
    \left( 1 - X_1^{-m} X_2^k \right)^k & = & q_2 X_1^{\Delta \kappa^{12}} X_2^{\Delta \kappa^{22}}.
\end{eqnarray}

Now, to compare to the conjecture above, specialize to ordinary quantum $K$-theory, for which
$\Delta \kappa^{ab} = 0$.  Following \cite[section 4.1]{Sharpe:2024ujm}, if we define $z = 1 - X_1^{-m} X_2^k$, then the ring relations can be written in the form
\begin{eqnarray}
    \left( 1 - X_1 \right)^n & = & q_1 z^m,
    \\
    z^k & = & q_2.
\end{eqnarray}
By redefining $z$ to absorb $q_2$, the second relation can be described as saying that $z$ is a $k$th root of unity, giving rise to $k$ copies of the ordinary quantum $K$-theory relation for ${\mathbb P}^{n-1}$.  Furthermore, since $z$ itself is determined by $X_2^k$, there is another $k$-fold choice in solutions, giving altogether $k^2$ copies of the quantum $K$-theory ring of ${\mathbb P}^{n-1}$.

For more general Chern-Simons levels, however, although there still appears to be one level of
$k$-fold ambiguity arising from the second relation, the interpretation otherwise appears more obscure.

\section{Conclusions}

In this paper, we have proposed a dictionary between two twistings of quantum $K$-theory:
\begin{itemize}
    \item In physical realizations, by modifications of the Chern-Simons level, within the `geometric window' of allowed values which do not introduce additional topological vacua and spoil the geometric interpretation,
    \item and in mathematics, by twistings \cite{cg,Tonita,wallcrossing} that include, as special cases, Ruan-Zhang level structures \cite{rz,rwz}.  
\end{itemize}
Under the dictionary, the physical geometric window corresponds 
to the mathematical geometric window, which in turn can be 
formulated in terms of mirror-triviality, i.e., the equality of 
the $I$ and $J$ functions of the relevant twisted theories, and the Coulomb branch equations correspond to symbols of operators annihilating the abelianized $I$-function.

We have described this dictionary for the cases of projective spaces, 
Grassmannians, and flag manifolds -- essentially, everything that we 
can currently compute in mathematics and which also has a GLSM realization.

Our predictions have natural generalizations to arbitrary GIT quotients that represent promising opportunities for further study.

\section{Acknowledgments}

We would like to thank C.~Closset, W.~Gu, and W.~Xu for useful discussions.  L.M. was partially supported by NSF grant DMS-2152294, and gratefully acknowledges the support of Charles Simonyi Endowment, which provided funding for the membership at the Institute of Advanced Study during the 2024-25 Special Year in `Algebraic and Geometric Combinatorics'. E.S.~was partially supported by NSF grant PHY-2310588.

\end{document}